\documentclass[useAMS,usenatbib]{mn2e}

\usepackage{epsfig}
\usepackage{longtable}
\usepackage{times}

\usepackage{amsmath}
\usepackage{amssymb}

\def \xmm {$XMM$-$Newton$}

\def \nustar {$NuSTAR$}

\def \degmark {^\circ}

\def \ferg {erg cm$^{-2}$ s$^{-1}$}
\def \hcm {\hbox {\ifmmode $ atom cm$^{-2}\else atom cm$^{-2}$\fi}}

\def \mdot {\dot{M}_{w}}

\def \apj {ApJ}

\def \apjl {ApJL}
\def \apjs {ApJS}
\def \aap {A\&A}

\def \apss {A\&Sp Sc.}

\def \pasj {PASJ}

\def \mnras {MNRAS}
\def \na {New Astronomy}
\def \nat {Nature}

\def \ssr {Space Science Reviews}
\def \iaucirc {IAU Circ.}
\def \aapr {A\&A Rev.}

\newcommand{\be}{\begin{equation}}

\newcommand{\ee}{\end{equation}}

\newcommand{\msun}{{~M_{\odot}}}

\newcommand\integral{\textsl{INTEGRAL}}

\newcommand\inte{\textsl{INTEGRAL}}

\title[HMXBs: an INTEGRAL overview]{An \integral\ overview of High Mass X--ray Binaries: classes or transitions?
}
\author[Sidoli \& Paizis]{L.~Sidoli,$^{1}$\thanks{E-mail: lara.sidoli@inaf.it} and A.~Paizis$^{1}$ \\
$^{1}$INAF, Istituto di Astrofisica Spaziale e Fisica Cosmica, Via E.\ Bassini 15,   I-20133 Milano,  Italy   \\
}

\begin{document}

\date{Accepted 2018 August 30. Received 2018 August 22; in original form 2018 May 16}

\pagerange{\pageref{firstpage}--\pageref{lastpage}} \pubyear{2018}

\maketitle

\label{firstpage}

\begin{abstract}
We analyzed in a systematic way the public \inte\ observations spanning from December 2002 to September 2016, 
to investigate the hard X-ray properties of about 60 High Mass X-ray Binaries (HMXBs). 
We considered both persistent and transient sources, hosting
either a Be star (Be/XRBs) or a blue supergiant companion (SgHMXBs, including
Supergiant Fast X-ray Transients, SFXTs), a neutron star or a black hole. 
\inte\ X-ray light curves (18--50 keV), sampled at a bin time of about 2~ks,
were extracted for all HMXBs to derive the cumulative distribution of their
hard X-ray luminosity, their duty cycle, the range of variability of their hard X-ray luminosity.  
This allowed us to obtain an overall and quantitative characterization of the long-term hard X--ray activity of the 
HMXBs in our sample.
Putting the phenomenology observed with \inte\ into context with
other known source properties (e.g. orbital parameters, pulsar spin periods)
together with observational constraints coming from softer X-rays (1-10 keV),
enabled the investigation of the way  the different HMXB sub-classes behave (and sometimes
overlap). For given source properties, the different sub-classes of massive binaries seem to cluster in a suggestive way.
However, for what concerns supergiant systems (SgHMXBs versus SFXTs), 
several sources with intermediate properties exist, suggesting a smooth transition between the two sub-classes.   
\end{abstract}


        %%%%%%%%%%%%%%%%%%%%%%%%%%%%%%%%%%%%%%%%%%%%%%%%%%%%%%%%%
        \section{Introduction} \label{intro}
        %%%%%%%%%%%%%%%%%%%%%%%%%%%%%%%%%%%%%%%%%%%%%%%%%%%%%%%%%

In recent years, the field of High Mass X-ray Binaries (HMXBs) revitalized thanks to the discoveries performed  
by the INTErnational Gamma-Ray Astrophysics Laboratory (\inte; \citealt{Winkler2003, Winkler2011}).
The number of HMXBs with supergiant companions tripled \citep{Krivonos2012, Bird2016} and new sub-classes of massive X--ray binaries were discovered:
the so-called ``highly obscured sources'' (the first was IGR~J16318--4545, \citealt{Courvoisier2003}) 
and the Supergiant Fast X--ray Transients (SFXTs; \citealt{Sguera2005, Negueruela2006}), a new sub-class of transient X--ray sources
associated with early-type supergiant stars. 
This latter new type of X--ray binaries (XRB) in particular, with its bright and short flaring emission, 
posed into question the standard picture of X--ray emission in supergiant HMXBs: Bondi-Hoyle accretion onto 
the neutron star (NS) of matter gravitationally captured from the wind of the massive donor.
Many physical mechanisms were proposed to explain the SFXT X-ray flares:
gated mechanisms (magnetic or centrifugal barriers; \citealt{Grebenev2007, Bozzo2008}) 
that halt the accretion most of the time (depending on the values of the NS spin period and surface magnetic field),
or the quasi-spherical settling accretion regime,
where the phenomenology of persistent supergiant HMXBs versus SFXTs is explained by different cooling regimes of the gravitationally 
captured matter above the NS magnetosphere \citep{Shakura2012, Shakura2014, Shakura2017}.
The characteristics of the (clumpy) wind from the companion are expected to play a fundamental role, acting as an external condition for the accreting pulsar.
However, the properties of the companion wind (clump density, velocity  and magnetic field) are largely unknown in HMXBs 
(see \citealt{Martinez-Nunez2017} for a comprehensive review
of the supergiant winds and their impact onto the accretion in HMXBs hosting NSs). 

A huge observational effort towards both the donor star and the X--ray source has been pushed forward in the recent years by many authors 
on specific  interesting sources (see \citealt{Sidoli2017review} for a review and references therein).
A few years ago we performed a systematic analysis of all publicly available \inte\ observations of all known SFXTs \citep{Paizis2014}, compared with
three classical persistent HMXBs.
We found that the cumulative luminosity distributions of the SFXT X--ray flares were power-law-like,
while the luminosity distributions of the three persistent HMXBs were more appropriately represented by 
log-normal functions. 
This suggests a different approach in obtaining observational information about the accretion mechanism 
and the production of the X-ray flares in SFXTs \citep{Paizis2014, Shakura2014, Sidoli2016}.

In this work, we widen our investigation to all HMXBs observed by \inte\ in 14 years,  
to obtain a global view of all sub-classes of HMXBs in accretion (gamma-ray binaries and colliding-wind massive binaries are excluded). 
We report here a full characterization of their hard X--ray phenomenology by means of time-integrated quantities, 
insensitive to  the temporal evolution  of their long-term X--ray light curves.
The extraction of cumulative luminosity distributions in the 18--50\,keV energy range allows us to obtain
quantitative information that offers a comprehensive picture of their behaviour at high energy, 
also put into context of other source properties.

The {\em paper} is structured as follows: 
in Sect.~\ref{anita} we outline the \inte\ archive, the data analysis and the selection of our sample of HMXBs;
in Sect.~\ref{sec:results} we report on our \inte\ results, the cumulative luminosity distributions and the source duty cycles; 
in Sect.~\ref{literature} we describe the search through the literature of other interesting properties for all members of our sample, producing 
an updated catalogue for the HMXBs of our interest;
in Sect.~\ref{sec:disc} we discuss the results on both the cumulative luminosity distributions and the hard X--ray results
into context of published properties.
Our conclusions are given in Sect.~\ref{sec:concl}.

%%%%%%%%%%%%%%%%%%%%%%%%%%%%%%%%%%%%%%%%%%%%%%%%%%%%%%%%%%%%%%%%%%%%%%%%%%%%%%%%%%%%%%%%%%%%%%%%%%%%%%%%%%%%%%%%%%%%%%%%%%%%%%%%%%%%%%%%%%%%%%%%%%%%%%%%%%%%%%%%%%%%%%%%%%%%%%%%%%%%%%%%%
%%%%%%%%%%%%%%%%%%%%%%%%%%%%%%%%%%%%%%%%%%%%%%%%%%%%%%%%%%%%%%%%%%%%%%%%%%%%%%%%%%%%%%%%%%%%%%%%%%%%%%%%%%%%%%%%%%%%%%%%%%%%%%%%%%%%%%%%%%%%%%%%%%%%%%%%%%%%%%%%%%%%%%%%%%%%%%%%%%%%%%%%%
%%%%-------------------------------------------------------------------------------------------------------------------------------------------------------------------------------------
 \begin{table*}
 \centering
  \caption{\inte\ IBIS/ISGRI results (18--50\,keV) for our sample of HMXBs. }
 \begin{tabular}{lccrcccr}
\hline
   Name$^{a}$ & Field exposure  &   Source activity   & Duty Cycle        &  Average L$_{X}$        &    Min  L$_{X}$         &    Max  L$_{X}$    &  Dynamic range        \\
	& (s)	           &      (s)             & (per cent)        &    (erg~s$^{-1}$)       &   (erg~s$^{-1}$)        &   (erg~s$^{-1}$)   &  Max  L$_{X}$ /  Min  L$_{X}$ \\	
\hline
\multicolumn{8}{c}{SgHMXBs} \cr
SMC X-1$^{*}$   &     3.7E+06    &     1.8E+06          &    49.05       &     1.7E+38   &     7.0E+37   &     3.0E+38  &        4.25   \\
3A 0114+650 &     7.0E+06   &     1.0E+06   &       14.63   &     2.1E+36   &     6.7E+35   &     1.2E+37  &       17.31   \\
Vela X-1$^{*}$ &     5.2E+06   &     4.1E+06   &       79.22   &     1.3E+36   &     6.1E+34   &     1.0E+37  &      165.74   \\
1E 1145.1-6141 &     6.4E+06   &     2.0E+06   &       31.95   &     3.0E+36   &     1.0E+36   &     1.4E+37  &       13.86   \\
GX 301-2 &     6.2E+06   &     5.8E+06   &       94.47   &     2.8E+36   &     2.3E+35   &     3.0E+37  &      127.46   \\
H 1538-522$^{*}$ &     8.4E+06   &     2.5E+06   &       30.15   &     9.2E+35   &     4.2E+35   &     4.3E+36  &       10.19   \\
IGR J16207-5129 &     9.2E+06   &     3.6E+04   &        0.39   &     1.1E+36   &     6.5E+35   &     2.1E+36  &        3.24   \\
IGR J16320-4751 &     1.0E+07   &     2.2E+06   &       21.32   &     5.9E+35   &     1.8E+35   &     2.5E+36  &       14.01   \\
IGR J16393-4643 &     1.1E+07   &     4.2E+04   &        0.40   &     3.4E+36   &     1.5E+36   &     6.2E+36  &        4.02   \\
OAO 1657-415$^{*}$ &     1.2E+07   &     7.3E+06   &       59.78   &     5.8E+36   &     1.0E+36   &     2.0E+37  &       19.04   \\
4U 1700-377$^{*}$ &     1.6E+07   &     1.2E+07   &       73.09   &     1.1E+36   &     8.5E+34   &     9.5E+36  &      111.43   \\
IGR J17252-3616$^{*}$ &     2.3E+07   &     1.1E+06   &        4.65   &     2.9E+36   &     1.3E+36   &     9.5E+36  &        7.34   \\
IGR J18027-2016$^{*}$ &     2.1E+07   &     1.1E+05   &        0.54   &     5.2E+36   &     2.7E+36   &     1.3E+37  &        4.92   \\
IGR J18214-1318 &     8.9E+06   &     5.7E+03   &        0.06   &     3.4E+36   &     2.0E+36   &     5.1E+36  &        2.60   \\
XTE J1855-026$^{*}$ &     8.4E+06   &     8.1E+05   &        9.64   &     4.2E+36   &     2.1E+36   &     1.9E+37  &        9.05   \\
H 1907+097 &     8.7E+06   &     1.8E+06   &       20.13   &     8.1E+35   &     3.6E+35   &     4.8E+36  &       13.48   \\
4U 1909+07 &     8.7E+06   &     2.2E+06   &       24.84   &     7.1E+35   &     3.5E+35   &     3.5E+36  &        9.96   \\
IGR J19140+0951 &     8.8E+06   &     1.2E+06   &       14.18   &     5.2E+35   &     2.0E+35   &     3.3E+36  &       16.36   \\
\hline
\multicolumn{8}{c}{giant HMXBs} \cr
LMC X-4$^{*}$ &     7.8E+06   &     3.7E+06   &       47.23   &     1.2E+38   &     4.4E+37   &     2.3E+38  &        5.11   \\
Cen X-3$^{*}$ &     6.1E+06   &     3.8E+06   &       62.79   &     4.0E+36   &     5.8E+35   &     1.4E+37  &       24.34   \\
\hline
\multicolumn{8}{c}{SFXTs} \cr
IGR J08408-4503 &     5.4E+06   &     4.9E+03   &        0.09   &     3.0E+35   &     2.7E+35   &     3.4E+35  &        1.27   \\
IGR J11215-5952 &     6.0E+06   &     3.9E+04   &        0.64   &     1.6E+36   &     8.6E+35   &     5.1E+36  &        5.92   \\
IGR J16328-4726 &     1.0E+07   &     2.9E+04   &        0.28   &     1.7E+36   &     7.9E+35   &     3.6E+36  &        4.56   \\
IGR J16418-4532$^{*}$ &     1.1E+07   &     1.3E+05   &        1.22   &     6.1E+36   &     3.0E+36   &     2.1E+37  &        7.07   \\
IGR J16465-4507 &     1.1E+07   &     1.9E+04   &        0.18   &     2.9E+36   &     2.1E+36   &     4.9E+36  &        2.29   \\
IGR J16479-4514$^{*}$ &     1.1E+07   &     3.6E+05   &        3.33   &     3.6E+35   &     1.6E+35   &     1.2E+36  &        7.47   \\
IGR J17354-3255 &     2.5E+07   &     3.5E+03   &        0.01   &     3.0E+36   &     1.9E+36   &     4.6E+36  &        2.38   \\
XTE J1739-302  &     2.5E+07   &     2.2E+05   &        0.89   &     4.8E+35   &     1.5E+35   &     1.5E+36  &        9.78   \\
IGR J17544-2619 &     2.5E+07   &     1.3E+05   &        0.54   &     5.6E+35   &     2.0E+35   &     5.3E+36  &       25.99   \\
SAX J1818.6-1703 &     1.1E+07   &     9.1E+04   &        0.81   &     2.9E+35   &     1.1E+35   &     1.4E+36  &       12.24   \\
IGR J18410-0535 &     7.9E+06   &     4.2E+04   &        0.53   &     3.8E+35   &     2.2E+35   &     8.5E+35  &        3.87   \\
IGR J18450-0435 &     8.0E+06   &     2.8E+04   &        0.35   &     1.5E+36   &     1.0E+36   &     2.7E+36  &        2.57   \\
IGR J18483-0311 &     8.2E+06   &     3.8E+05   &        4.63   &     5.2E+35   &     2.3E+35   &     1.5E+36  &        6.63   \\
\hline
\multicolumn{8}{c}{Be/XRBs} \cr
H 0115+634 &     7.1E+06   &     6.8E+05   &        9.55   &     1.5E+37   &     9.9E+35   &     3.7E+37  &       36.90   \\
RX J0146.9+6121 &     4.7E+06   &     5.0E+03   &        0.11   &     1.1E+35   &     1.0E+35   &     1.1E+35  &        1.11   \\
EXO 0331+530 &     2.8E+06   &     7.1E+05   &       25.10   &     2.4E+37   &     9.4E+35   &     6.4E+37  &       68.01   \\
X Per &     2.1E+06   &     1.6E+06   &       76.96   &     2.5E+34   &     1.2E+34   &     6.3E+34  &        5.19   \\
1A 0535+262 &     7.2E+06   &     8.9E+05   &       12.34   &     4.4E+36   &     9.9E+34   &     1.5E+37  &      153.68   \\
GRO J1008-57 &     4.8E+06   &     4.3E+05   &        8.87   &     2.4E+36   &     5.6E+35   &     9.8E+36  &       17.68   \\
4U 1036-56 &     5.1E+06   &     1.8E+04   &        0.35   &     7.5E+35   &     5.2E+35   &     9.9E+35  &        1.88   \\
IGR J11305-6256 &     6.3E+06   &     2.5E+04   &        0.41   &     1.9E+35   &     1.3E+35   &     3.0E+35  &        2.30   \\
IGR J11435-6109 &     6.4E+06   &     1.7E+05   &        2.68   &     1.4E+36   &     9.9E+35   &     2.1E+36  &        2.08   \\
H 1145-619 &     6.3E+06   &     6.8E+04   &        1.07   &     1.2E+35   &     8.1E+34   &     1.9E+35  &        2.34   \\
XTE J1543-568 &     7.0E+06   &     9.9E+03   &        0.14   &     2.7E+36   &     2.1E+36   &     3.3E+36  &        1.58   \\
AX J1749.1-2733 &     2.5E+07   &     4.2E+04   &        0.17   &     8.1E+36   &     4.1E+36   &     1.3E+37  &        3.11   \\
GRO J1750-27 &     2.5E+07   &     1.2E+06   &        4.88   &     2.9E+37   &     7.8E+36   &     8.5E+37  &       10.95   \\
AX J1820.5-1434 &     9.5E+06   &     1.4E+04   &        0.15   &     2.1E+36   &     1.7E+36   &     3.8E+36  &        2.24   \\
Ginga 1843+009 &     9.3E+06   &     3.2E+05   &        3.39   &     5.8E+36   &     1.5E+36   &     1.6E+37  &       10.33   \\
XTE J1858+034 &     1.0E+07   &     5.3E+05   &        5.34   &     8.8E+36   &     1.7E+36   &     1.5E+37  &        8.91   \\
4U 1901+03 &     9.9E+06   &     1.0E+06   &       10.44   &     1.2E+37   &     2.7E+36   &     1.9E+37  &        6.81   \\
KS 1947+300 &     9.9E+06   &     9.3E+05   &        9.41   &     6.8E+36   &     1.7E+36   &     1.7E+37  &       10.20   \\
EXO 2030+375 &     1.2E+07   &     3.5E+06   &       28.99   &     7.8E+36   &     9.2E+35   &     6.4E+37  &       69.34   \\
SAX J2103.5+4545 &     8.1E+06   &     9.0E+05   &       11.14   &     2.0E+36   &     7.5E+35   &     8.4E+36  &       11.23   \\
\hline
\end{tabular}
\label{tab:inte}
\end{table*}

%%%%%%%%%%%%%%-------------------------------------  
\setcounter{table}{0}
\begin{table*}
 \centering
 \caption{\inte\ IBIS/ISGRI results (18--50\,keV) for our sample of HMXBs.  {\it (continued).}}
  \begin{tabular}{lccrcccr}
%%----------------------------------------------------------------------------------------
\hline
   Name & Field exposure  &   Source activity   & Duty Cycle        &  Average L$_{X}$        &    Min  L$_{X}$         &    Max  L$_{X}$    &  Dynamic range        \\
	& (s)	           &      (s)             & (per cent)        &    (erg~s$^{-1}$)       &   (erg~s$^{-1}$)        &   (erg~s$^{-1}$)   &  Max  L$_{X}$ /  Min  L$_{X}$ \\	
\hline
%%----------------------------------------------------------------------------------------
\multicolumn{8}{c}{Other HMXBs} \cr
IGR J16318-4848 &     9.9E+06   &     3.5E+06   &       35.17   &     7.4E+35   &     2.2E+35   &     4.1E+36  &       18.73   \\
3A 2206+543 &     4.7E+06   &     3.0E+05   &        6.41   &     2.5E+35   &     1.3E+35   &     6.7E+35  &        4.99   \\
Cyg X-1 &     1.1E+07   &     1.1E+07   &       99.88   &     2.5E+36   &     9.1E+34   &     8.2E+36  &       89.51   \\
Cyg X-3 &     1.2E+07   &     1.1E+07   &       93.49   &     1.0E+37   &     1.3E+36   &     2.5E+37  &       19.82   \\
SS 433$^{*}$ &     9.2E+06   &     1.4E+06   &       14.97   &     8.5E+35   &     4.5E+35   &     1.9E+36  &        4.25   \\
\hline
\multicolumn{8}{c}{Symbiotic binary} \cr
XTE J1743-363 &     2.4E+07   &     3.1E+04   &        0.13   &     1.1E+36   &     9.2E+35   &     1.5E+36  &        1.62   \\
\hline
\end{tabular}
\flushleft{
 $^{a}$Sources marked with an asterisk are eclipsing binaries.
}
\end{table*}
%%%%-------------------------------------------------------------------------------------------------------------------------------------------------------------------------------------

%%%%%%%%%%%%%%%%%%%%%%%%%%%%%%%%%%%%%%%%%%%%%%%%%%%%%%%%%%%%%%%%%%%%
\section{\textit{INTEGRAL}: data analysis and the selection of the HMXB sample} \label{anita}
%%%%%%%%%%%%%%%%%%%%%%%%%%%%%%%%%%%%%%%%%%%%%%%%%%%%%%%%%%%%%%%%%%%%

{\it INTEGRAL} \citep{Winkler2003, Winkler2011} is a medium size ESA mission launched in October 2002. 
It comprises two main gamma-ray instruments - the spectrometer SPI \cite[15\,keV -- 10\,MeV][]{Vedrenne2003} 
and the imager IBIS \cite[][]{Ubertini2003} -, two X--ray monitors, JEM--X \cite[4--35\,keV][]{Lund2003} 
and an optical camera, OMC  \cite[500--600\,nm,][]{Mashesse2003}. 
The imager IBIS consists of two layers, IBIS/ISGRI \cite[15\,keV -- 1\,MeV, ][]{Lebrun2003} 
and IBIS/PICsIT \cite[0.175--10\,MeV, ][]{Labanti2003}. 

The long-standing activity of \integral, its wide field of view (hereafter, FoV, 30$^{\circ}$x30$^{\circ}$ for the imager) 
together with the good angular resolution in hard X-rays, 
essential in the crowded Galactic Plane and Centre regions, make \integral\ a very powerful instrument 
to study the wide sample of hard X-ray sources as a class.

\subsection{The \textit{INTEGRAL} archive} \label{arc}

We have built an \integral\ archive, 
described in detail in \citet{Paizis2013}, providing  scientific results for IBIS/ISGRI public data.
The scripts used to build the archive are online\footnote{http://www.iasf-milano.inaf.it/$\sim$ada/GOLIA.html}. 
This database approach enabled an easy access to the long-term behaviour of a large sample of sources in the hard X-ray range, 
allowing us to explore the properties of a few members of the class of HMXBs, 
e.g. \citet[][]{Paizis2014}, \citet[][]{Sidoli2015vela, Sidoli2016b, Shakura2014, Shakura2015}. 
We investigate here a much larger sample of data and of HMXBs observed by \inte (see below for the selection of the sample).

We have recently developed a second generation archive, named ANITA (A New InTegral Archive), with important 
improved hardware and software issues/performances. Details of ANITA are given in \citet{Paizis2016} and in Paizis et al. (in prep.).
For completeness, we recall here only the basic information relative to the results used in this work. 

The public data analyzed span a period of fourteen years (December 2002 - September 2016, revolution 0026 to 1729), 
for a total of $\sim$130000 Science Windows (pointings, hereafter ScW, with a duration of $\sim$2\,ks each). 
This corresponds to a total exposure time of $\sim$200\,Msec. 
The standard \integral\ Off-line Scientific Analysis (OSA) version 10.2 software package has been used for the data analysis. 
For each ScW, images together with the list of detected sources are created in four 
energy bands: 18--50, 50--100, 100--150, 22--50\,keV. 
Hereafter we focus only on the results obtained in the 18--50\,keV energy band that, 
notwithstanding the degradation of the low energy threshold of IBIS/ISGRI with time, 
provides the best detection statistics among all the aforementioned energy bands. 
The 22--50\,keV band provides consistent results/trends with what is shown here. 
In this work, ScWs with durations smaller than 1\,ks have been ignored, to avoid non-standard snapshot contaminations. 
Furthermore, only ScWs with the sources within 12$^{\circ}$ from the centre of the FoV have been considered. 

Our selection (ScW exposure $>$1\,ks and off axis angle $<$12$^{\circ}$) provides the source final \emph{field exposure}. 
Within this field exposure, a source is considered detected, i.e. \emph{active} at the ScW level in the 18--50\,keV band, 
when the detection significance is higher than 5\,sigma in the ScW. 
We thus obtain the source duty cycle, DC$_{18-50~keV}$, as the percentage of detections at ScW level: activity time over field exposure time.
The HMXBs that have been detected (i.e. are active) at ScW level  are the object of this work. 
For each source in our sample, we list in Table~\ref{tab:inte} the \inte\ \emph{field exposure} (in units of seconds; col.~2),
the  \emph{source activity} (col.~3) and the source DC$_{18-50~keV}$ in the energy range 18--50 keV  (col.~4).

 \inte\ has performed a very thorough coverage of the Galactic plane where HMXBs reside and 
the selected $<$12$^{\circ}$ radius region results in a wide serendipitous source activity sampling. 
Indeed, unless an external bias is introduced, the sources will display a given flux regardless their position in the FoV.  
 However, there are some important aspects that need to be taken into account. 
First of all, we note that using only the fully coded FoV (source $\la$5$^{\circ}$ from the centre, 
where the sensitivity is maximal and approximately constant) results in a much smaller data sample, 
wasting an important fraction of the FoV coverage. 
Furthermore, it introduces an important bias in the DC$_{18-50~keV}$ estimation: Be X--ray Transients (Be/XRTs, hereafter), 
for which several target of opportunities (ToOs) have been performed, end up with a much higher DC$_{18-50~keV}$ when only 
a small portion of the FoV is considered. This is because ToOs are on-source observations. 
We note that no ToO observation has ever been performed on SFXTs except one, on IGR~J11215-5952, during which
the source was not detected \citep{Sidoli2007}. Indeed, the SFXT flares detectable by IBIS/ISGRI
are typically shorter than the ToO reaction time of the satellite. 
In a similar way, phase resolved observations, 
and in general time-constrained ones, provide an artificially 
high DC$_{18-50~keV}$ when only the fully coded FoV is considered. 
These aspects are clearly mitigated using a larger portion of the FoV.
This effect is maximum in the case of three Be/XRTs (H~0115$+$634, EXO~0331$+$530 and 1A~0535$+$262) that have been the target of several ToOs. 
Indeed, for these sources the DC$_{18−-50\,keV}$ decreases, respectively, from about 37\%, 56\% and 25\% when the 5$^{\circ}$ fully coded FoV is considered, 
to 10\%, 25\% and 12\% when the region up to  12$^{\circ}$ is included. Hence, while moving from the on-target ToO observations 
(with the source in the fully coded FoV) to the partially coded FoV (that includes serendipitous observations) the overall DC$_{18−-50\,keV}$ decreases. 
This does not mean that the bias introduced by the on-target ToO observations is completely eliminated using a 12$^{\circ}$ radius selection, but it is highly diluted: indeed, a large serendipitous coverage area is included in the duty cycle estimate. Unfortunately, a bigger fraction of the IBIS/ISGRI detector cannot be considered: the usage of the whole FoV would result in a DC$_{18−-50\,keV}$ of about 8~per cent, 19~per cent and 12~per cent for the three sources, respectively, but the (noisy) outer regions of the detector ($>$12$^{\circ}$) produce spurious detections, hence these latter percentages are highly unreliable.

On the other side, the inclusion of a portion of the partially coded FoV ($>$5$^{\circ}$, 
where the sensitivity decreases towards the instrumental edge) includes a detection bias in the 
sources that emit mostly at the IBIS/ISGRI detection threshold. 
Indeed, a - constant - source that is detected at 5\,sigma in the totally coded FoV 
will be undetected in the partially coded FoV (the further out, the lower the detection significance). 
We have seen, however, that with our choice of FoV ($<$12$^{\circ}$) this bias is important 
only for three sources: X~Per (a Be/XRB that is seen as a persistent source given its proximity - 0.8\,kpc), 
H~1538-522 (SgHMXB) and 1E~1145.1-6141 (SgHMXB). Indeed in these sources the DC$_{18-50~keV}$ shifts dramatically 
to lower values when a radius of 12$^{\circ}$ is considered. 
For example X~Per would have a DC$_{18-50~keV}\sim$90 instead of $\sim$77 per cent if only data $<$5$^{\circ}$ were to be considered. 
Similarly for the case of H~1538-522 and 1E~1145.1-6141. 
Hence, we may be losing detections and in reality these sources may be more persistent than what considered here. 
However, as it will appear clear throughout the text, this has no influence on our  conclusions: notwithstanding 
the detection loss, the sources are already amongst the ones with the highest DC$_{18-50~keV}$ and their shift 
towards even higher DC values does not change our final considerations.
For the remaining sources of the sample this bias has no effect, i.e. when active the sources are bright enough 
to be detected within 12$^{\circ}$ from the centre.

Finally, in order to quantify the effect of the IBIS/ISGRI degradation throughout the years, 
the overall Crab results will also be shown. In this respect, the Crab can be considered as the ``point spread function'' of our results.  

We believe that our selection criteria are the best trade-off currently available to maximize the 
scientific output of the \integral\ archive, while minimizing observational biases.

%%%%%%%%%%%%%%%%%%%%%%%%%%%%%%%%%%%%%%%%%%%%%%%%%%%%%%%%%%%%%%%%%%%%%%%%%%%%%%%%%%%%%%%%%%%%%%%%%%%%%%%%%%%%%%%%%%%%%%%%%%%%%%%%%%%%%%%%%%%%%%%%%%%%%%%%%%%%%%%%%%%%%%%%%%%%%%%%%
\begin{table*}
 \centering
  \caption{Summary of the source properties extracted from the literature for the different types of HMXBs of our sample.}
 \begin{tabular}{lcccccccl}
\hline
Name & Dist  &  P$_{orb}$  &  ecc     &     P$_{spin}$   &  F$_{min}$ (1--10\,keV)   &  F$_{max}$ (1--10\,keV)           &  DR$_{1-10~keV}$                          &   References$^{a}$        \\
     & (kpc) &  (d)        &          &        (s)        & (\ferg)                &   (\ferg)               &   (F$_{max}$/F$_{min}$)                &                     \\
%----------------------------------------------------------------------------------------
\hline
\multicolumn{9}{c}{SgHMXBs} \cr
%-----------------------------------------------------
SMC X-1    &      61$\pm{1}$      &       3.89 &     0.0002 &       0.71 &       2.80E-10 &       2.15E-09 &          7.7   &  14, 4, 4, 4, 7, 11 \\    
3A 0114+650 &       7.2$\pm{3.6}$  &      11.60 &     0.18   &      10008 &       1.00E-10 &          $-$   &            $-$ &  1, 1, 4, 6, 12, $-$ \\      
Vela X-1    &       1.9$\pm{0.2}$  &       8.96 &     0.09   &     283.5  &       7.50E-10 &       1.27E-09 &          1.7   &  7, 1, 5, 1, 7, 7 \\ 
1E 1145.1-6141 &     8.5$\pm{1.5}$ &      14.36 &     0.2    &     296.6  &       6.00E-11 &          $-$   &            $-$ &  1, 1, 4, 1, 13,  $-$   \\
GX 301-2    &       3.5$\pm{0.5}$  &      41.49 &     0.46    &     675   &       1.20E-09 &       3.10E-09 &          2.6   &  1, 1, 4, 1, 7, 7 \\ 
H 1538-522  &       5.0$\pm{0.5}$  &       3.73 &     0.18   &     526.8  &       4.10E-12 &       1.35E-10 &         32.9   &  1, 1, 5, 1, 7, 7 \\ 
IGR J16207-5129 &   6.0$\pm{3.5}$  &       9.73 &      $-$   &      $-$   &       4.30E-12 &       4.00E-11 &          9.3   &  1, 1, $-$, $-$, 15, 16   \\  
IGR J16320-4751     &  3.5         &       8.99 &      $-$   &    1309    &       5.30E-11 &       7.80E-10 &         14.7   &  1, 1, $-$, 1, 7, 7 \\ 
IGR J16393-4643    &  10.6         &       4.24 & 0.0$^{b}$  &     912    &       2.80E-11 &       9.00E-11 &          3.2   &  18, 1, $-$, 1, 19, 20 \\ 
OAO 1657-415 &      7.1$\pm{1.3}$  &      10.45 &     0.103  &      38.2  &      8.00E-11 &       8.00E-10 &             10 &  21, 1, 5, 1, 6, 6 \\  
4U 1700-377    & 1.9$\pm{0.3}$     &       3.41 & 0.0$^{b}$  &    $-$     &       3.00E-10 &       3.60E-09 &          12.0   &  1, 1, $-$,  $-$, 7, 7 \\
IGR J17252-3616 &     8$\pm{2}$   &       9.74 &     0.0    &     413.89 &       4.00E-12 &       6.90E-11 &         17.3   &  21, 1, 5, 1, 6, 7 \\
IGR J18027-2016 & 12.4$\pm{0.1}$  &       4.57 &      $-$   &     139.61 &       4.00E-12 &       1.50E-09 &          375   &  1, 5, $-$, 1, 22, 22 \\   
IGR J18214-1318 &      8$\pm{2}$  &        $-$ &      $-$   &    $-$     &       5.80E-11 &         $-$    &          $-$   &  21, $-$, $-$,  $-$, 23, $-$ \\  
XTE J1855-026   &   10.8$\pm{1.0}$&       6.07 &     0.04   &     360.7  &       1.10E-10 &         $-$   &           $-$   &  24, 1, 5, 1, 6, $-$\\ 
H 1907+097  &       5.0$\pm{1.2}$ &       8.36 &     0.28   &     437.5  &       9.70E-12 &       5.30E-09 &         546   &  1, 1, 6, 1, 72, 6 \\ 
4U 1909+07  &       4.85$\pm{0.5}$ &       4.4  &     0.021  &     605    &       2.60E-10 &       3.00E-09 &         11.5   &   1, 1, 6, 1, 25, 26 \\  
IGR J19140+0951 &    3.6$\pm{1.0}$ &      13.55 &     $-$    &      5900  &       2.60E-12 &       2.00E-09 &        769   &  27, 1, $-$, 28, 29, 30 \\       
%----------------------------------------------------------------------------------------
\hline
%%--------
\multicolumn{9}{c}{giant HMXBs} \cr
LMC X-4      &      50$\pm{1}$   &       1.4  &     0.006  &      13.5  &       7.40E-11 &       2.50E-10 &          3.4 &  68, 5, 5, 5, 12, 12   \\  
Cen X-3      &       6.5$\pm{1.5}$ &       2.09 &  0.0$^{b}$ &       4.82 &       4.40E-10 &       2.20E-09 &          5.0 &  3, 3, $-$, 3, 7, 69    \\  
%----------------------------------------------------------------------------------------
\hline
\multicolumn{9}{c}{SFXTs} \cr
%\hline
IGR J08408-4503 &       2.7          &       9.54 &  0.63  &    $-$   &       4.00E-13 &       2.70E-09 &       6750   &   2, 2, 2, 2, 8, 9    \\     
IGR J11215-5952 &      7.0$\pm{1.0}$ &     164.6  &  $>$0.8 &   187   &    $<$5.00E-13 &       2.40E-10 &      $>$480   &    2, 2, 2, 2, 10, 17 \\   
IGR J16328-4726 &      7.2$\pm{0.3}$ &      10.07 &  $-$   &    $-$   &       5.00E-12 &       1.50E-09 &        300   &    2, 2, 2, 2, 31, 32 \\     
IGR J16418-4532 &      13            &       3.75 &  0.0   &    1212  &       6.50E-12 &       2.00E-09 &        308 &    2, 2, 2, 2, 33 , 32 \\    
IGR J16465-4507 &      9.5$\pm{5.7}$ &      30.24 &  $-$   &     228  &       4.00E-12 &       1.50E-10 &         37.5 &    2, 2, 2, 2,  7, 32  \\   
IGR J16479-4514 &       2.8$\pm{1.7}$ &       3.32 &  0.0   &    $-$   &       6.00E-12 &       1.00E-08 &       1667 &    2, 2, 2, 2, 34, 35 \\   
IGR J17354-3255 &       8.5           &       8.45 &  $-$   &    $-$   &    $<$1.40E-13 &       1.30E-10 &    $>$929 &    2, 2, 2, 2, 36, 37  \\   
XTE J1739-302   &       2.7           &      51.47 &  $-$   &    $-$   &    $<$2.50E-12 &       5.10E-09 &     $>$2040 &    2, 2, 2, 2, 32, 9\\    
IGR J17544-2619 &       3.0$\pm{0.2}$ &       4.93 &  $<$0.4&   71.49(?)$^{b}$  &       6.00E-14 &       1.00E-07 & 1.67$\times10^6$ &  2, 2, 2, 2, 38, 39 \\   
SAX J1818.6-1703 &      2.1$\pm{0.1}$ &      30    &  $-$   &    $-$   &    $<$1.10E-13 &       1.50E-10 &     $>$1364 &     2, 2, 2, 2, 40, 41 \\          
IGR J18410-0535 &       3$\pm{2}$    &       6.45 &  $-$   &    $-$   &       9.00E-14 &       1.00E-09 & 1.1$\times10^4$ &   2, 2, 2, 2, 42, 32 \\     
IGR J18450-0435 &       6.4           &       5.7  &  $-$   &    $-$   &       3.90E-12 &       2.00E-09 &        513 &   2, 2, 2, 2, 31, 32 \\          
IGR J18483-0311 &       3.5$\pm{0.5}$ &      18.52 &$\sim$0.4 &  21.05 &       8.90E-13 &       8.00E-10 &        899 &   2, 2, 2, 2, 7, 32 \\       
%----------------------------------------------------------------------------------------
\hline
\multicolumn{9}{c}{Be/XRBs} \cr
%\hline
H~0115+634      &       8$\pm{1}$     &      24.32 &       0.34 &       3.61 &       2.80E-14 &       4.00E-09 & 1.4$\times10^5$  &  43, 4, 4, 4, 12, 44  \\  
RX~J0146.9+6121 &       2.3$\pm{0.5}$ &     330    &       $-$  &    1400    &       1.70E-11 &          $-$   &       $-$        &  43, 43, 43, 43, 7, $-$   \\   
EXO~0331+530    &       7.0$\pm{1.5}$  &      36.5  &       0.42 &       4.38 &       2.80E-14 &       3.00E-08 & 1.07$\times10^6$ &  43, 4, 4, 4, 7, 45 \\    
X~Per           &       0.8$\pm{0.14}$ &     250    &       0.11 &     837.7  &       1.00E-10 &       1.00E-09 &         10       &   21, 4, 4, 4, 46, 46 \\ 
1A~0535+262     &       2.0$\pm{0.7}$  &     111    &       0.47 &     103.5  &       3.70E-12 &       1.00E-07 & 2.7$\times10^4$  &  43, 43, 43, 4,  47 \\  
GRO~J1008-57    &       5.8            &     249.48 &       0.68 &      93.5  &       2.21E-11 &       4.00E-09 &       181      &  48, 48, 48, 48,  56, 49  \\    
4U~1036-56      &       5              &      60.9  &      $-$   &     853.4  &       4.00E-12 &       2.40E-10 &         60     &   43, 50, $-$, 51, 52, 52 \\    
IGR~J11305-6256 &       3              &        $-$ &      $-$   &    $-$     &       4.30E-11 &          $-$   &        $-$     &   3, 3, $-$, 3, 23, $-$ \\   
IGR~J11435-6109 &       8$\pm{2}$      &      52.4  &      $-$   &     161.76 &       8.30E-12 &          $-$   &         $-$    &    3, 3, $-$, 3,  23, $-$  \\   
H~1145-619      &       2.0$\pm{1.5}$  &     187.5  &    $>$0.5  &     292    &       1.00E-11 &       2.50E-09 &        250     &     3, 3, 3, 3, 53, 53 \\      
XTE~J1543-568   &      10              &      75.56 &    $<$0.03 &      27.12 &       1.00E-10 &       8.00E-10 &          8     &   6, 6, 6, 6, 54, 54 \\     
AX~J1749.1-2733 &      14.5$\pm{1.5}$ &       $-$  &      $-$   &     132    &       9.60E-12 &          $-$   &        $-$     &   3, $-$, $-$, 3,  7, $-$   \\   
GRO~J1750-27    &      18             &      29.8  &       0.36 &       4.45 &    $<$2.00E-10 &       2.00E-09 &       $>$10    &   6, 6, 6, 6, 55, 55 \\ 
AX~J1820.5-1434 &       8.2$\pm{3.5}$ &      54    &       $-$  &     152.26 &       1.50E-12 &           $-$  &        $-$     &   21, 3, $-$, 3, 7, $-$  \\    
Ginga~1843+009  &      10             &       $-$  &       $-$  &      29.5  &       1.06E-13 &       6.0E-10 &        5660  &   6, $-$, $-$, 6, 56, 57 \\     
XTE~J1858+034   & 10$^{c}$            &     380    &       $-$  &     221    &       4.80E-10 &            $-$ &        $-$     &   $-$, 3, $-$, 3, 6,    $-$     \\  
4U~1901+03      &      10             &      22.58 &       0.04 &       2.76 &       1.00E-11 &       1.00E-08 &       1000     &   6, 6, 6, 6, 58, 58 \\         
KS~1947+300     &       9.5$\pm{1.1}$  &      40.41 &       0.03 &      18.76 &       3.75E-12 &       3.00E-09 &        800    &  3, 6, 6, 6, 56, 59 \\   
EXO~2030+375    &      7.1$\pm{0.2}$   &      46.02 &       0.41 &      42    &    $<$9.70E-12 &       2.70E-08 &       $>$2784 & 6, 6, 6, 6, 6, 6  \\   
SAX~J2103.5+4545 &      6.5$\pm{2.3}$   &      12.67 &       0.41 &     358.6  &       2.20E-13 &       1.40E-09 &       6364   &   6, 6, 6, 6, 60, 61 \\    
\hline
\end{tabular}
\label{tab:literature}
\end{table*}

%%%%%%%%%%%%%%-------------------------------------  
\setcounter{table}{1}
\begin{table*}
 \centering
 \caption{Summary of the source properties extracted from the literature, for the different types of HMXBs of our sample {\it (continued).}}
  \begin{tabular}{lcccccccc}
\hline
Name & Dist  &  P$_{orb}$  &  ecc     &   P$_{spin}$   &  F$_{min}$             &  F$_{max}$              &  DR$_{1-10~keV}$                        &  References        \\
     & (kpc) &  (d)        &          &     (s)        & (\ferg)                &   (\ferg)               &   (F$_{max}$/F$_{min}$)                &                     \\
%----------------------------------------------------------------------------------------
\hline
\multicolumn{9}{c}{Other HMXBs} \cr
IGR J16318-4848 &       3.6$\pm{2.6}$  &    80.09     &    $-$     &    $-$ &       2.30E-12 &       7.50E-12 &          3.3 &  7,  70\&71,  $-$,  $-$, 7, 7 \\  
3A~2206+543     &       2.6            &    9.57    &     0.3   &    5588 &       1.20E-11 &       3.00E-09 &        250 &  3, 62, 62, 63, 6, 64 \\  
Cyg~X-1         &       1.86$^{+0.12}_{-0.11}$ &       5.6  &     0.018 &    $-$  &       5.10E-09 &       1.90E-08 &          3.7 &  3, 3,  65, $-$, 7, 7 \\ 
Cyg~X-3         &       7.4$\pm{1.1}$  &       0.2  &   0.0$^{d}$ &    $-$  &       1.70E-09 &       8.30E-09 &          4.9 &  66, 3, $-$, $-$, 6, 6 \\   
SS~433          &       5.5$\pm{0.2}$  &      13.08 &     $-$   &    $-$  &       4.00E-11 &       2.00E-10 &          5.0  &   6, 6, $-$, $-$, 6, 6 \\     
\hline
%----------------------------------------------------------------------------------------
\multicolumn{9}{c}{Symbiotic binary} \cr
%\hline
XTE~J1743-363   &       5    &        $-$ &       $-$ &     $-$ &       7.80E-12 &       4.80E-11 &          6.2 &     67,   $-$,  $-$,   $-$, 67, 67     \\    
\hline
\end{tabular}
\flushleft{
 $^{a}$For each source, the six numbers indicate the six references for the source distance, orbital period, eccentricity, spin period, minimum and maximum fluxes (1--10\,keV) reported in columns from 2 to 7, respectively:  
(1)~\citet{Martinez-Nunez2017}, 
(2)~\citet{Sidoli2017review},
(3)~\citet{Walter2015},
(4)~\citet{Townsend2011},
(5)~\citet{Falanga2015},
(6)~\citet{Liu2006},
(7)~\citet{Gimenez2015},
(8)~\citet{Sidoli2010igr08408},
(9)~\citet{Sidoli2009out},
(10)~\citet{Romano2009},
(11)~\citet{Inam2010},
(12)~\citet{Rosen2016},
(13)~\citet{Saxton2008},
(14)~\citet{Hilditch2005},
(15)~\citet{Bodaghee2010},
(16)~\citet{Tomsick2009},
(17)~\citet{Romano2007},
(18)~\citet{Islam2016},
(19)~\citet{Islam2015},
(20)~\citet{Bodaghee2016},
(21)~\citet{Bodaghee2012},
(22)~\citet{Aftab2016},
(23)~\citet{Tomsick2008},
(24)~\citet{Coleiro2013},
(25)~\citet{Jaisawal2013},
(26)~\citet{Fuerst2012},
(27)~\citet{Torrejon2010},
(28)~\citet{Israel2016},
(29)~\citet{Sidoli2016},
(30)~\citet{Rodriguez2005},
(31)~\citet{Bozzo2017},
(32)~\citet{Romano2015},
(33)~\citet{Sidoli2012},
(34)~\citet{Sidoli2013suz},
(35)~\citet{Romano2008},
(36)~\citet{Bozzo2012},
(37)~\citet{Ducci2013},
(38)~\citet{zand2005},
(39)~\citet{Romano2015giant},
(40)~\citet{Bozzo2008_atel},
(41)~\citet{Boon2016},
(42)~\citet{Bozzo2011},
(43)~\citet{Reig2011},
(44)~\citet{Nakajima2017},
(45)~\citet{Doroshenko2017},
(46)~\citet{Lutovinov2012},
(47)~\citet{Ballhausen2017}, 
(48)~\citet{Kuhnel2017},
(49)~\citet{Evans2014},
(50)~\citet{Cusumano2013},
(51)~\citet{LaPalombara2009},
(52)~\citet{Li2012},
(53)~\citet{Stevens1997},
(54)~\citet{zand2001},
(55)~\citet{Shaw2009},
(56)~\citet{Tsygankov2017},          
(57)~\citet{Piraino2000},    
(58)~\citet{Reig2016},
(59)~\citet{Ballhausen2016},
(60)~\citet{Reig2010},
(61)~\citet{Reig2014},
(62)~\citet{Stoyanov2014},
(63)~\citet{Wang2013},
(64)~\citet{Wang2010},
(65)~\citet{Orosz2011},
(66)~\citet{McCollough2016},
(67)~\citet{Bozzo2013},
(68)~\citet{Neilsen2009},
(69)~\citet{Rodes2017},
(70)~\citet{Jain2009igr16318},
(71)~\citet{Iyer2017},
(72)~\citet{Roberts2001}
\\
$^{b}$  the spin period of IGR~J17544--2619 needs  confirmation, being obtained from $RXTE$/PCA observations; it is possible that X--ray
pulsations come from a different transient source within the field of view.  \\
$^{c}$A distance of 10~kpc for XTEJ~1858+034 is assumed. \\
$^{d}$Here we assume a circular orbit, given the short orbital period.
}
\end{table*}

%%%%%%%%%%%%%%%%%%%%%%%%%%%%%%%%%%%%%%%%%%%%%%%%%%%%%%%%%%%%%%%%%%%%%%%%%%%%%%%%%%%%%%%%%%%%%%%%%%%%%%%%%%%%%%%%%%%%%%%%
\subsection{The HMXB sample}

Given our detection criteria (ScW exposure $>$1\,ks, off axis angle $<$12$^{\circ}$ and ScW detection  $\geqq$5\,sigma), 
we obtain a sample of 58 HMXBs  (plus one symbiotic X--ray transient, XTE~J1743-363). 
All but two (LMC~X--4 and SMC~X-1) are Galactic sources, that represent about half of the total number of HMXBs known in our Galaxy \citep{Liu2006}.
They belong to different sub-classes:  persistent HMXBs with supergiant companions (SgHMXBs, hereafter); 
HMXBs with early-type giant donor stars (Roche lobe overflow systems, like Cen~X--3 and LMC~X--4); 
the Supergiant Fast X--ray Transients (SFXTs); 
Be X--ray binaries (hereafter Be/XRBs, meant to  include both persistent and transients Be sources - in case 
we only mean the X--ray transient Be systems, we will use the acronym Be/XRTs); 
black-hole binaries and other peculiar sources (Cyg~X--1, Cyg~X--3, SS~433);
the source 3A~2206+543 
(where the companion has an anomalous wind, \citealt{Blay2009});
the highly obscured source IGR~J16318--4848 \citep{Courvoisier2003}, 
where the companion is a B[e] supergiant star \citep{Filliatre2004, Chaty2012}, that is 
a supergiant star that shows the B[e] phenomenon (forbidden emission
lines in its optical spectrum, \citealt{Lamers1998}).
Finally, our sample includes also a symbiotic X--ray transient, XTE~J1743--363 \citep{Bozzo2013}, 
where a compact object accretes matter from the wind of an M8~III giant, in order to compare its 
 behaviour with other wind-fed massive X--ray binaries with OB-type stars, from the point of view of
its hard X--ray emission.

          %%%%%%%%%%%%%%%%%%%%%%%%%%%%%%%%%%%%%%%%%%%%%%%%%%%%%%%%%%%%%%%%%%%%
          \section{\textit{INTEGRAL} results} \label{sec:results}
          %%%%%%%%%%%%%%%%%%%%%%%%%%%%%%%%%%%%%%%%%%%%%%%%%%%%%%%%%%%%%%%%%%%%

Table~\ref{tab:inte} summarizes our IBIS/ISGRI results (see Sect.~\ref{arc} for the definition of the columns).

          %%%%%%%%%%%%%%%%%%%%%%%%%%%%%%%%%%%%%%%%%%%%%%%%%%%%%%%%%%%%%%%%%%%%
          \subsection{From count-rate to luminosity} \label{sec:spectra}
          %%%%%%%%%%%%%%%%%%%%%%%%%%%%%%%%%%%%%%%%%%%%%%%%%%%%%%%%%%%%%%%%%%%%

Conversion factors from IBIS/ISGRI count-rates to X--ray luminosities (18--50 keV) 
have been derived from the analysis of IBIS/ISGRI spectra extracted from a subsample of HMXBs, fitted with
models typical for accreting pulsars, like power laws with high energy cutoff. 
%----------- 
The subsample adopted for the IBIS/ISGRI spectroscopy included the following sources: Vela X-1, 4U~1700-377, H~1907+097, IGR~J08408-4503, IGR~J11215-5952, IGR~J16418-4532, IGR~J16465-4507, IGR~J16479-4514, XTE~J1739-302, IGR~J17544-2619, SAX~J1818.6-1703, IGR~J18410-0535, IGR~J18450-0435 and IGR~J18483-0311.
For each source, we extracted an average spectrum from a subsample of ScWs where the sources were detected above 5\,sigma, within 12$^{\circ}$ from the FoV centre. We verified that within the considered observations no strong evidence for spectral variability was present. For each source, we fitted the average spectra in {\sc xspec} using power law models with exponential cutoffs ({\sc cutoffpl} or {\sc pow*highecut} models in {\sc xspec}).   
%-----------
From this spectroscopy, we derived an average conversion factor of 
4.5$\times10^{-11}$~erg~cm$^{-2}$~count$^{-1}$ to obtain the source fluxes.
Source distances listed in Table~\ref{tab:literature} were then used to calculate the luminosities in hard X--rays (18--50~keV). 

For the Crab, that has a different spectrum and 
we consider as our standard candle to get an idea of the luminosity variability of the different sources, 
an average spectrum was extracted from a subsample of ScWs as well. Fitting it with a power law model, we obtained
a conversion factor of 4.7$\times10^{-11}$~erg~cm$^{-2}$~count$^{-1}$ (18--50~keV). A distance of 2~kpc was adopted to obtain the luminosity.

For each detected source $j$, we calculate the fluence over $i$, over the total number of ScWs where the source has been detected:

\begin{equation}
Fluence_{j} = \sum\nolimits_{ScW_i} count~rate (ScW_i) \times dur  (ScW_i) 
\end{equation}

The fluence (total number of counts for each detected source {\em j}) was used to calculate an
average rate,  as $Rate_{j} = Fluence_{j} / \Delta t_{j}$,
where  $\Delta~t_{j}$ is the {\em source activity} (defined in Sect.~\ref{anita} and reported in Table~\ref{tab:inte}, third column).
This average rate, $Rate_{j}$ was then converted into an average luminosity, L$_{X}$, 
in the energy range 18--50\,keV (reported in Table~\ref{tab:inte}, fifth column). 
For transient sources, this L$_{X}$ is the average luminosity observed by \inte\ when the source is in outburst.
In Table~\ref{tab:inte} we show also the minimum and maximum 18--50\,keV luminosities displayed by each source 
(always at ScW level, bin time of $\sim$2~ks), as observed by \inte\ (cols. 6 and 7),
together with their ratio (col.~9), that represent the maximum source variability observed by \inte\ in 14~years. 
We note that for a highly variable source, the average hard X-ray luminosity might be, in principle, more affected by the source distance than the maximum luminosity.
However, the average value is more representative of the source global behavior, while the maximum luminosity,
being derived from a single 2 ks  \inte\ observation, might be  reached during a rare short flare, brighter than the usual source intensity.

No statistical error is given on the average rate (or L$_{X}$), because
the largest source of uncertainty in the hard X--ray luminosities is due to the source distance,
reported in Table~\ref{tab:literature}. 
When the uncertainty is not listed, is because we could not find it in the literature. For these sources we have assumed  $\pm{1}$~kpc.
However, the uncertainty on the source luminosities has no impact on our conclusions.

         %%%%%%%%%%%%%%%%%%%%%%%%%%%%%%%%%%%%%%%%%%%%%%%%%%%%%%%%%%%%%%%%%%%%
         \subsection{Cumulative luminosity distributions (18--50\,keV)} \label{sec:dis_inte}
         %%%%%%%%%%%%%%%%%%%%%%%%%%%%%%%%%%%%%%%%%%%%%%%%%%%%%%%%%%%%%%%%%%%%

The long-term IBIS/ISGRI light curves ($\sim$2 ks bin time) were used  to build the  
\textit{complementary cumulative luminosity distribution functions} (hereafter, 
only cumulative luminosity distributions, CLDs) of the hard X--ray emission for our sample of HMXBs. 
In \citet{Paizis2014} we investigated the CLDs of the SFXT flares caught by \inte\ in 9 years of observations,
and compared them to three classical HMXBs. 
Here we enlarge our investigation on a larger dataset (14 years) to different types of massive X--ray binaries, to compare their behaviour 
in terms of source activity and variability amplitude in hard X--rays.

In Figs~\ref{fig:cumdis_sfxt}, \ref{fig:cumdis_sg}, \ref{fig:cumdis_be} and  \ref{fig:cumdis_other}, we show the normalized 
CLDs for  different sub-classes of HMXBs: SFXTs, SgHMXBs, Be/XRBs and ``other sources'' including BH binaries.
The normalization of each curve has been performed with respect to the total exposure time for which the source 
was within $12\degmark$. In this way, the source DC$_{18-50~keV}$ (percentage of time in which the source is active at ScW level)
can be derived directly from the CLD as the highest value in the y-axis (and it is also reported in Table~\ref{tab:inte}).
Not only can the variability amplitude of the source luminosity  (shown in 14 years of \inte\ mission)  be read from the x-axis in each source CLD,
but also the  percentage of time spent by each source in different luminosity states can be derived.

Transient versus persistent behaviour depends both on the intrinsic variability and on the instrumental sensitivity threshold. 
In principle also the source distance might contribute (but see  Fig.~\ref{lsfig:dyn_1_10_dist} and Sect.~\ref{sect:source_dc_dyn_1_10}).
As can be seen from the CLDs (and from the Table~\ref{tab:inte}), only the Crab and Cyg~X-1 are always detected. 
Persistent, eclipsing HMXBs  (marked in Table~\ref{tab:inte} with an asterisk), 
show a lower DC$_{18-50~keV}$ because \inte\ is not able to detect the sources in eclipse on timescales of a single ScW. 
The less active (more transient) the source, the lower the DC$_{18-50~keV}$. 
Furthermore, the more variable the hard X--ray luminosity (at $\sim$2~ks timescale), the flatter the CLD.
A persistent, bright and constant source seen by an ideal detector  
would display a CLD as a vertical straight line. 
In real life, the Crab shows a deviation (Figs~\ref{fig:cumdis_sfxt}, \ref{fig:cumdis_sg}, \ref{fig:cumdis_be} and  \ref{fig:cumdis_other}) 
that  includes both the intrinsic intensity decline, observed in the Crab by several instruments \citep{Wilson-Hodge2011}, 
and the IBIS/ISGRI loss of sensitivity. These two effects cannot be disentangled in the CLDs. 
However the Crab CLD indicates the amount of variability beyond which a source can be safely considered as intrinsically variable.

%%%%%%%%%%%%%%%%%%%%%%%%%%%%%%%%%%%%%%%%%%%%%%%%%%%%%%%%%%%%%%%%%%%%%%%%%%%%%%%%%%%%%%%%%%%%%%%%%%%%%%%%%%%%%%%%%%%%%%%%%%%%%%%%%%%%%%%%
%%%%%%%%%%%%%%%%%%%%%%%%%%%%%%%%%%%%%%%%%%%%%%%%%%%%%%%%%%%%%%%%%%%%%%%%%%%%%%%%%%%%%%%%%%%%%%%%%%%%%%%%%%%%%%%%%%%%%%%%%%%%%%%%%%%%%%%%

 	 %%%%%%%%%%%%%%%%%%%%%%%%%%%%%%%%%%%%%%%%%%%%%%%%%%%%%%%%%%%%%%%%%%%%
 	 \section{HMXBs properties collected from the literature } \label{literature}
 	 %%%%%%%%%%%%%%%%%%%%%%%%%%%%%%%%%%%%%%%%%%%%%%%%%%%%%%%%%%%%%%%%%%%%

For each HMXB in our sample, we searched through the literature to collect available information about
the source distance, spin and orbital periodicities, orbital eccentricity, minimum and maximum flux in
soft X--rays (1--10\,keV, corrected for the absorption). 
We list these quantities in Table~\ref{tab:literature}.
We relied  on a number of  review papers. 
This implies that the references reported in Table~\ref{tab:literature} (last column) 
for a particular parameter is not, in most cases, the original discovery paper, but 
a more recent review article collecting previous literature about a large number of sources. 
This has also the advantage that up to date values for these quantities are reported in Table~\ref{tab:literature}.

We caution that the  minimum and maximum  unabsorbed fluxes (1--10\,keV) taken from the literature have been
obtained with different instruments and are integrated over different timescales. 
For instance, some of the soft X-ray fluxes have been derived from dedicated monitoring campaigns (as in the case of a sample of SFXTs monitored by Swift/XRT,
\citet{Romano2015}), while others have been taken from papers reporting 
on deep exposures performed with high throughput instruments,
like EPIC on-board \xmm\ \citep{Gimenez2015}. 
Moreover, an obvious bias arises when collecting values from the literature, as only a new result is usually reported in a paper. 
It is possible that rarer intensity states (implying a larger dynamic range than the one
 listed in Table~\ref{tab:literature}) await discovery.

Normally, fluxes integrated over different soft energy ranges are reported in the literature.
Hence, for consistency's sake, we extrapolated them to the 1--10\,keV band, 
using \textsc{WebPIMMS}\footnote{https://heasarc.gsfc.nasa.gov/cgi-bin/Tools/w3pimms/w3pimms.pl} 
and the appropriate spectral model reported in the same paper. 
If  more complex best-fit models than the ones present in \textsc{WebPIMMS}
are reported in the literature, we used  {\sc xspec} \citep{Arnaud1996} to extrapolate the soft X--ray fluxes.
We never extrapolated hard X--ray fluxes (E$>$20 keV) to the 1--10\,keV energy band,
since these values would be largely unreliable, given the presence of a 
high energy cutoff in HMXB pulsars spectra, around 10-30 keV.

For X-ray pulsars, we only considered spin-phase-averaged fluxes.
Since we are interested in the intrinsic flux variability, we report here out-of-eclipse values, 
for well-known eclipsing HMXBs.
Sometimes this information is not clearly reported in the literature, especially when dealing with X--ray catalogues. 
In this respect, the review paper of Fe line properties by \citet{Gimenez2015} is notable,
in clearly flagging the \xmm\ observations performed during X-ray eclipses in their large sample of HMXBs.

From the maximum and minimum 1--10 keV fluxes, we then calculated their ratio, i.e. the dynamic range ``DR$_{1-10~keV}$'' 
 (see Table~\ref{tab:literature}, col.~8). 
This variability amplitude is indeed one of the properties used to define SFXTs 
(see \citealt{Sidoli2017review} for the most recent review) compared with other, more steady, SgHMXBs.
If only a single measurement of the soft X--ray flux has been found in the literature, we assigned it to the minimum 1--10 keV flux.
In this case, we did not calculate the DR$_{1-10~keV}$.

%%%%%%%%%%%%%%%%%%%%%%%%%%%%%%%%%%%%%%%%%%%%%%%%%%%%%%%%%
\section{Discussion}\label{sec:disc}
%%%%%%%%%%%%%%%%%%%%%%%%%%%%%%%%%%%%%%%%%%%%%%%%%%%%%%%%%

We discuss here the hard X--ray results obtained from the \inte\ archive spanning 14 years (Sect.~\ref{discussion:cld} 
and Table~\ref{tab:inte}) and put these results into context of soft X-ray dynamic ranges and 
other interesting source properties  (Sect.~\ref{discussion:properties} and Table~\ref{tab:literature}).

     %%%%%%%%%%%%%%%%%%%%%%%%%%%%%%%%%%%%%%%%%%%%%%%%%%%%%%%%%
     \subsection{Characterizing the cumulative hard X--ray luminosity distributions } \label{discussion:cld}
     %%%%%%%%%%%%%%%%%%%%%%%%%%%%%%%%%%%%%%%%%%%%%%%%%%%%%%%%%

For all the sources of our sample we extracted the CLDs,
expanding our previous investigation of the CDLs of a number of SFXTs with \inte\ \citep{Paizis2014}.
They are reported in Figs~\ref{fig:cumdis_sfxt}, \ref{fig:cumdis_sg}, \ref{fig:cumdis_be} and  \ref{fig:cumdis_other},
for the different sub-classes of HMXBs.

     %%%%%%%%%%%%%%%%%%%%%%%%%%%%%%%%%%%%%%%%%%%%%%%%%%%%%%%%%
     \subsubsection{General remarks on CLDs}  \label{cld:general}
     %%%%%%%%%%%%%%%%%%%%%%%%%%%%%%%%%%%%%%%%%%%%%%%%%%%%%%%%%

Before discussing the single CLDs, it is important to remark 
that there are two features in the CLDs  that  need to be taken into account: 
a turn-over is often present at both low and high luminosities.
The low luminosity one is due to the difficulty to detect faint X--ray emission, near the 
sensitivity threshold of the detector. This means that the sampling is not complete approaching the faintest luminosity level of each source,
but it is only above a so-called {\em truncation point} (see \citealt{Paizis2014} for more details).
A high luminosity cutoff can be observed as well in some CLDs, because of either a real presence of a maximum X-ray luminosity or because of the
fact that the long-term \inte\ monitoring is still not long enough to observe the most luminous, rare, X--ray flaring activity. 
This implies a large uncertainty in the high luminosity
part of the CDL of the most variable and transient sources (SFXTs). 
These two  effects combined imply  that the most robust part of a CLD is in-between these two cutoff luminosities, especially for SFXTs.

%%%%%%%%%%%%%%%%%%%%%%%%%%%%%%%%%%%%%%%%%%%%%%%%%%%%%%%%%%%%%%%%%%%%%%%% 
\begin{figure*}
\begin{center}
\centerline{\includegraphics[width=16cm]{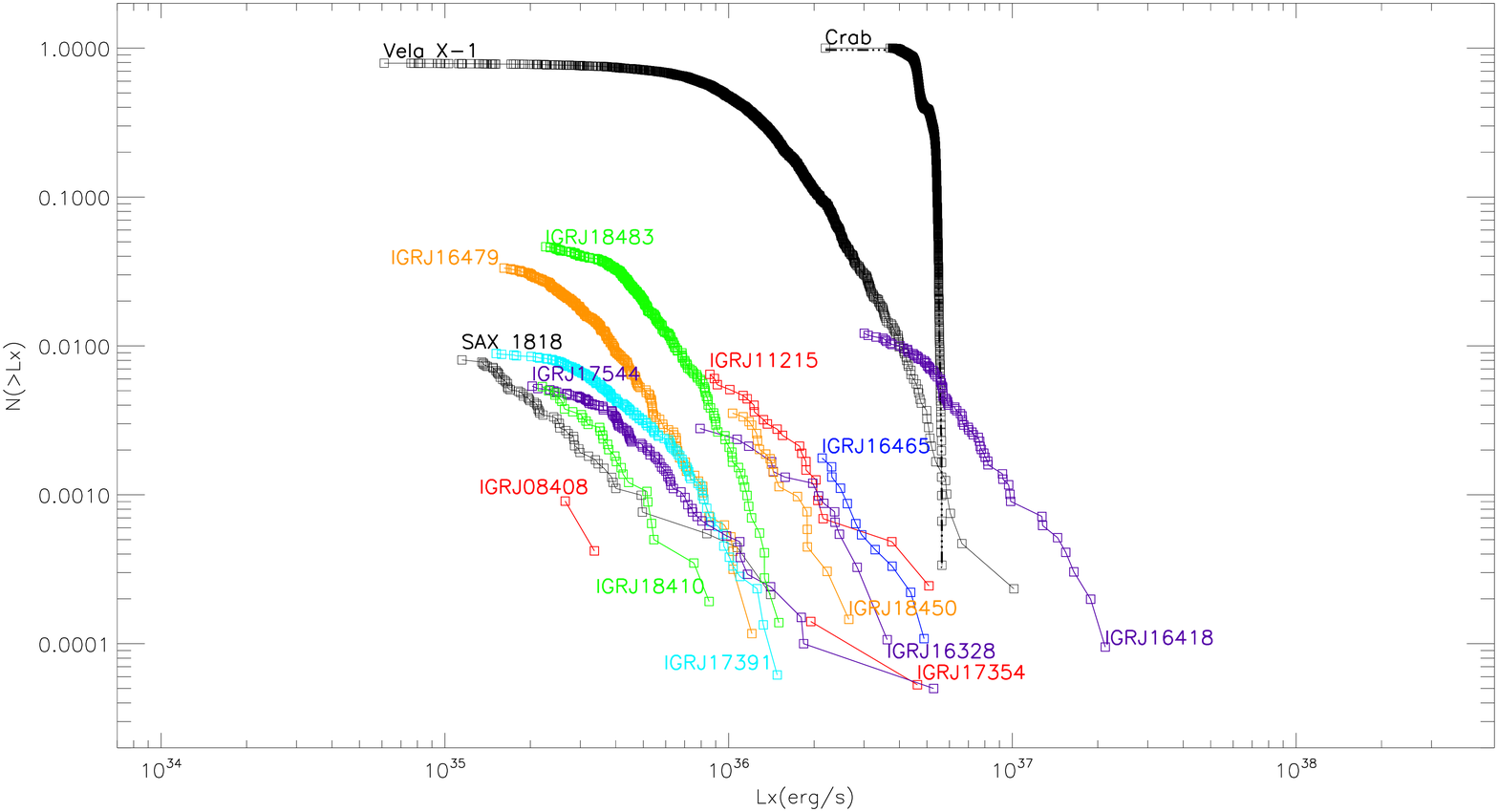}}
\caption{Cumulative luminosity (18--50\,keV) distributions of SFXT flares. 
Each data point is a ScW based detection (detection significance $\ge$5). Vela~X--1 and the Crab are shown for comparison.
}
\label{fig:cumdis_sfxt}
\end{center}
\end{figure*}
%%%%%%%%%%%%%%%%%%%%%%%%%%%%%%%%%%%%%%%%%%%%%%%%%%%%%%%%%%%%%%%%%%%%%%%%

%%%%%%%%%%%%%%%%%%%%%%%%%%%%%%%%%%%%%%%%%%%%%%%%%%%%%%%%%%%%%%%%%%%%%%%% 
\begin{figure*}
\begin{center}
\centerline{\includegraphics[width=16cm]{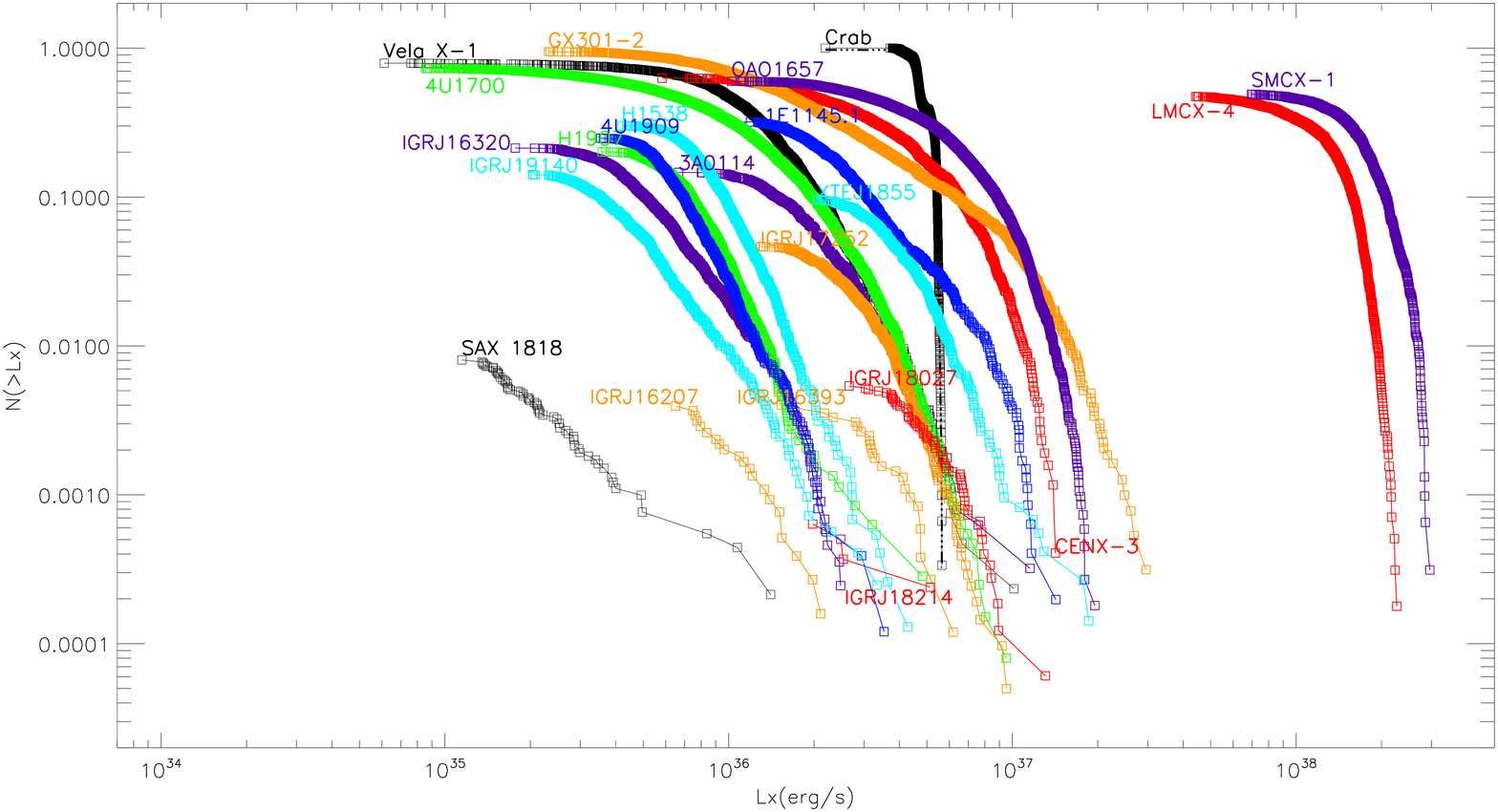}}
\caption{Cumulative luminosity (18--50\,keV) distributions of SgHMXBs, compared with one SFXT (SAX~J1818.6-1703 ) and the Crab.
}
\label{fig:cumdis_sg}
\end{center}
\end{figure*}
%%%%%%%%%%%%%%%%%%%%%%%%%%%%%%%%%%%%%%%%%%%%%%%%%%%%%%%%%%%%%%%%%%%%%%%%

%%%%%%%%%%%%%%%%%%%%%%%%%%%%%%%%%%%%%%%%%%%%%%%%%%%%%%%%%%%%%%%%%%%%%%%% 
\begin{figure*}
\begin{center}
\centerline{\includegraphics[width=16cm]{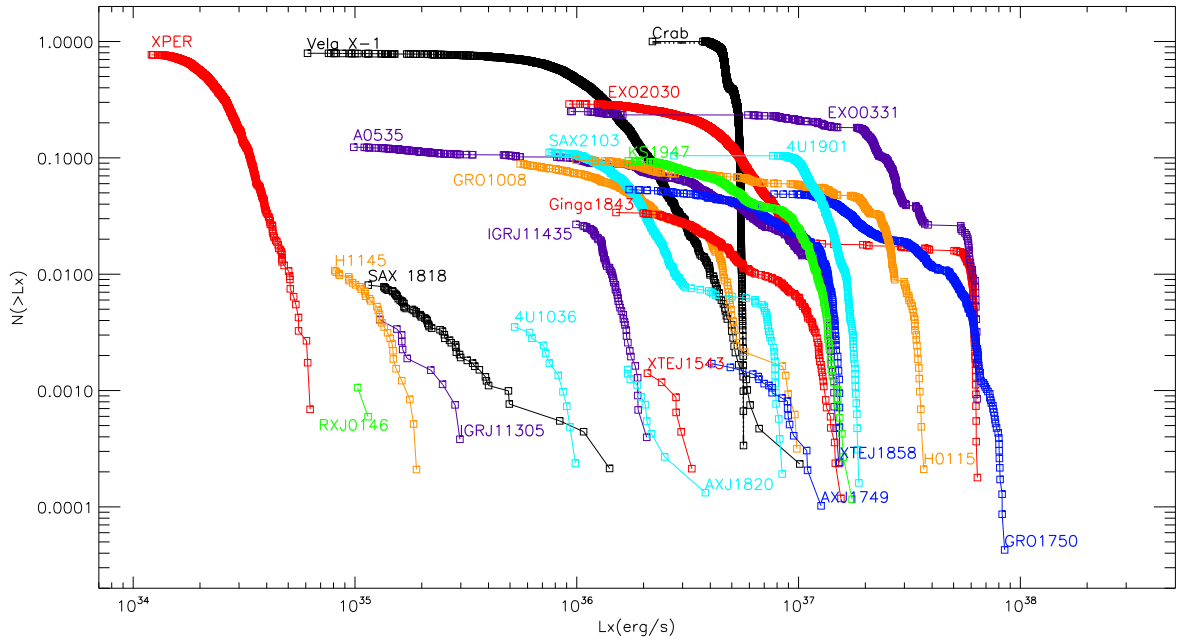}}
\caption{Cumulative luminosity (18--50\,keV) distributions of Be/XRBs, compared with one SFXT (SAX~J1818.6-1703), one persistent SgHMXB (Vela X--1) and the Crab.
}
\label{fig:cumdis_be}
\end{center}
\end{figure*}
%%%%%%%%%%%%%%%%%%%%%%%%%%%%%%%%%%%%%%%%%%%%%%%%%%%%%%%%%%%%%%%%%%%%%%%%

%%%%%%%%%%%%%%%%%%%%%%%%%%%%%%%%%%%%%%%%%%%%%%%%%%%%%%%%%%%%%%%%%%%%%%%% 
\begin{figure*}
\begin{center}
\centerline{\includegraphics[width=16cm]{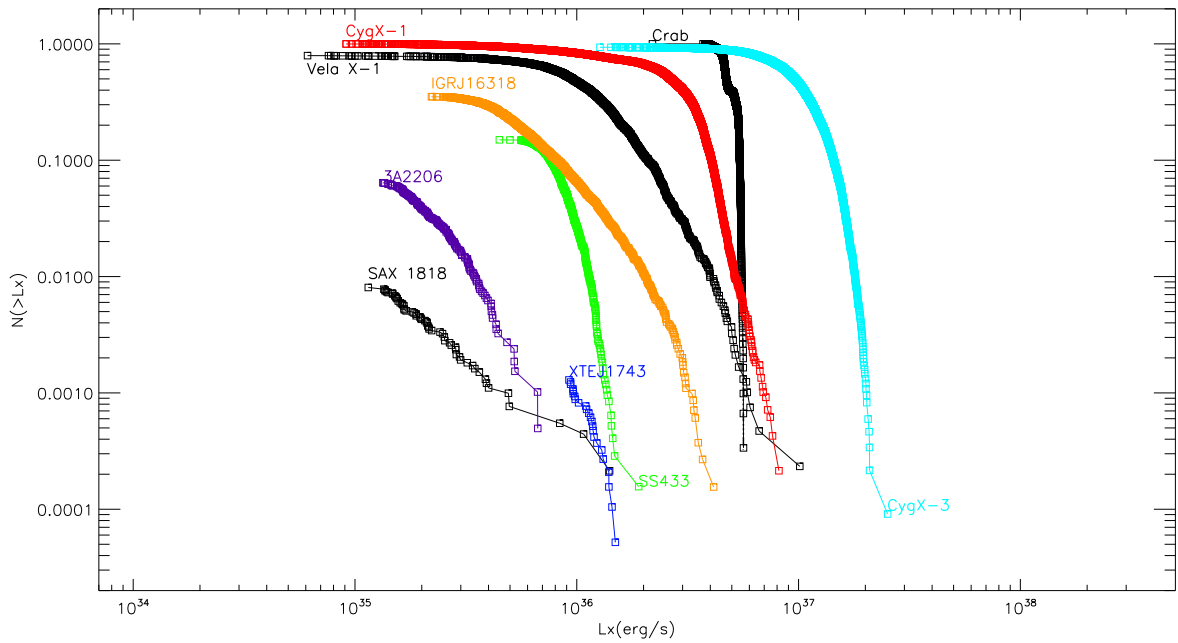}}
\caption{Cumulative luminosity (18--50\,keV) distributions of BH and peculiar sources, compared with one SFXT (SAX~J1818.6-1703), one persistent SgHMXB (Vela X--1)  and the Crab.
}
\label{fig:cumdis_other}
\end{center}
\end{figure*}
%%%%%%%%%%%%%%%%%%%%%%%%%%%%%%%%%%%%%%%%%%%%%%%%%%%%%%%%%%%%%%%%%%%%%%%%

In Figs~\ref{fig:cumdis_sfxt}, \ref{fig:cumdis_sg}, \ref{fig:cumdis_be} and  \ref{fig:cumdis_other}
the CLDs for different types of HMXBs are displayed. Note that we always display, as reference on each plot, 
the CLDs of the Crab, of Vela X--1 (prototype of 
persistent SgHMXBs) and of the prototypical SFXT SAX~J1818.6--1703. 
These plots report  remarkably different CLD shapes depending on the source sub-class.
 
\begin{figure*}
\centering
\begin{tabular}{cc}
\includegraphics[height=6.5cm, angle=0]{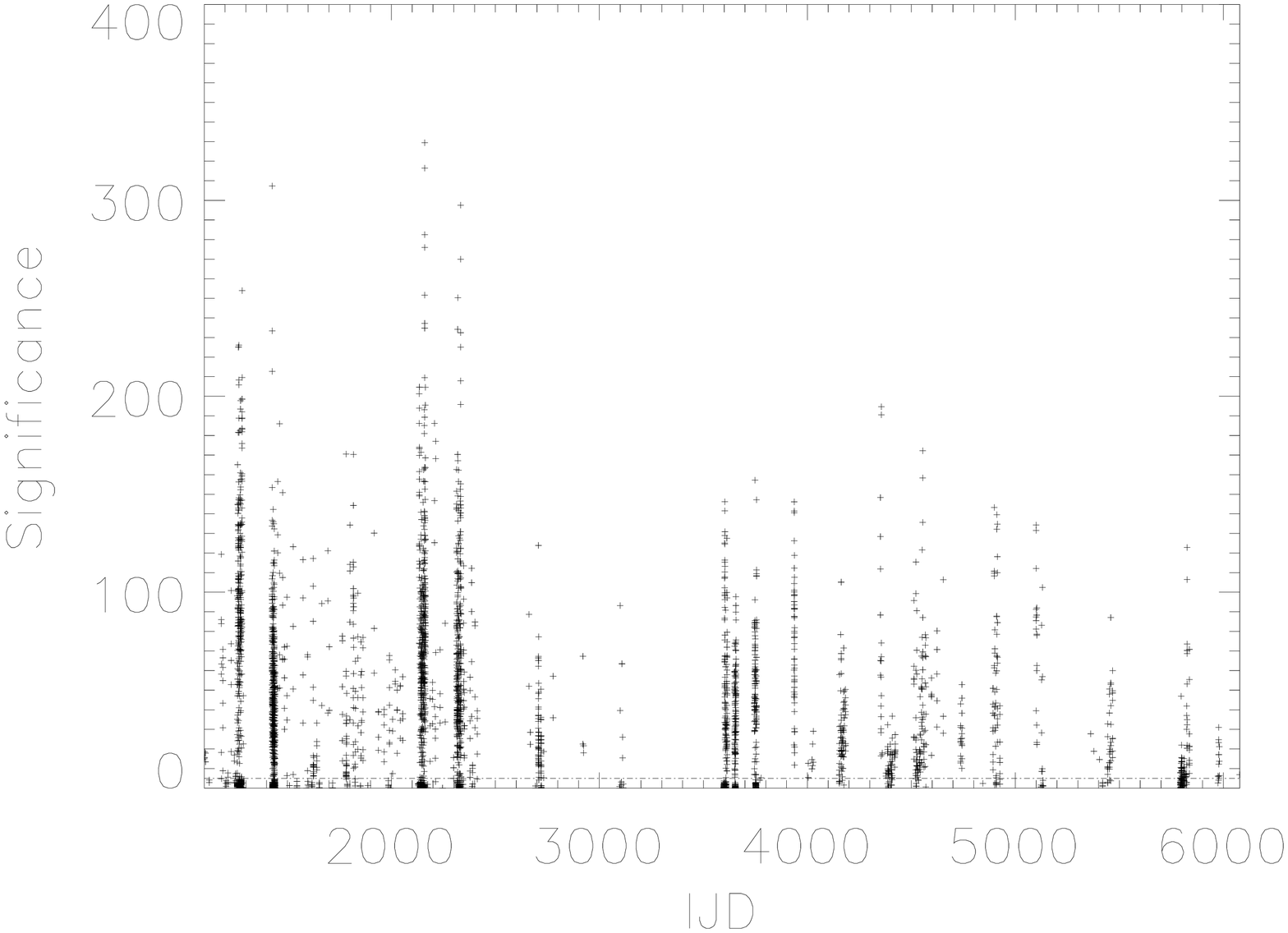} & 
\includegraphics[height=6.5cm, angle=0]{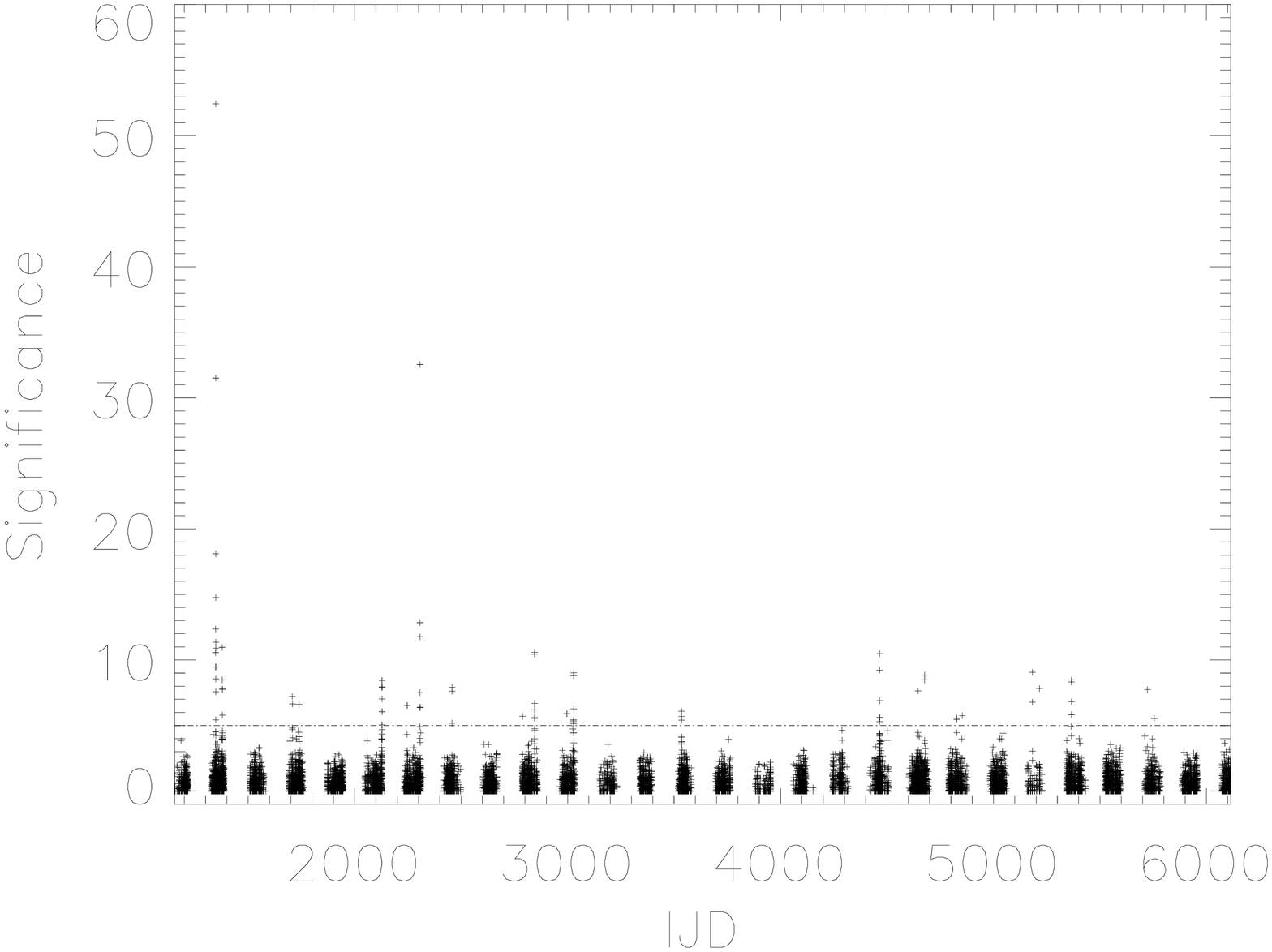} \\
\includegraphics[height=6.5cm, angle=0]{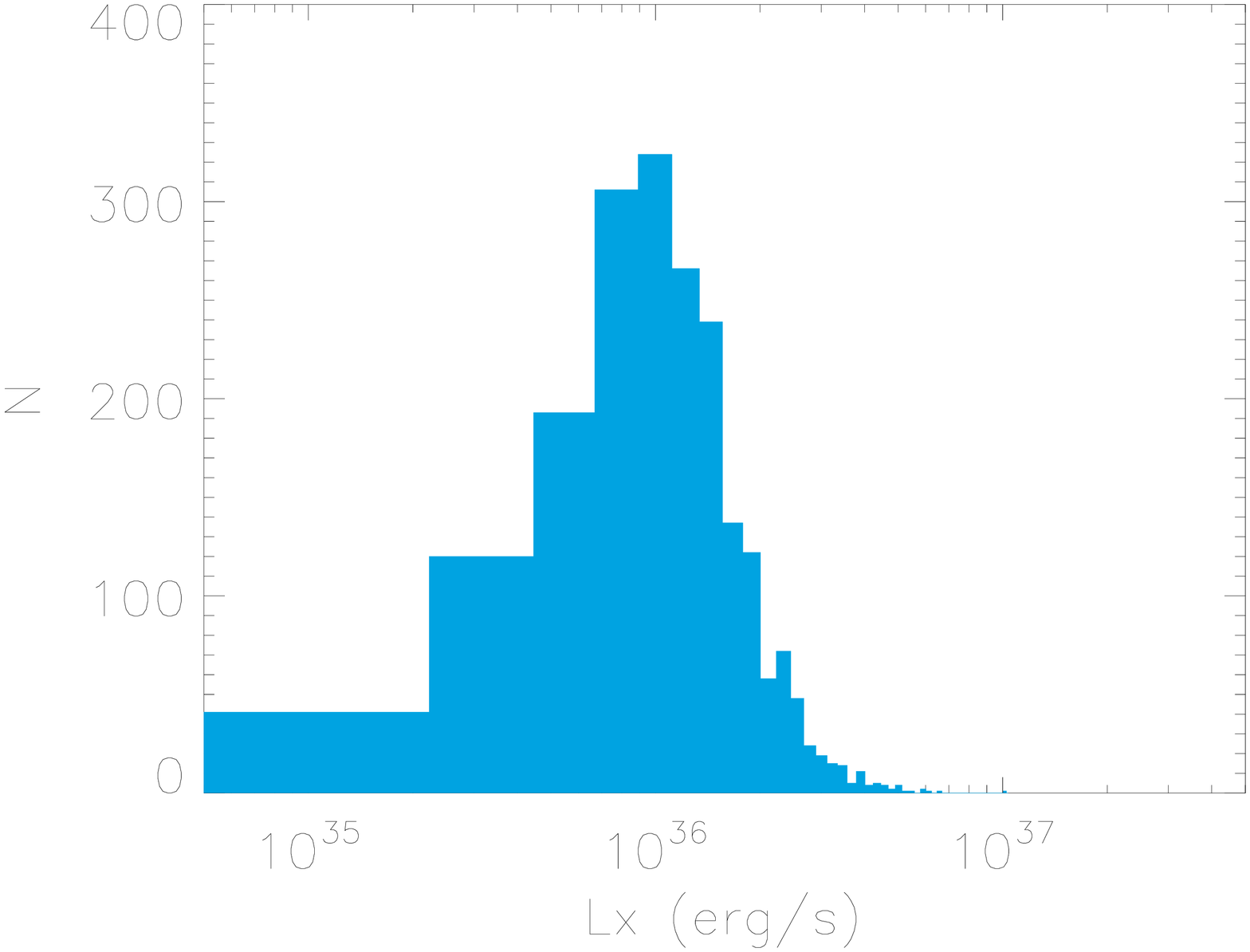} &  
\includegraphics[height=6.5cm, angle=0]{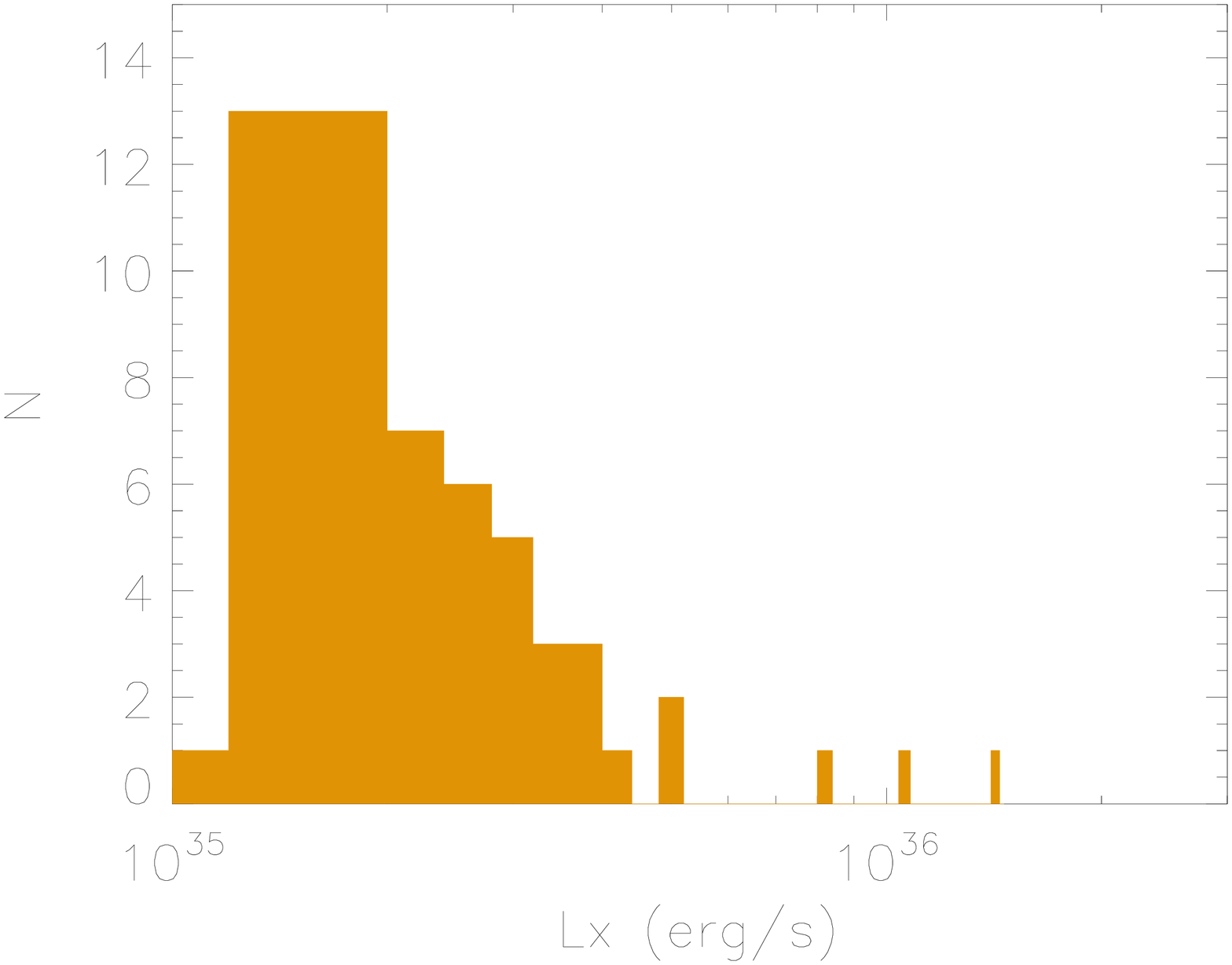}  \\
\includegraphics[height=6.5cm, angle=0]{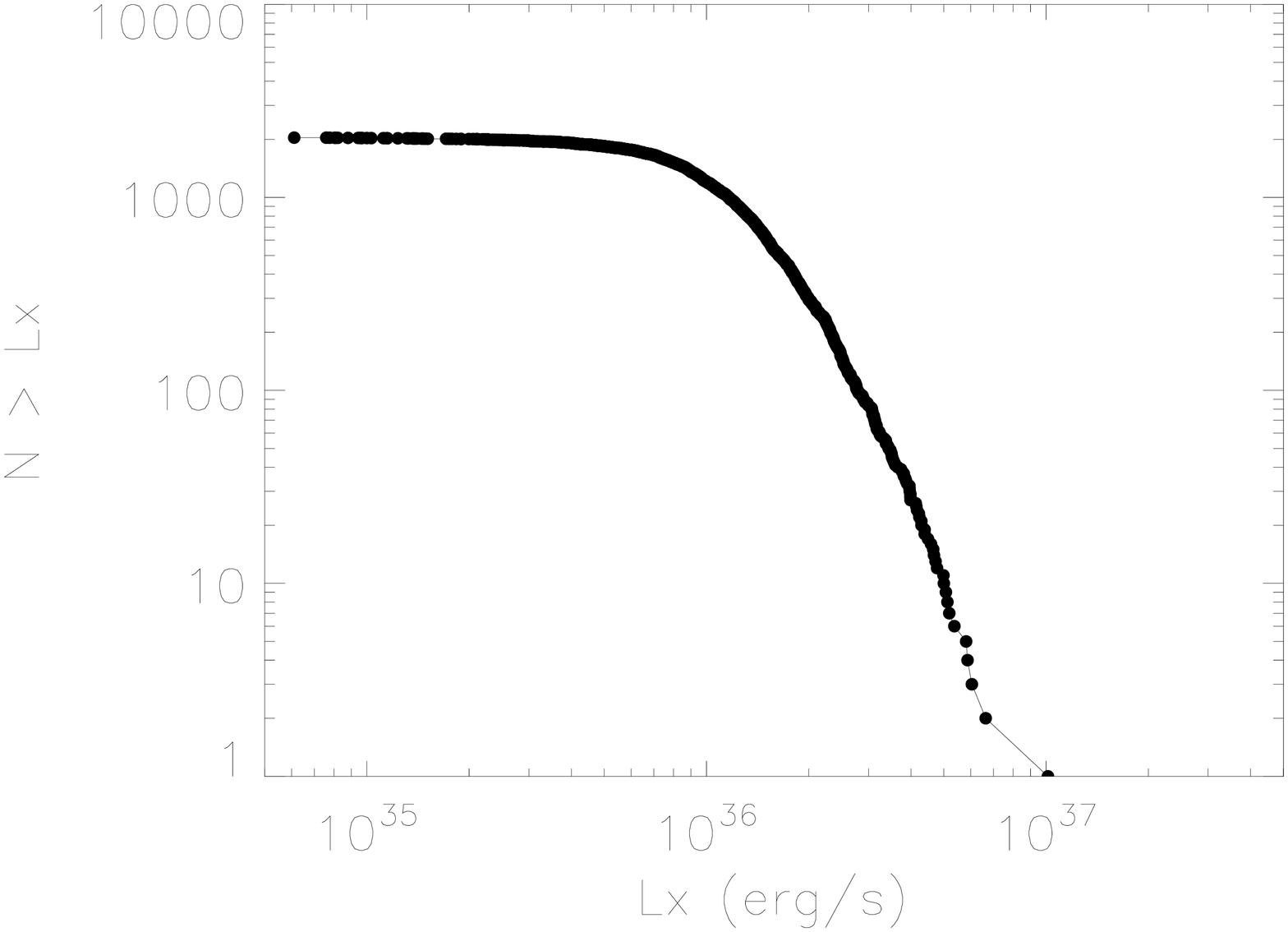} & 
\includegraphics[height=6.5cm, angle=0]{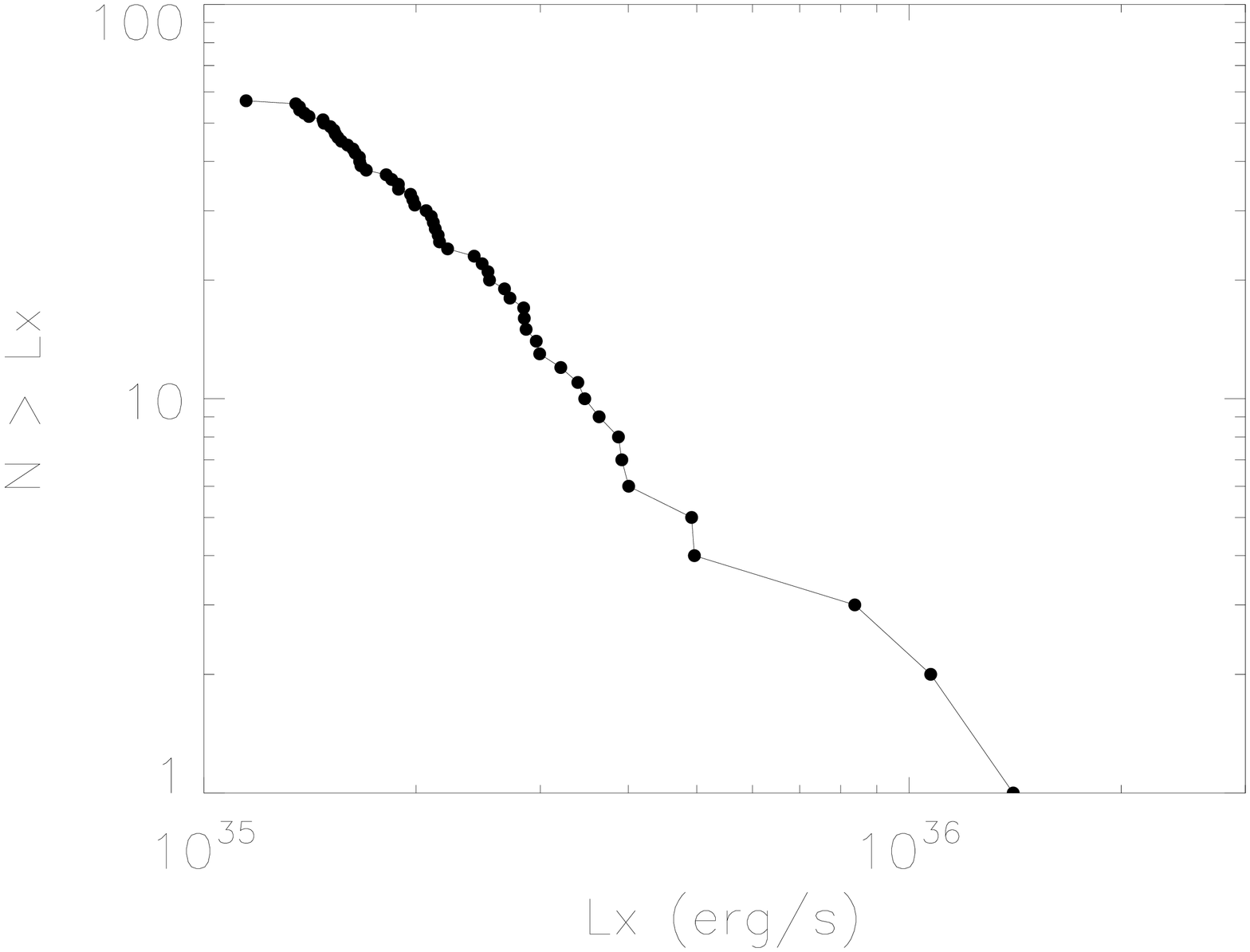} 
\end{tabular}
\caption{{\em Left panels:} Vela X--1 behaviour in the \inte\ archive, as representative of SgHMXBs. From top to bottom: 
source light curve (detection significance in the energy range 18--50\,keV, 
versus time in units of \inte\ JD -- IJD=MJD-51544, together with a 5$\sigma$ horizontal line),
histogram of the detection occurrence and, last panel, its non-normalized CLD. 
{\em Right panels:}
the same for SAX~J1818.6--1703, as an example of SFXTs.
}
\label{fig:lchistocum_vela_sax1818}
\end{figure*}

\begin{figure*}
\centering
\begin{tabular}{cc}
\includegraphics[height=6.5cm, angle=0]{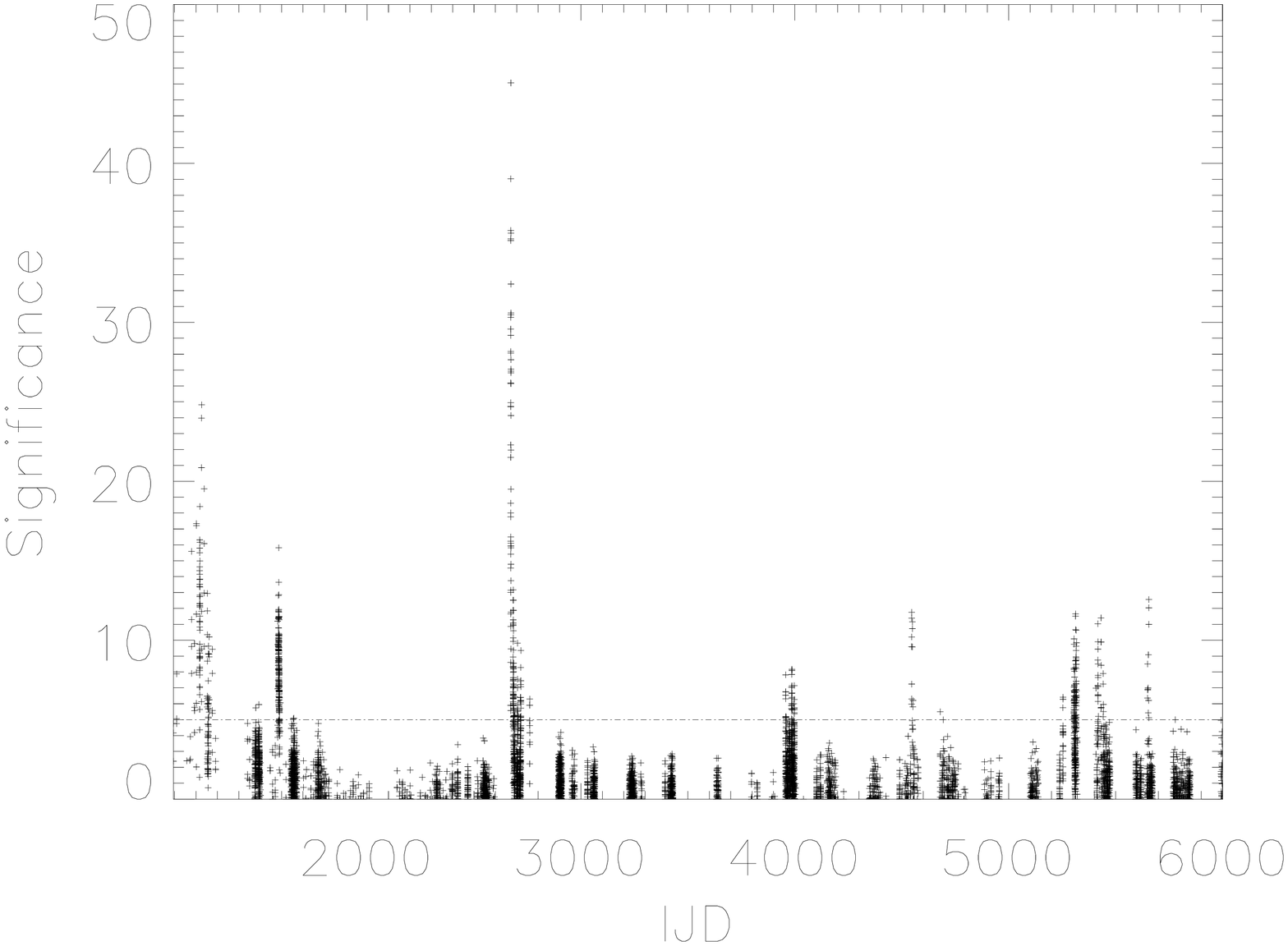} & 
\includegraphics[height=6.5cm, angle=0]{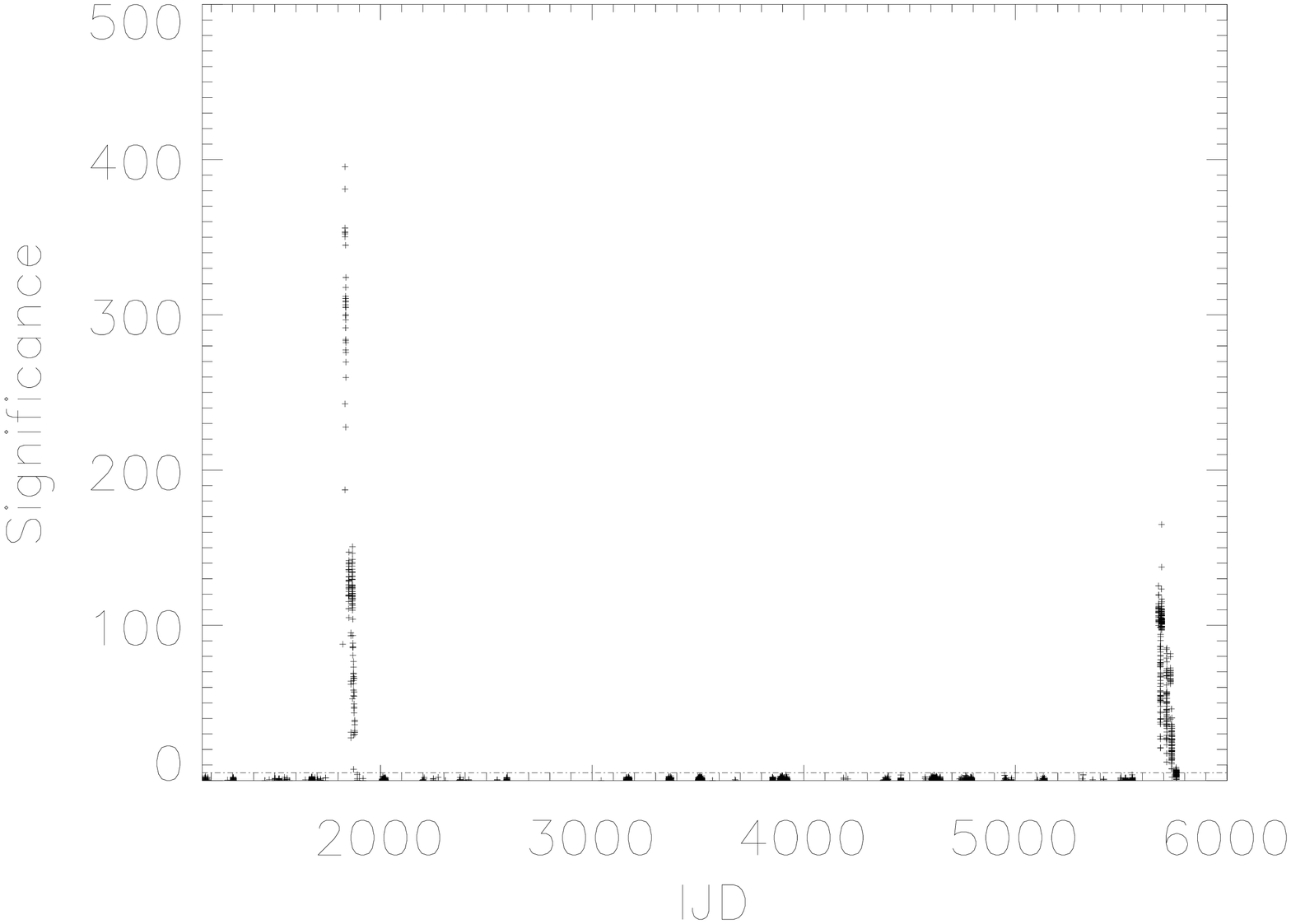} \\
\includegraphics[height=6.5cm, angle=0]{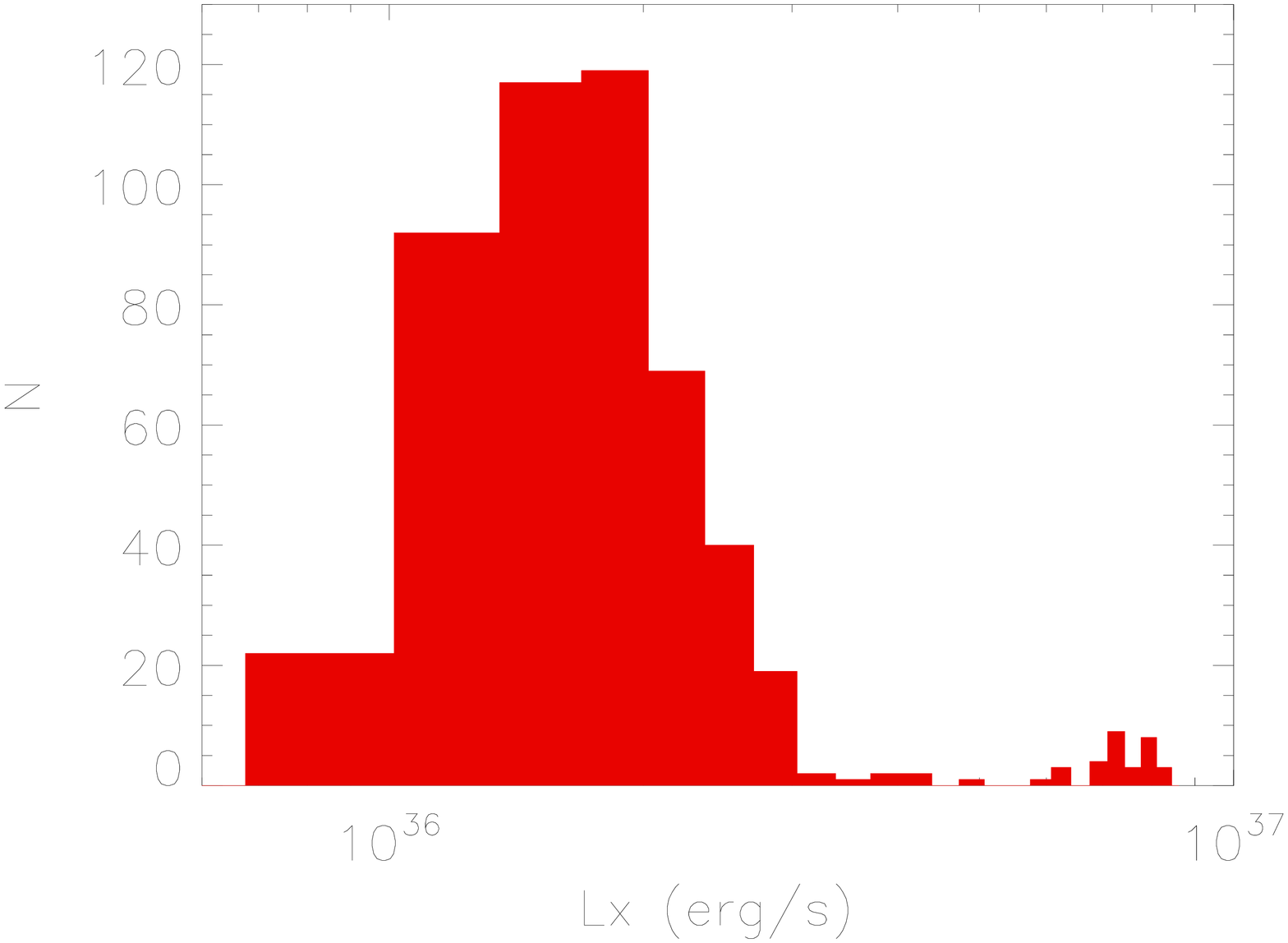} &  
\includegraphics[height=6.5cm, angle=0]{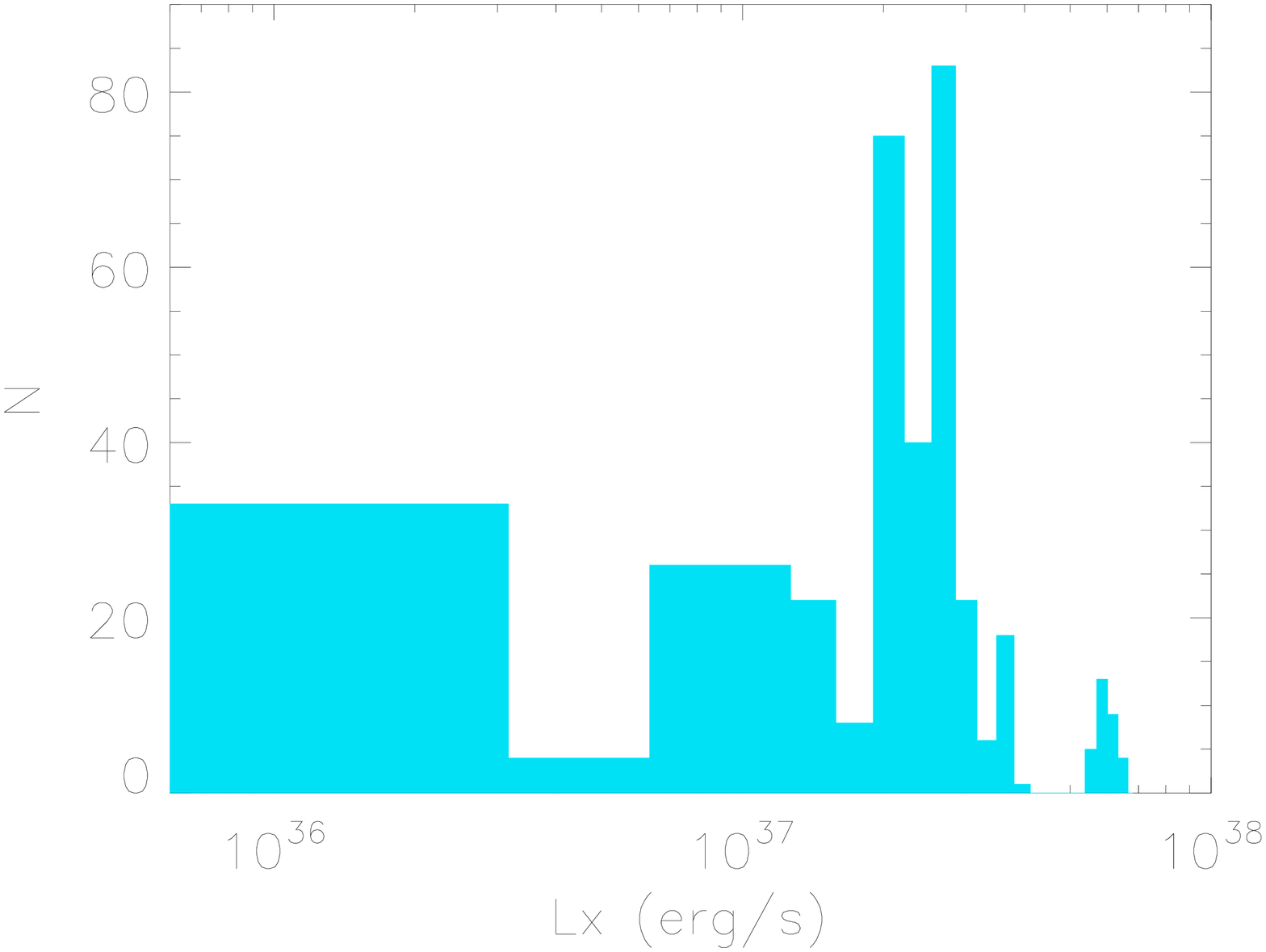}  \\
\includegraphics[height=6.5cm, angle=0]{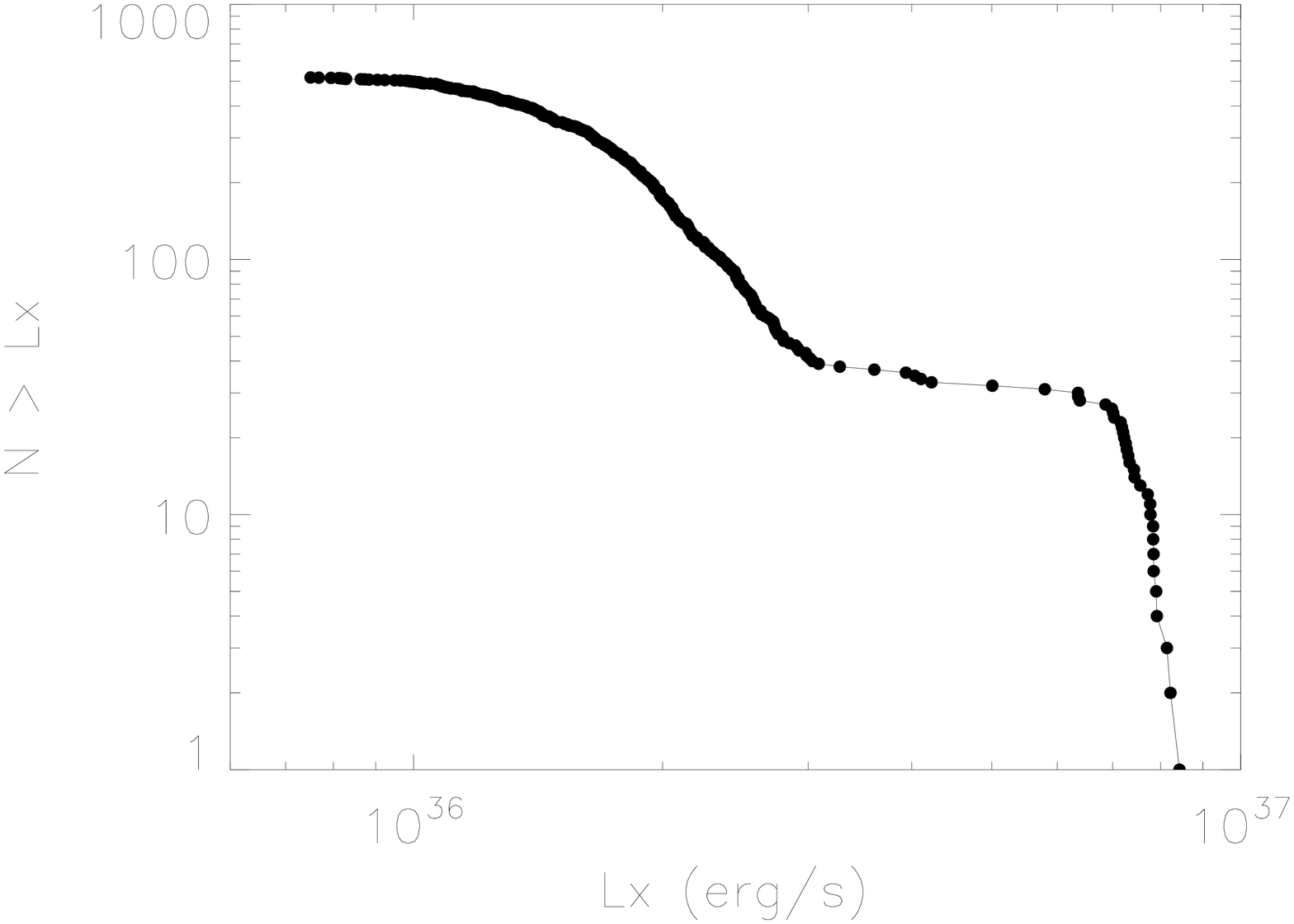} & 
\includegraphics[height=6.5cm, angle=0]{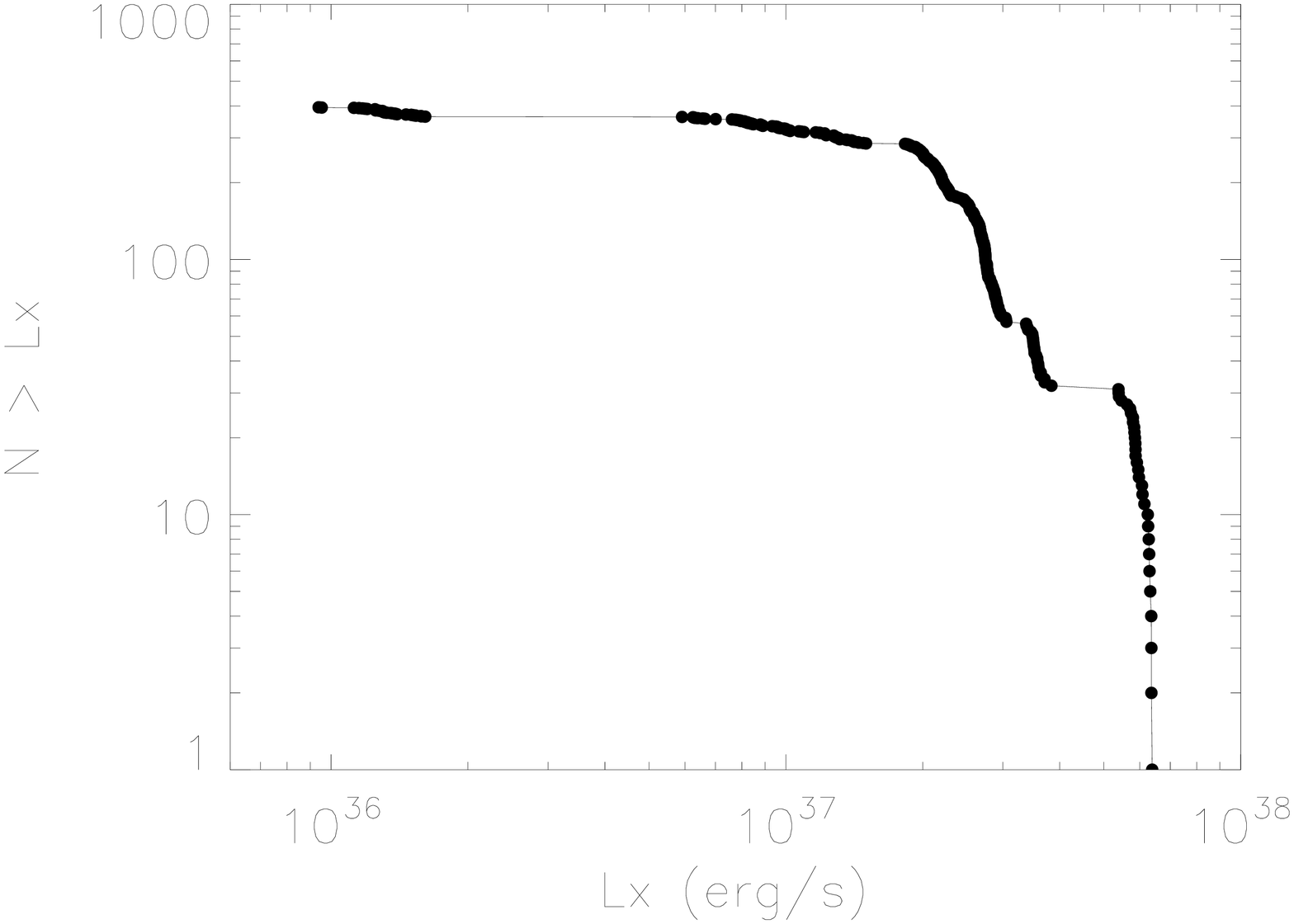} 
\end{tabular}
\caption{Same as Fig.~\ref{fig:lchistocum_vela_sax1818} for two Be/XRTs: 
SAX~J2103.5+4545 (on the left) and EXO~0331+530 (on the right).
}
\label{fig:lchistocum_sax2103_exo0331}
\end{figure*}

We  plot in Figs.~\ref{fig:lchistocum_vela_sax1818} and \ref{fig:lchistocum_sax2103_exo0331},
as an example, the differential luminosity distributions of four sources, to show 
how the features characterizing the CLDs are related with the histograms of the source luminosity, 
detected at ScW level, and to their long-term light curves.
In particular, when a peak is present in the histogram, the correspondent CLD shows a curved feature (a more or less pronounced  knee). 
The narrower the peak in the histogram, the steeper the knee in the CLD. 
A multimodal histogram translates into multiple knees in the CLD (as the one shown by Be/XRTs, that we will discuss below).
From this comparison it is also clear the advantage of using the cumulative, instead of differential, distributions: 
in the cumulative distribution there is no need to arbitrarily bin the data, so that all information is retained.
This is especially crucial when the number of the detections is not very large (SFXTs). 
Moreover, from the normalized CLDs it is immediately possible to visualize the source DC$_{18-50~keV}$, 
the luminosity range of variability, the median luminosity.
Since we focus here on  global (integrated over time) characteristics of HMXB sub-classes, 
we will not discuss the source X--ray light curves further.

%%%%%%%%%%%%%%%%%%%%%%%%%%%%%%%%%%%%%%%%%%%%%%%%%%%%%%%%
%%%%%%%%%%%%%%%%%%%%%%%%%%%%%%%%%%%%%%%%%%%%%%%%%%%%%%%%

%%%%%%%%%%%%%%%%%%%%%%%%%%%%%%%%%%%%%%%%%%%%%%%%%%%%%%%%
%%%%%%%%%%%%%%%%%%%%%%%%%%%%%%%%%%%%%%%%%%%%%%%%%%%%%%%%

     %%%%%%%%%%%%%%%%%%%%%%%%%%%%%%%%%%%%%%%%%%%%%%%%%%%%%%%%%
     \subsubsection{SFXT CLDs} \label{cld:sfxt}
     %%%%%%%%%%%%%%%%%%%%%%%%%%%%%%%%%%%%%%%%%%%%%%%%%%%%%%%%%

In Fig.~\ref{fig:cumdis_sfxt} we show the CLD of the SFXTs, where it is evident the very low (a few per cent) DC$_{18-50~keV}$ 
of the flaring activity from these transients.
We remark that the SFXT CLDs reported here can be considered as the CLDs of the SFXT flares, given the similar timescale 
of both the ScWs and X--ray flare duration. We have already  discussed this point in \citet{Paizis2014}, to which we refer the reader. 
In \citet{Paizis2014} we compared the CLDs of the SFXT flares extracted from about 9 years of \inte\ data, 
with the CLD shown by three classical HMXB systems: two persistent (Vela X--1 and 4U~1700--377) 
and one (likely transient) SgHMXB with highly variable X-ray emission (H~1907$+$09). 
The CLDs of the SFXT X-ray flares were found to be well represented by power-law models,
while  log-normal-like functions were a more plausible representation of the CLD of the other three classical SgHMXBs. 
The SFXT CLDs are confirmed to be well described by power-law models also in the larger data-base we are reporting here (14 years of \inte\ mission instead of 9), showing 
slopes consistent with the ones reported in our previous investigation  \citep{Paizis2014}.  
Indeed, following the same procedure as in \citet{Paizis2014}, 
we characterized the CLDs of SFXT flares
adopting a power-law model. In Table~\ref{tab:powfit} we report the results, quoting
the luminosity of the truncation point adopted for each SFXT flare luminosity distribution, 
the resulting power-law slope and the corresponding Kolmogorov-Smirnov (KS) probability.
The power-law slopes are consistent, within their uncertainties, with
previous values reported by \citet{Paizis2014} analysing 9~yrs of \inte\ data. 
Only in case of IGR~J16418--4532 is the power-law steeper (slope of 2.28$\pm{0.40}$, compared to previous value of 1.31$\pm{0.31}$). 
The steeper slope of the CLD is produced by the most recent flares detected from this source (later than April 2012). These flares are not included in \citet{Paizis2014} - who considered data up to April 2012 - and lie in a narrow range of intermediate luminosities, from 5$\times$10$^{36}$~erg~s$^{-1}$ to about 8$\times$10$^{36}$~erg~s$^{-1}$, resulting in a different overall slope of the distribution.
Moreover, we can now add to our sample two more SFXT sources, given the larger \inte\ database (IGR~J16328--4726 and IGR~J16465--4507).

SFXTs show similar, very small DC$_{18-50~keV}$ ($<$5 per cent). As previously observed, the SFXTs with the highest DC$_{18-50~keV}$, 
in the range from 1 to 5 per cent, are confirmed to be 
the so-called ``intermediate SFXTs'' (IGR~J16418--4532, IGR~J16479--4514 and IGR~J18483--0311), which are more active than other, prototypical, SFXTs.
In the present work two more SFXTs are included in our sample: IGR~J16328--4726 and IGR~J17354--3255, 
that resulted in very low DC$_{18-50~keV}$, as well (0.28 and 0.01 per cent, respectively).

%%%%%%%%%%%%%%%%%%%%%%%%%%%%%%%%%%%%%%%%%%%%%%%%%%%%%%%%%%%%%%%%%%%%%%%%%%%%%%%%%%%%%%%%%%%%%%%%%%%%%%%%%%%%%%%%%%%%%%%
\begin{table}
 \centering
  \caption{  Maximum Likelihood Estimation of the power-law  parameters of the CLD (18--50\,keV) of SFXT flares
}
  \begin{tabular}{@{}lccc@{}}
\hline
   Name             &  Truncation point                 &  Power-law      &   KS prob.  \\
                    &  L$_{X}$ (10$^{35}$ erg~s$^{-1}$) &         slope              &            \\
\hline
XTE~J1739--302      &      6.3                          & 2.85$\pm{0.73}$ &   0.999   \\ 
IGR~J17544--2619    &      4.1                          & 2.12$\pm{0.32}$ &   0.947    \\
SAX~J1818.6--1703   &      1.5                          & 1.76$\pm{0.29}$ &   0.993    \\
IGR~J16418--4532    &     50.0                          & 2.28$\pm{0.40}$ &  0.934     \\ 
IGR~J16479--4514    &      3.4                          & 2.56$\pm{0.32}$ &  0.999     \\
IGR~J18483--0311    &      4.1                          & 2.24$\pm{0.23}$ &  0.999     \\  
IGR~J18483--0311    &      9.7                          & 3.25$\pm{1.14}$ &  0.999     \\ 
IGR~J18450--0435    &      11.2                         & 2.12$\pm{0.76}$ &  0.989     \\
IGR~J18410--0535    &      3.6                          & 2.65$\pm{1.04}$ &  0.985      \\
IGR~J11215--5952    &      11.0                         & 1.76$\pm{0.60}$ &  0.805     \\
IGRJ~16328--4726   &      20.0                          & 3.2$\pm{1.7}$   &  0.968      \\
IGR~J16465--4507   &      21.0                          & 1.5$\pm{0.6}$  &  0.996     \\
\hline
\end{tabular}
\label{tab:powfit}
\end{table}
%%%%%%%%%%%%%%%%%%%%%%%%%%%%%%%%%%%%%%%%%%%%%%%%%%%%%%%%%%%%%%%%%%%%%%%%%%%%%%%%%%%%%%%%%%%%%%%%%%%%%%%%%%%%%%%%%%%%%%%

     %%%%%%%%%%%%%%%%%%%%%%%%%%%%%%%%%%%%%%%%%%%%%%%%%%%%%%%%%
     \subsubsection{SgHMXB CLDs} \label{cld:sg}
     %%%%%%%%%%%%%%%%%%%%%%%%%%%%%%%%%%%%%%%%%%%%%%%%%%%%%%%%%

SgHMXB CLDs are shown in Fig.~\ref{fig:cumdis_sg}, together with two sources
hosting an early-type giant companion, LMC~X-4 and Cen~X--3.
Most of these  CLDs show a completely different behaviour with respect to SFXTs, 
not only in their  high DC$_{18-50~keV}$ (indeed, many SgHMXB are classified in the literature as persistent X--ray sources) 
but also in the shape of their CLD, that appears
closer to a log-normal function (e.g. Vela X--1; see also Fig.~\ref{fig:lchistocum_vela_sax1818}, left panels). 
We did not attempt to perform an analytical description of these CLDs, which display many different shapes.
Indeed, in these brightest sources,  many effects, like  orbital and super-orbital modulations of the X-ray light curves, as well as 
aperiodic trends, can modify the shape of the  CLDs.

Globally, the CLDs of classical HMXBs hosting giant or supergiant donors appear unimodal and can be 
more (like in SMC~X--1 and LMC~X--4) or less (like in OAO~1657--415, Cen~X--3) peaked.
The highest luminosities are reached by SMC~X--1 and LMC~X--4, that are Roche lobe overflow (RLO) systems,
allowing the formation of an accretion disc, sustaining a much higher mass transfer rate onto the pulsar than in wind-fed systems.
Excluding these two  systems, the HMXBs shown in Fig.~\ref{fig:cumdis_sg} reaching the highest luminosity
are GX~301--2, OAO~1657-415 and Cen X--3. These findings are in agreement with what is known about these sources. 
GX~301--2 is the only system with a B-type hypergiant donor (Wray~977) that  displays a huge mass loss rate ($\mdot$=10$^{-5}$~$\msun$~yr$^{-1}$; \citealt{Kaper2006})
and a slow wind (terminal velocity, v$_{\infty}$, $\sim$300 km~s$^{-1}$), explaining the high X-ray luminosity compared to other wind accretors.
Cen X--3 is an eclipsing, persistent HMXBs, where RLO is thought to take place, and 
the accretion occurs through a disc  \citep{Bonnet-Bidaud1979}. 
OAO~1657-415 is an eclipsing HMXBs that is located between the Be and SgHMXB systems 
on the Corbet diagram (spin versus orbital period), 
suggested to alternate wind acretion with disc accretion phases, sustaining a higher X--ray luminosity \citep{Audley2006}.
It is interesting to note that the shape of the CLDs in  OAO~1657-415 and Cen X--3 appears very similar, 
probably because of the presence of an accretion disc most of the time  in OAO~1657-415 too.

The eclipsing HMXBs  show a lower DC$_{18-50~keV}$, reduced by the non detection with IBIS/ISGRI during  X--ray eclipses.
Also short, intrinsic drops of flux (the so-called off-states), might led to a lower DC$_{18-50~keV}$ at 2~ks time bin \citep{Kreykenbohm2008, Sidoli2015vela}. 
In other sources, the low DC$_{18-50~keV}$  is to be ascribed only to an intrinsic X--ray intensity variability.
This is the case of IGR~J19140+0951, where the large variability  of its X--ray flux is  confirmed at softer energies \citep{Sidoli2016}. 
The presence of the SgHMXB H1907+097 in this group, known to be in-between persistent and transient sources \citep{Doroshenko2012}, 
strengthens the interpretation of these sources as truly highly variable SgHMXBs.

There is also a group of SgHMXBs in Fig.~\ref{fig:cumdis_sg} which displays a very low DC$_{18-50~keV}$ 
(below 1~per~cent, IGR~J16207-5129, IGR~J16393-4643, IGR~J18027-2016 and IGR~J18214-1318), 
similar to SFXTs, indicative of either  transient emission, or  persistent but faint emission, just
below the threshold of detectability of IBIS/ISGRI, that might be detected only during some flaring activity.
This phenomenology might suggest that within  HMXBs with supergiant companions there is a  smooth transition  from persistent sources to SFXTs.

\begin{figure*}
\centering
\begin{tabular}{cc}
\hspace{-1.0cm}
\includegraphics[height=7.5cm, angle=0]{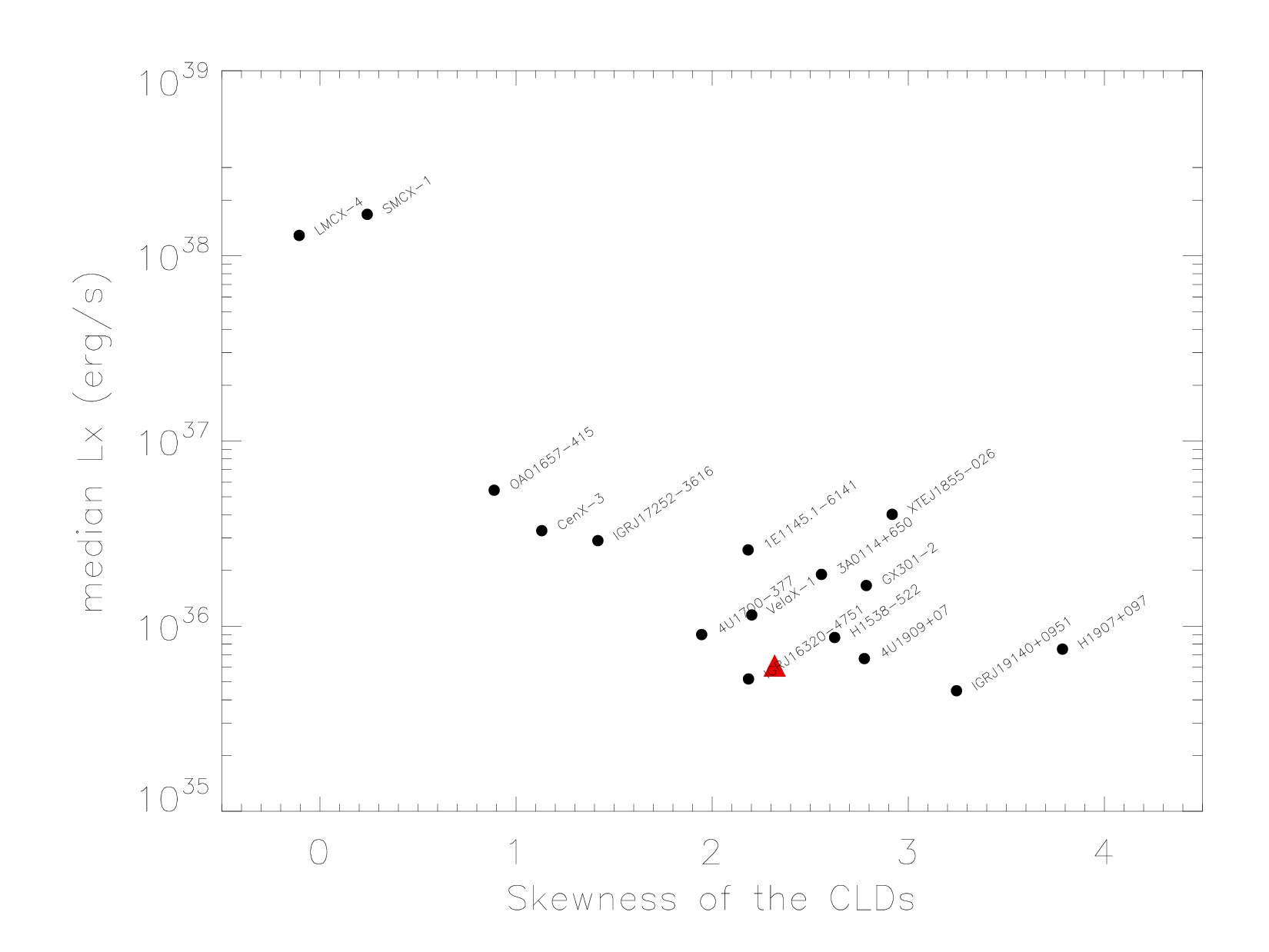} & 
\hspace{-1.0cm}
\includegraphics[height=7.5cm, angle=0]{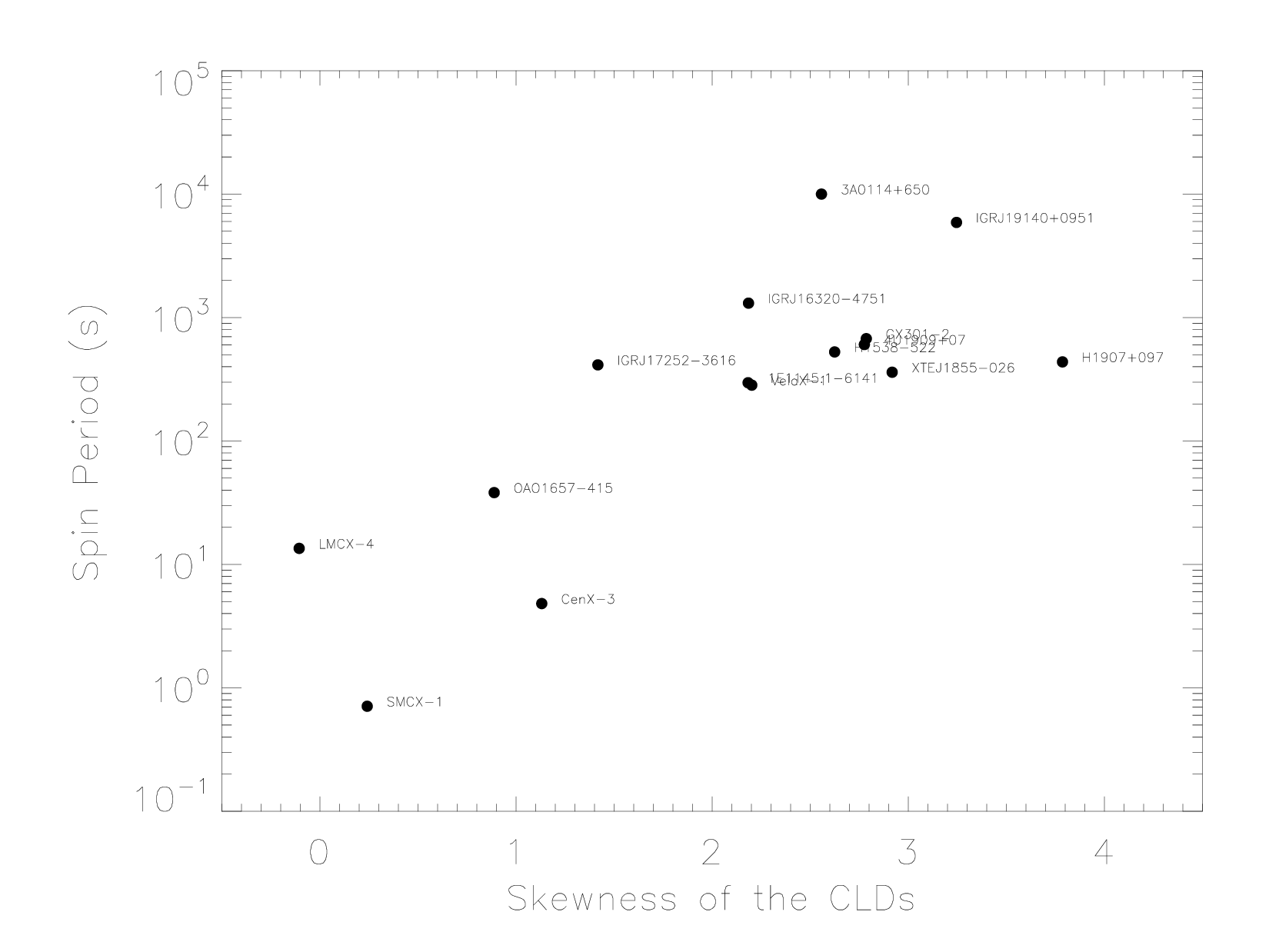}
\end{tabular}
\caption{Median luminosity (18--50\,keV; left panel)  and pulsar spin period (right panel)  versus
 the skewness of the CLDs, for giant and supergiant HMXBs with DC$_{18-50~keV}$ larger than 1~per~cent.
The red triangle on the left panel indicates the position of the SgB[e] binary IGR~J16318--4848 (see Sect.~\ref{cld:pec}).
}
\label{lsfig:corr_skew}
\end{figure*}

From the visual inspection of Fig.~\ref{fig:cumdis_sg} it is tempting to see an evolution 
in the shape of the CLDs, from steeper (at high DC$_{18-50~keV}$ and high median luminosity) 
to flatter ones (at lower DC$_{18-50~keV}$ and lower luminosity).
In particular, there seems to be a transition from LMC X--4 and SMC X--1 CLDs, which share a very similar shape 
(high DC$_{18-50~keV}$, more peaked, less variabile and with a higher median luminosity)
through Cen X--3 and OAO~1657--415 (less peaked CLDs, at lower luminosity) 
down to  IGR~J16320--4751  and IGR~J19140+0951 (with a much flatter CLD,  lower 
luminosity and lower DC$_{18-50~keV}$, almost announcing the power-law-like CLDs of the SFXT flares).
To investigate this apparent trend in a more quantitative way, we calculated the skewness of the SgHMXB luminosity distributions, 
defined as follows:
\begin{equation}
Skewness_j = \frac{1}{N}\sum_{i=0}^{N-1} \left(\frac{x_i - \bar{x}}{\sqrt{Variance_j}}\right)^{3}
\end{equation}
where, for each source j, N is the total number of detections, $x_i$ is the source luminosity during the i-th observation, and $\bar{x}$ 
is the mean luminosity. 
We selected the best defined and populated distributions,  
with a DC$_{18-50~keV}$ larger than 1~per~cent,
and we explored their dependence on the median luminosity (Fig.~\ref{lsfig:corr_skew}, left panel).
We found an anticorrelation,  confirmed by the adoption of the Spearman's correlation  coefficient, r$_s$  (r$_s$=$-$0.685, n=16, p-value=0.00339).
Other relations  between   the  skewness of the luminosity distributions and the source parameters were investigated, 
finding a significant correlation with the pulsar spin period  (Fig.~\ref{lsfig:corr_skew}, right panel),
confirmed by running the Spearman's correlation  
(r$_s$=0.693, n=15, p-value=0.0042).  The source 4U1700-37 was excluded, since the pulsation is not known.
This dependence extends over several orders of magnitude and
might be partly explained by the known anticorrelation between the (maximum) X--ray luminosity 
and the  rotational period  (\citealt{Stella1986}; in faster pulsars, the centrifugal barrier enables accretion only at higher accretion rates).
However, the new finding here is that, not only does the spin anticorrelate with the median of the X-ray luminosity in SgHMXBs,
but also the shape of the luminosity distribution is important: 
the fastest SgHMXB pulsars are found not only in the most luminous sources,
but also where the X-ray luminosity distribution is more symmetric and peaked. 
This is the case of RLO-systems, where an accretion disc forms around the neutron star (LMC~X--4, SMC~X--1, Cen X--3). 
More and more skewed CLDs are present in more variable
(or transient) wind-fed accretors, hampering an effective pulsar spin-up.
Also longer orbital periods and  more eccentric orbits play a role,
resulting in a stronger correlation of the CLD's skewness with the binary
eccentricity [r$_s$=0.606, n=14, p=0.0217] than with the orbital period  [r$_s$=0.376, n=16, p=0.151].
This implies that the periodic modulation of the X--ray emission expected in wind-fed eccentric SgHMXBs 
produces a significantly skewed CLD.

%%%%%%%%%%%%%%%%%%%%%%%%%%%%%%%%%%%%%%%%%%%%%%%%%%%%%%%%
%%%%%%%%%%%%%%%%%%%%%%%%%%%%%%%%%%%%%%%%%%%%%%%%%%%%%%%%

     %%%%%%%%%%%%%%%%%%%%%%%%%%%%%%%%%%%%%%%%%%%%%%%%%%%%%%%%%
     \subsubsection{Be/XRB CLDs} \label{cld:be}
     %%%%%%%%%%%%%%%%%%%%%%%%%%%%%%%%%%%%%%%%%%%%%%%%%%%%%%%%%

Transient Be/XRBs show significantly different CLDs (Fig.~\ref{fig:cumdis_be}), in many respects. 
Many Be systems have DC$_{18-50~keV}$ around $\sim$10~per~cent and show bi- or multi-modal CLDs: wave-like structures in the CLDs 
correspond to peaks in the histograms (see  Fig.~\ref{fig:lchistocum_sax2103_exo0331}),
indicative of different populations of outbursts, or multi-peaked outbursts, and/or transitions to different regimes.
In the CLD of two transients, EXO~2030+375 and SAX~J2103.5+4545, an almost horizontal plateaux (more or less populated by detections) 
is located at intermediate luminosities, a bridge between two different kind of outbursts, 
the faintest (and more frequent) ones and the most luminous (and less frequent), ones. 
This step-like CLDs shape is the signature of Be/XRTs showing both Type~I and Type~II outbursts. 

Be/XRTs are known to display two kinds of outbursts (\citealt{Stella1986, Negueruela1998}, 
see also \citealt{Reig2011} and \citealt{Kuhnel2015} for reviews).
Type I (also named ``normal'') outbursts are periodic and triggered near the periastron passage, by an enhanced accretion rate
onto the neutron star (NS) by material captured from the circumstellar decretion disc, 
expelled by the Be companion \citep{Negueruela2001, Negueruela2001b, Okazaki2001}. 
Their duration is usually a fraction (0.2--0.3) of the orbital period.
The physical mechanism driving the formation of the Be disc is still unclear, although it 
might be due to both rapid rotation and nonradial pulsations \citep{Baade2017}.
Type II (also named ``giant'') outbursts can be observed at any orbital phase and are more luminous (and rarer) than
Type I outbursts, sometimes reaching the Eddington luminosity. They can show a duration much larger than one orbital cycle 
and are thought to be caused by major changes in the structure of the Be decretion disc, likely linked to some warping episodes \citep{Okazaki2013}, 
although the details of the accretion mechanism remain elusive. In any case, there is observational evidence that the Be circumstellar
disc undergoes a quasi-cyclic build-up and desruption \citep{Negueruela2001b}.
Also intermediate-luminosity outbursts have been observed \citep{Caballero2016},
making the situation more complicated. 

Different spectral states have been recognized in Be/XRTs during giant outbursts \citep{Reig2008, Reig2013}, from two different
patterns  observed in the hardness-intensity diagrams:
a horizontal branch, which corresponds to a low-intensity state, and a diagonal branch, emerging only 
when the X--ray luminosity exceeds a critical limit, L$_{crit}$.
This critical value depends on the NS magnetic field strength as 
L$_{crit}$ $\sim$ 1.28$\times$10$^{37}$ (E$_{cyc}$~/~10~keV)$^{16/15}$~erg~s$^{-1}$, where E$_{cyc}$ is the centroid energy of the
 cyclotron  resonance  scattering feature (CRSF; \citealt{Becker2012}). 
Two different accretion regimes, above and below L$_{crit}$ (that represents a threshold for the formation
of a radiative-dominated shock that decelerates the accretion flow above the NS surface), also explain the 
 bimodal variation (correlation or anti-corrrelation) of the energy of the CRSF 
with the X--ray luminosity  \citep{Becker2012}. 
In Table~\ref{tab:propeller} (col. 4) we list the critical luminosities for sources (a few Be/XRBs, SgHMXBs, and one SFXT) 
where cyclotron lines have been discovered.
Sometimes, spin-up episodes during Be/XRBs outbursts have been observed, indicating the
formation of an accretion disc around the NS \citep{Baykal2002}.

In the CLD of the Be transient EXO~0331+53 two empty horizontal plateaux are present, characterized by luminosity jumps. 
Since we are dealing with cumulative luminosity distributions that are integrated quantities where the temporal behaviour is lost,
we cannot say if the non-detections are real or if it is because there are no \inte\ observations covering these luminosity states. 
The luminosity jumps might indicate the transition to a different regime,  possibly the on-set of a propeller state 
(e.g., a centrifugal inhibition of the accretion, \citealt{Campana2017}).
The  limiting X--ray luminosity for the onset of the propeller  depends on the pulsar spin period and its magnetic field,
as follows  \citep{Campana2002}:

\begin{equation} \label{eq:propeller}
L_{\rm propeller} \simeq 3.9 \times 10^{37} \xi^{7/2} B_{12}^2 P_{\rm spin}^{-7/3} M_{1.4}^{-2/3} R_{6}^5 \, \,  {\rm erg~s^{-1}}
\end{equation}

\noindent where  the NS magnetic field, B$_{12}$, is in units of
$10^{12}$ G, the pulsar spin period, P$_{\rm spin}$, is in  seconds;
M$_{1.4}$ and R$_{6}$ are the NS mass and radius, in units of 1.4$\msun$ and $10^6$ cm. 
We here assume $\xi$=1, appropriate for spherical accretion (note that $\xi$=0.5 is for disc accretion).

Assuming the NS magnetic field strength reported by \citet{Revnivtsev2015} 
(their Table~1, as estimated from the CRSF centroid energy), 
and the pulsar spin periods, we obtain the
lowest X--ray luminosity listed in Table~\ref{tab:propeller}.
However, L$_{\rm propeller}$  is a bolometric luminosity. 
Although we have never attempted to extrapolate hard X-ray luminosities to a broader (bolometric) energy range,
we can obtain a rough idea of the conversion factor from the energy range 18--50 keV to a wider band  (0.1--100 keV),
re-analysing the \xmm\ plus \nustar\ simultaneous spectrum of an accreting pulsar  during an outburst, 
the SFXT IGR~J11215--5952 \citep{Sidoli2017}. 
We found that  L$_{\rm 0.1-100~keV}$$\sim$3$\times$L$_{\rm 18-50~keV}$.
Comparing (one third of) the values reported in Table~\ref{tab:propeller} with the CLDs, we find that only in two sources 
is there a possible overlap with  L$_{\rm propeller}$: Cen~X-3 and EXO~0331+530.
While in the persistent pulsar Cen X--3 there is no evidence for luminosity jumps in the CLD (possibly indicative of a more 
appropriate $\xi$ value of 0.5 in Eq.~\ref{eq:propeller}), the low luminosity gap
observed in the transient EXO~0331+530 is in agreement with the onset of a propeller state.
However, from the inspection of its hard X--ray light curve, 
we found that this low luminosity gap is actually due to
a missing \inte\ coverage of the declining part of the second outburst, visible in Fig.~\ref{fig:lchistocum_sax2103_exo0331} (top right panel).

Adopting the same conversion factor (of 3) between hard X--ray luminosities and critical luminosities calculated in Table~\ref{tab:propeller},
we conclude that all sources reported in this Table experienced sub-critical regimes, except four systems, where L$_{crit}$ 
might fall in-between the variability range of their X--ray luminosity: GX~301-2 and the three Be/XRTs H~0115+634, EXO~0331+530 and KS~1947+300.

%%%%%%%%%%%%%%%%%%%%%%%%%%%%%%%%%%%%%%%%%%%%%%%%%%%%%%%%%%%%%%%%%%%%%%%%%%%%%%%%%%%%%%%%%%%%%%%%%%%%%%%%%%%%%%%%%%%%%%%%%%%%%%%%%%%%%%%%%%%%%%%%%%%%%%%%%%%%%%%%
%%%%%%%%%%%%%%%%%%%%%%%%%%%%%%%%%%%%%%%%%%%%%%%%%%%%%%%%%%%%%%%%%%%%%%%%%%%%%%%%%%%%%%%%%%%%%%%%%%%%%%%%%%%%%%%%%%%%%%%%%%%%%%%%%%%%%%%%%%%%%%%%%%%%%%%%%%%%%%%%
 \begin{table}
 \centering
\caption{Expected lowest X--ray luminosity  L$_{\rm propeller}$, at the onset of 
the propeller effect (see Eq.~\ref{eq:propeller}), together
with the critical luminosity,   L$_{\rm crit}$.
B is the NS magnetic field in units of $10^{12}$ G, z is the NS gravitational redshift (values taken from \citealt{Revnivtsev2015}).
}
\begin{tabular}{lccc}
\hline
Source       & B/(1+z)         &   L$_{\rm propeller}$         &   L$_{\rm crit}$  \\
         & (10$^{12}$ G)	   &       (erg~s$^{-1}$)       &    (erg~s$^{-1}$) \\	
\hline
Vela~X-1       &       2.1  &  3.3$\times10^{32}$      &        3.4$\times10^{37}$       \\ 
GX~301-2       &       3.0  &  8.8$\times10^{31}$      &        4.9$\times10^{37}$   \\
H~1538-522     &       2.0  &  7.0$\times10^{31}$      &        3.1$\times10^{37}$  \\
H~1907+097      &      1.5  &  6.0$\times10^{31}$      &        2.4$\times10^{37}$   \\
Cen~X-3         &      2.6  &  6.7$\times10^{36}$      &        4.1$\times10^{37}$ \\
IGR~J17544-2619 &      1.5  &  4.1$\times10^{33}$      &        2.3$\times10^{37}$  \\
H~0115+634      &      1.0  &  2.0$\times10^{36}$      &        1.4$\times10^{37}$  \\
EXO~0331+530    &      2.2  &  6.0$\times10^{36}$      &        3.5$\times10^{37}$  \\
X~Per           &      2.5  &  3.7$\times10^{31}$      &        4.0$\times10^{37}$   \\
1A~0535+262     &      4.0  &  1.2$\times10^{34}$      &        6.5$\times10^{37}$  \\
GRO~J1008-57    &      6.7  &  4.4$\times10^{34}$      &        1.1$\times10^{38}$   \\
KS~1947+300     &      1.0  &  4.2$\times10^{34}$      &        1.6$\times10^{37}$   \\
\hline
\end{tabular}
\label{tab:propeller}
\end{table}
%%%%%%%%%%%%%%%%%%%%%%%%%%%%%%%%%%%%%%%%%%%%%%%%%%%%%%%%%%%%%%%%%%%%%%%%%%%%%%%%%%%%%%%%%%%%%%%%%%%%%%%%%%%%%%%%%%%%%%%%%%%%%%%%%%%%%%%%%%%%%%%%%%%%%%%%%%%%%%%%

In other Be/XRBs there is a number of different behaviours: in some sources the bi-modality in the CLD is present, 
but with  no plateaux in clearly  separating  two population of outbursts (Ginga~1843+009).
The same is true for the multi-modality in the CLDs apparent in other Be transients (GRO~J1750--27, 
 H~0115+634, 1A~0535+262, 4U~1901+03, XTE~J1858+034), where outbursts reaching different peak luminosity have been caught 
by \inte\ or where multi-peaked giant outbursts were observed.
In this respect, the \inte\ archival observations suggest a more complex behaviour of Be/XRT outbursts, 
than a simple distinction in ``normal'' and ``giant'' outbursts. 

Another remarkable feature in the Be/XRT CLDs is the sharp, almost vertical, maximum luminosity, indicative of a
constant, flat-topped, giant outburst peak (GRO~J1750--27, EXO~0331+530, EXO~2030+375, H~0115+634, 4U~1901+03, SAX~J2103.5+4545, KS1947+300).

In Fig~\ref{fig:cumdis_be} there are also Be sources located in the lower luminosity
range of the plot. 
A distinctive behaviour is shown by X~Per, which is almost always detected by \inte\ 
and displays a low luminosity with a median value of 2.4$\times$10$^{34}$~erg~s$^{-1}$.
X~Per is  the prototype of a small sub-class of Be~XRBs showing persistent and low luminosity ($\sim$10$^{34}$~erg~s$^{-1}$),
with long spin periods and wide, low eccentricity, orbits \citep{Reig1999, Pfahl2002}. 
Other Be/XRBs belonging to this group are  RX~J0146.9+6121, 4U~1036-56 (aka RX~J1037.5-564), RX~J0440.9+4431 \citep{LaPalombara2009, Reig2011}, 
with the first two sources also belonging to our sample.
However, as observed by \inte,  RX~J0146.9+6121 and 4U~1036-56 appear as transients, 
with a very low duty cycle (0.11 and 0.35 per cent, respectively), likely the
high luminosity tail of a flaring  variable X--ray emission.

     %%%%%%%%%%%%%%%%%%%%%%%%%%%%%%%%%%%%%%%%%%%%%%%%%%%%%%%%%
     \subsubsection{Other sources: BH and peculiar source CLDs} \label{cld:pec}
     %%%%%%%%%%%%%%%%%%%%%%%%%%%%%%%%%%%%%%%%%%%%%%%%%%%%%%%%%

\begin{figure}
\begin{center}
\centerline{\includegraphics[width=8.cm]{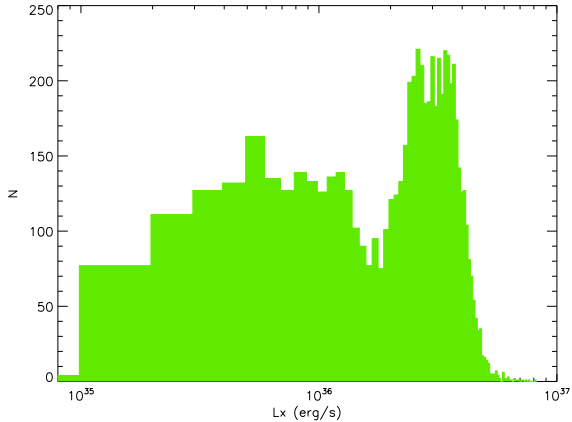}}
\caption{Cyg X--1 luminosity distribution (18--50\,keV). Here the bimodality is more evident than in the correspondent CLD shown in 
Fig.~\ref{fig:cumdis_other}.
}
\label{fig:histo_cygx1}
\end{center}
\end{figure}

%%%%%%%%%%%%%%%%%% %%%%%%%%%%%%%%%%%%%%%%%%%%%%%%%%%%%%%%%%%%%%%%%%%%%%%%%%%%%%%%%%%%%%%%%%%%%%%%%%%%%%%%%
In Fig.~\ref{fig:cumdis_other} the CLDs of other massive binaries,  black hole binaries (BHB) together with some peculiar sources  
not included in the three main sub-classes, are shown.
The highest DC$_{18-50~keV}$ and the largest variability is shown by the persistent BHB Cyg X-1. It is characterized by 
a bimodal distribution of its hard X--ray luminosity distribution, more clearly evident 
in the histogram of the luminosity detections reported in Fig.~\ref{fig:histo_cygx1}.
In fact, this bimodality is not very clear from its normalized cumulative distribution, since the logarithmic scale compresses the curve, close to one.
A more frequent source state peaks at 3$\times$10$^{36}$~erg~s$^{-1}$, together a second peak 
at a luminosity a factor of $\sim$5 fainter (Fig.~\ref{fig:histo_cygx1}). 
We  ascribe this behaviour to its well known spectral variability between two main states, a low-hard  versus a high-soft state. 
A transition to the soft state  was indeed reported with \inte\ in middle of 2010, 
after a long period spent by Cyg~X--1 in hard state \citep{Jourdain2014}.
Also an intermediate state is known in the source (see, e.g., \citealt{Rodriguez2015} and references therein).
Moreover, from its CLD another feature (an ankle) is evident: an excess of hard X-ray detections 
above 5$\times$10$^{36}$~erg~s$^{-1}$, with respect to the extrapolation of the distribution at high luminosity. 
Cyg X-1 is also known to show orbital modulation of the X--ray flux, hence also this kind of variability is expected
to affect its CLD.
Note that we used the same conversion factor from count rate to unabsorbed flux, for all sources, including Cyg~X--1, although
it is known that its power-law photon index, $\Gamma$, ranges from 1.4 (hard state) to 2 in the soft state, through an intermediate
state with $\Gamma$=1.7 \citep{Grinberg2013}. An in-depth investigation of the different spectral states of Cyg~X--1 with \inte\
is beyond the scope of the paper and has already been performed \citep{Grinberg2013b}. 
%
%%%%%%%%%%%%%%%%%%%%%%%%%%%%%%%%%%%%%%%%%%%%%%%%%%%%%%%%%%%%%%%%%%%%%%%% 
%%%%%%%%%%%%%%%%%%%%%%%%%%%%%%%%%%%%%%%%%%%%%%%%%%%%%%%%%%%%%%%%%%%%%%%%

Cyg~X-3 is a persistent source  with one of the highest DC$_{18-50~keV}$ (93.5 per cent). 
It is the only HMXB in our Galaxy known to host a Wolf-Rayet star as a donor \citep{Kerkwijk1996}. 
The nature of the compact object is still unknown \citep{Zdziarski2013}, 
although in the literature Cyg X--3 is often discussed as a WR-BH binary system \citep{Esposito2013}.
The amplitude of its hard X--ray variability is a factor of $\sim$20, at most. 
An orbital modulation of its X--ray flux is known, as well as its state dependence \citep{Weng2013}. 
Its CLD at 18--50 keV appears unimodal and peaks at 10$^{37}$~erg~s$^{-1}$.

SS~433 is another peculiar, unique source in our Galaxy, thought to be in an evolutionary stage 
with supercritical accretion onto a BH \citep{Cherepashchuk2013}.
It appears as a transient source as observed by \inte, with a DC$_{18-50~keV}$ of 15~per~cent.
The CLD shape is unimodal,  with a median hard X--ray luminosity L$_{X}$$\sim$8$\times$10$^{35}$~erg~s$^{-1}$.

At lower DC$_{18-50~keV}$ are located the wind-fed accretor 3A~2206+543 \citep{Wang2013} 
and the symbiotic X--ray transient XTE~J1743-363 \citep{Bozzo2013}.
The former shows an unimodal CLD that above 3$\times$10$^{35}$~erg~s$^{-1}$ can be described by 
a steep power-law, similar to the high luminosity
part  (L$_{X}>$4$\times$10$^{36}$~erg~s$^{-1}$) of the CLD of Vela X--1 \citep{Paizis2014}. 
XTE~J1743-363 was included in our sample of sources, although it is not an HMXB, since symbiotic binaries sometimes show flaring activity that
can resemble the SFXT flares, and can be misclassified as SFXTs, before the nature of the companion is unveiled.
Its long-term behaviour is peculiar: the DC$_{18-50~keV}$ is very low, similar to many SFXTs, but 
the shape of its CLDs is  different, much steeper than in SFXTs, and closer to the 
high luminosity part of CLDs of wind-fed HMXBs, like Vela~X--1.

The last source in Fig.~\ref{fig:cumdis_other} is IGR~J16318--4848 \citep{Courvoisier2003}, 
harboring a B[e] supergiant companion \citep{Filliatre2004}, a rare object in our Galaxy (similar to XTE~J0421+56/CI~Cam, \citealt{Boirin2002}).
From the point of view of its CLD, it is similar to what displayed by the SgHMXBs.
Also its CLD skewness and median luminosity fits well within 
the anticorrelation shown by SgHMXBs (Fig~\ref{lsfig:corr_skew}, left panel). 
From this similarity, the correlation of the pulsar spin periods with the skewness (marked with a red triangle in Fig~\ref{lsfig:corr_skew}, right panel),
allow us to predict a rotational period in IGR~J16318--4848 ranging from a few hundreds to a few thousands seconds.

%%%%%%%%%%%%%%%%%%%%%%%%%%%%%%%%%%%%%%%%%%%%%%%%%%%%%%%%%%%%%%%%%%%%%%%%%%%%%%%%%%%%%%%%%%%%%%%%%%%%%%%%%%%%%%%%%
%%%%%%%%%%%%%%%%%%%%%%%%%%%%%%%%%%%%%%%%%%%%%%%%%%%%%%%%%%%%%%%%%%%%%%%%%%%%%%%%%%%%%%%%%%%%%%%%%%%%%%%%%%%%%%%%%

    %%%%%%%%%%%%%%%%%%%%%%%%%%%%%%%%%%%%%%%%%%%%%%%%%%%%%%%%%
     \subsection{HMXBs: hard X--rays into  context of other  properties} \label{discussion:properties}
     %%%%%%%%%%%%%%%%%%%%%%%%%%%%%%%%%%%%%%%%%%%%%%%%%%%%%%%%%

In Table~\ref{tab:inte} we have listed our results on HMXBs from the \inte\ archive, while in  
Table~\ref{tab:literature} some properties of the sample are reported
(distance, orbital period and eccentricity, pulsar spin period), 
together with information about the soft X--ray fluxes (see Sect.~\ref{literature}),
minimum and maximum values in the energy band 1-10~keV, and their ratio (DR$_{1-10~keV}$).

The second aim of our investigation is to put our long-term \inte\ results into context with the overall known source
properties and to characterize the different HMXB phenomenology from the point of view of time-integrated quantities.
Although the variable of time is lost, we gain a global, wider, view of the whole behaviour of our sample.

There are two quantities that have been often used in the literature to characterize SFXTs with respect to other SgHMXBs: 
the dynamic range at soft X--rays (DR$_{1-10~keV}$=F$_{max}$/F$_{min}$) and the source duty cycle at 18--50 keV  (DC$_{18-50~keV}$, \citealt{Sidoli2017review}).
The former energy band choice is due to the fact that instruments observing the sky in the soft X--rays are much more sensitive than 
in the hard X--rays, reaching the true quiescence of these objects (10$^{32}$~erg~s$^{-1}$, e.g. \citealt{zand2005}). 
The latter is because SFXTs were discovered by \inte\  \citep{Sguera2005, Negueruela2006}, 
and IBIS/ISGRI can work as a sort of high-pass filter, catching bright and short flares.

\subsubsection{Source DC$_{18-50~keV}$ and DR$_{1-10~keV}$ } 
\label{sect:source_dc_dyn_1_10}

\begin{figure*}
\centering
\begin{tabular}{cc}
\includegraphics[height=6.3cm, angle=0]{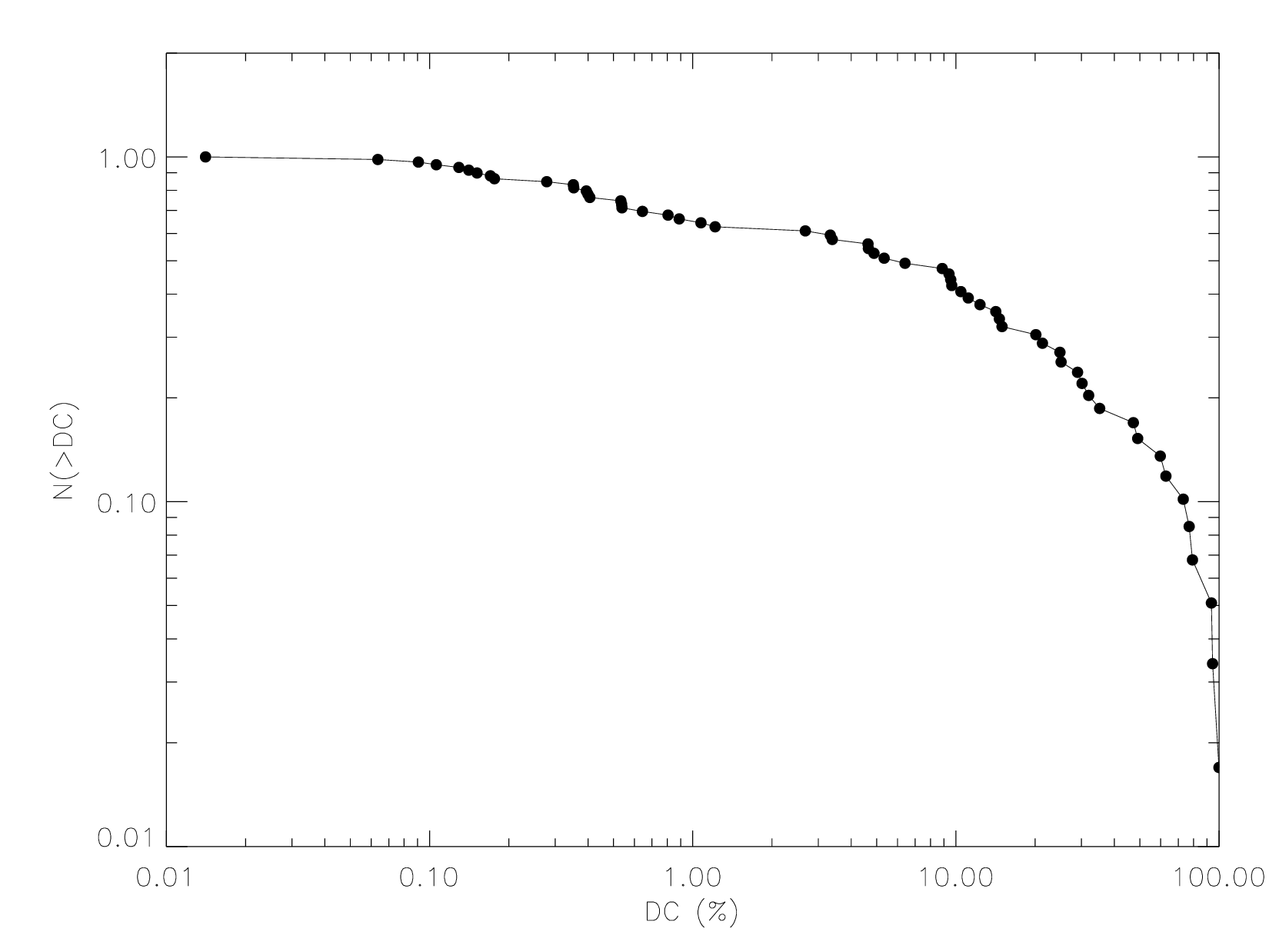} & 
\includegraphics[height=6.3cm, angle=0]{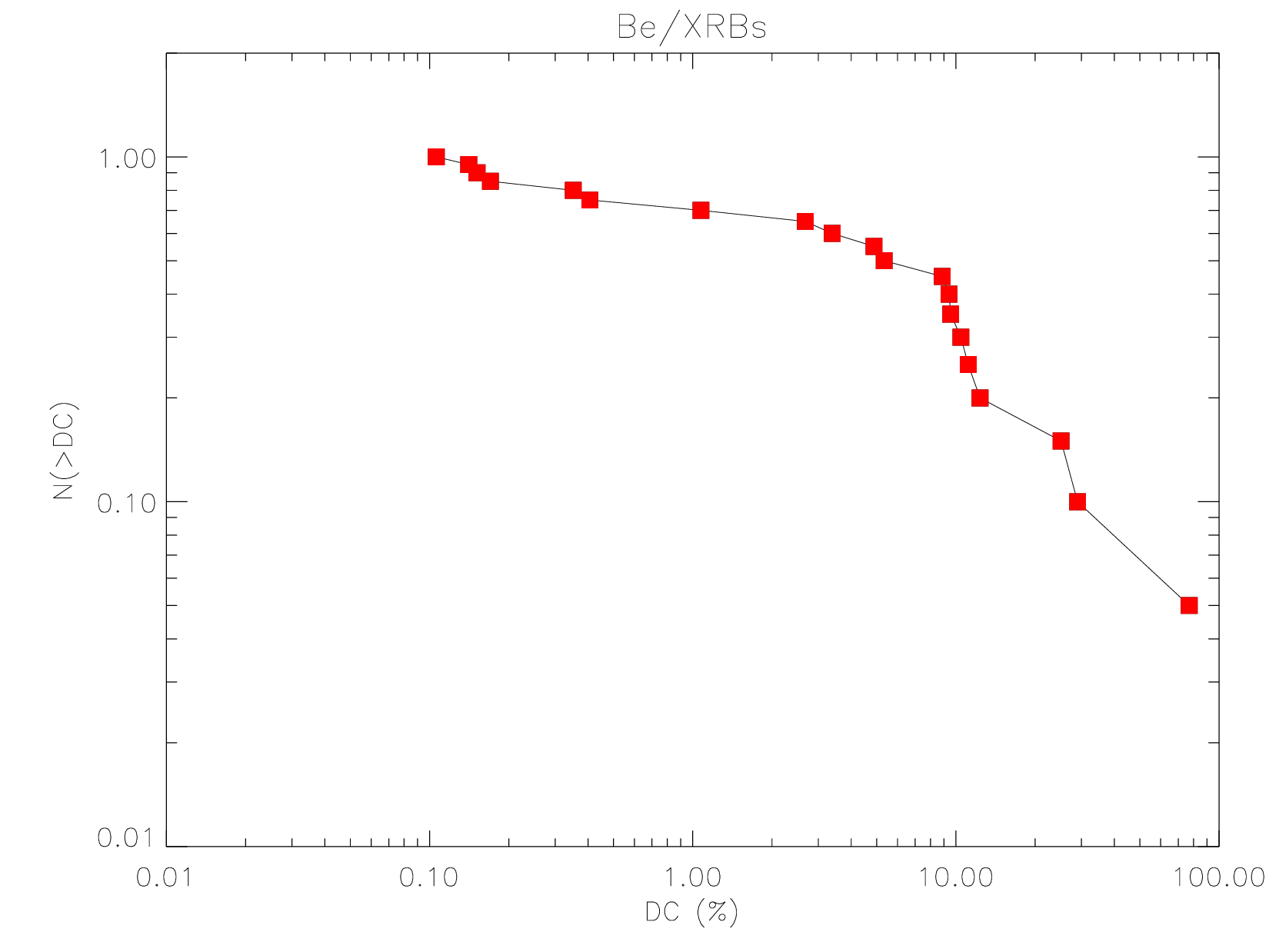} \\
\includegraphics[height=6.3cm, angle=0]{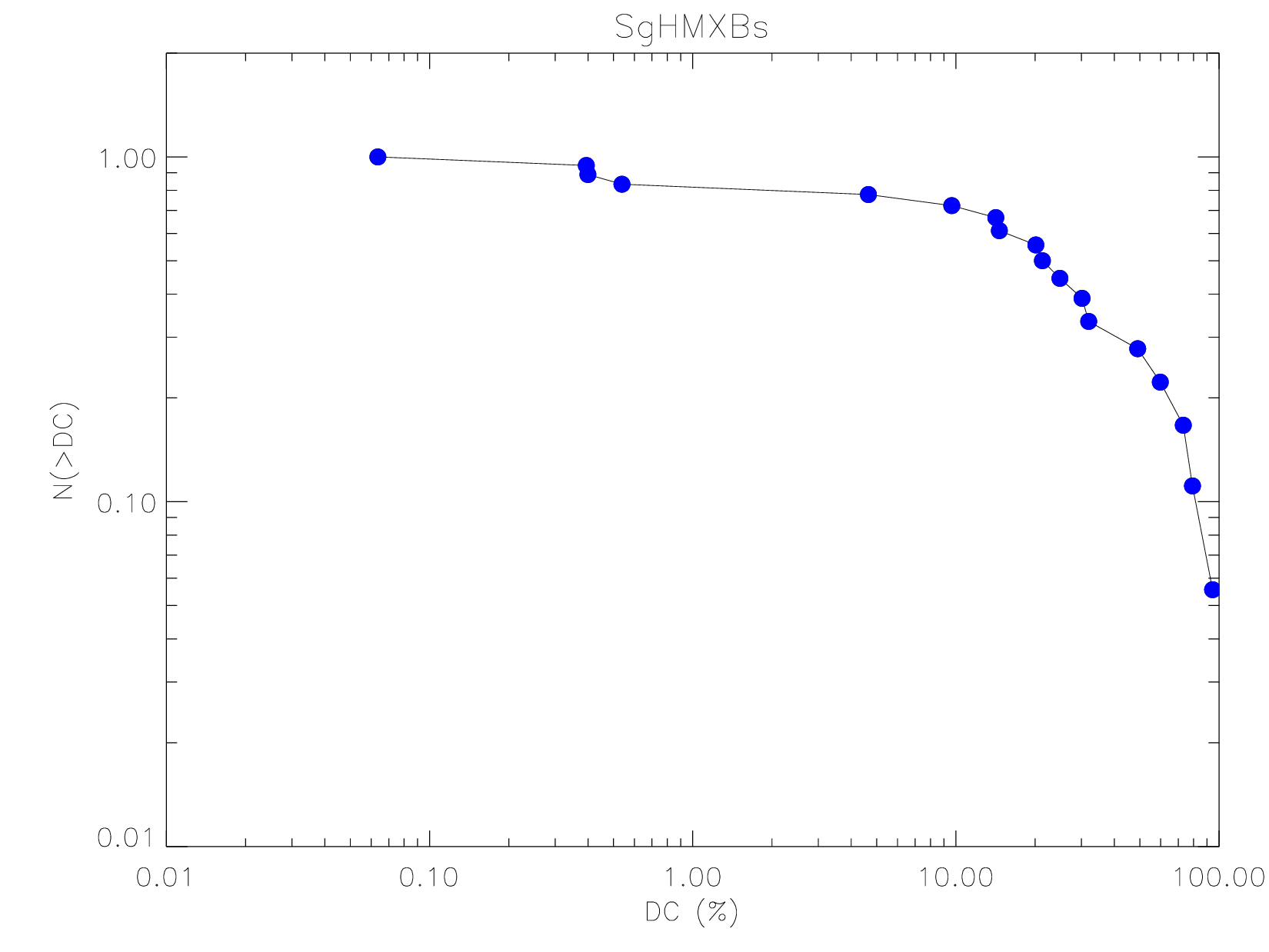} & 
\includegraphics[height=6.3cm, angle=0]{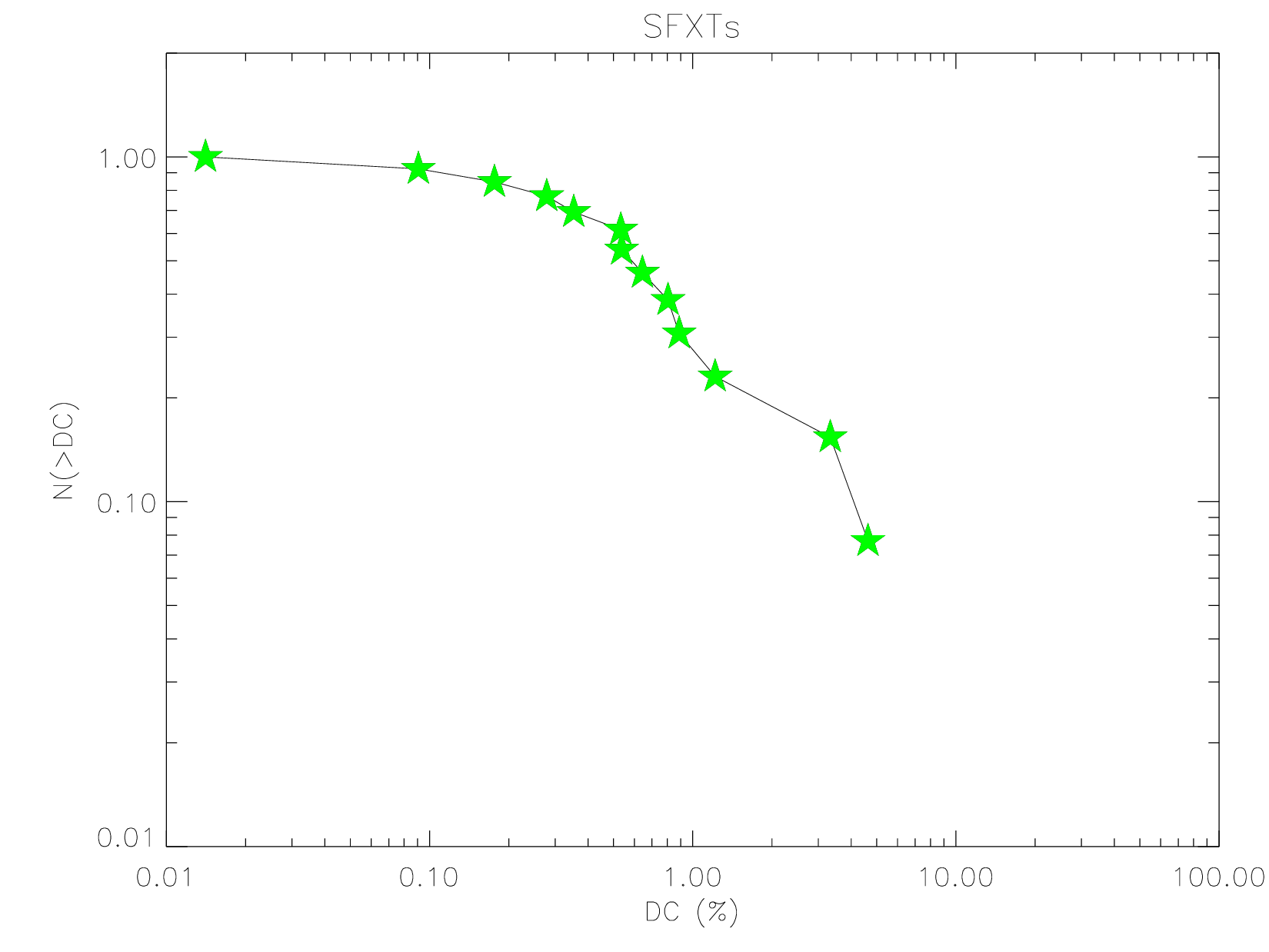}
\end{tabular}
\caption{Normalized cumulative distributions of the 18--50\,keV duty cycles, for all sources (black) 
and for the three sub-classes of Be/XRBs (red), SgHMXBs (blue) and SFXTs (green).}
\label{lsfig:cum_source_dc2}
\end{figure*}

We have investigated the DC$_{18-50~keV}$ and the DR$_{1-10~keV}$ for all sources in our sample, 
to enable a full characterization of the three types of HMXBs.
In Fig.~\ref{lsfig:cum_source_dc2} we plot the cumulative distribution of source DC$_{18-50~keV}$ (considering
SgHMXBs, SFXTs and Be/XRBs)  and then for the three sub-classes, separately. 
This plot clearly demonstrates that SFXTs show a flaring activity for less than 5 per cent of the time,
while the great majority of the SgHMXBs cluster around much higher DC$_{18-50~keV}$ (SgHMXBs include persistent or almost persistent sources). 
However a few SgHMXBs exist that show a very low  DC$_{18-50~keV}$,
suggesting that some of them are truly transient systems, probably SFXTs mis-classified as more ``standard'' 
supergiant systems (IGR~J16207--5129, IGR~J16393--4643, IGR~J18027--2016 and IGR~J18214--1318). Alternatively, they are
faint sources  (probably because of their large distance), just under the sensitivity threshold of the instrument, 
that rarely underwent some short flaring activity, on a timescale of $\sim$2~ks.
Be/XRBs cover a wide range of values, from an almost persistent behaviour ($\sim$77~per~cent, in X~Per), 
through a very frequent duty cycle around 10 per cent displayed by several transient Be sources, down
to very low values of $\sim$0.1~per~cent (RX~J0146.9+6121, AX~J1820.5--1434, XTE~J1543--568 and AX~J1749.1--2733).
%
%%%%%%%%%%%%%%%%%%%%%%%%%%%%%%%%%%%%%%%%%%%%%%%%%%%%%%%%
%%%%%%%%%%%%%%%%%%%%%%%%%%%%%%%%%%%%%%%%%%%%%%%%%%%%%%%%

\begin{figure*}
\centering
\begin{tabular}{cc}
\includegraphics[height=6.3cm, angle=0]{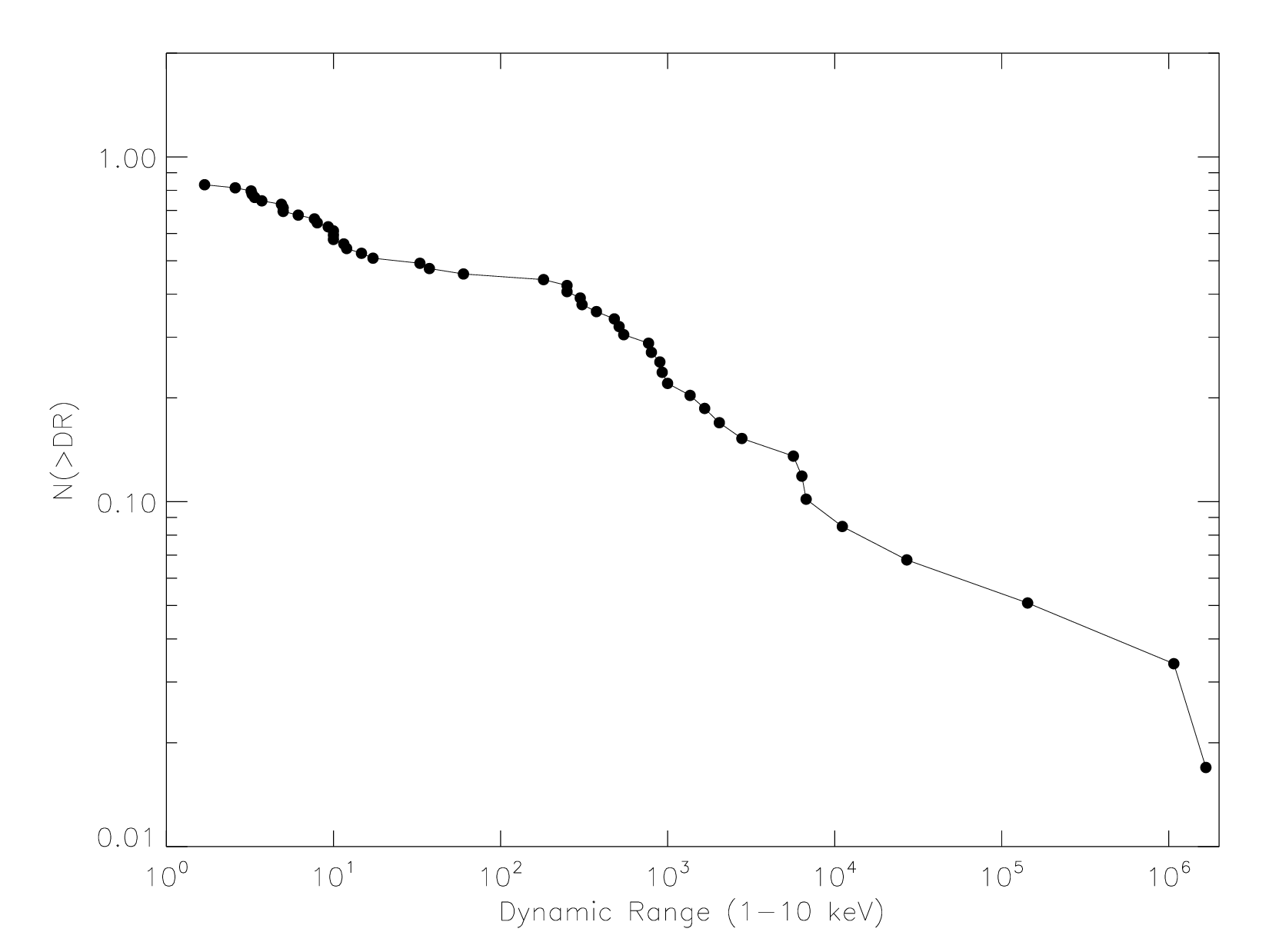} & 
\includegraphics[height=6.3cm, angle=0]{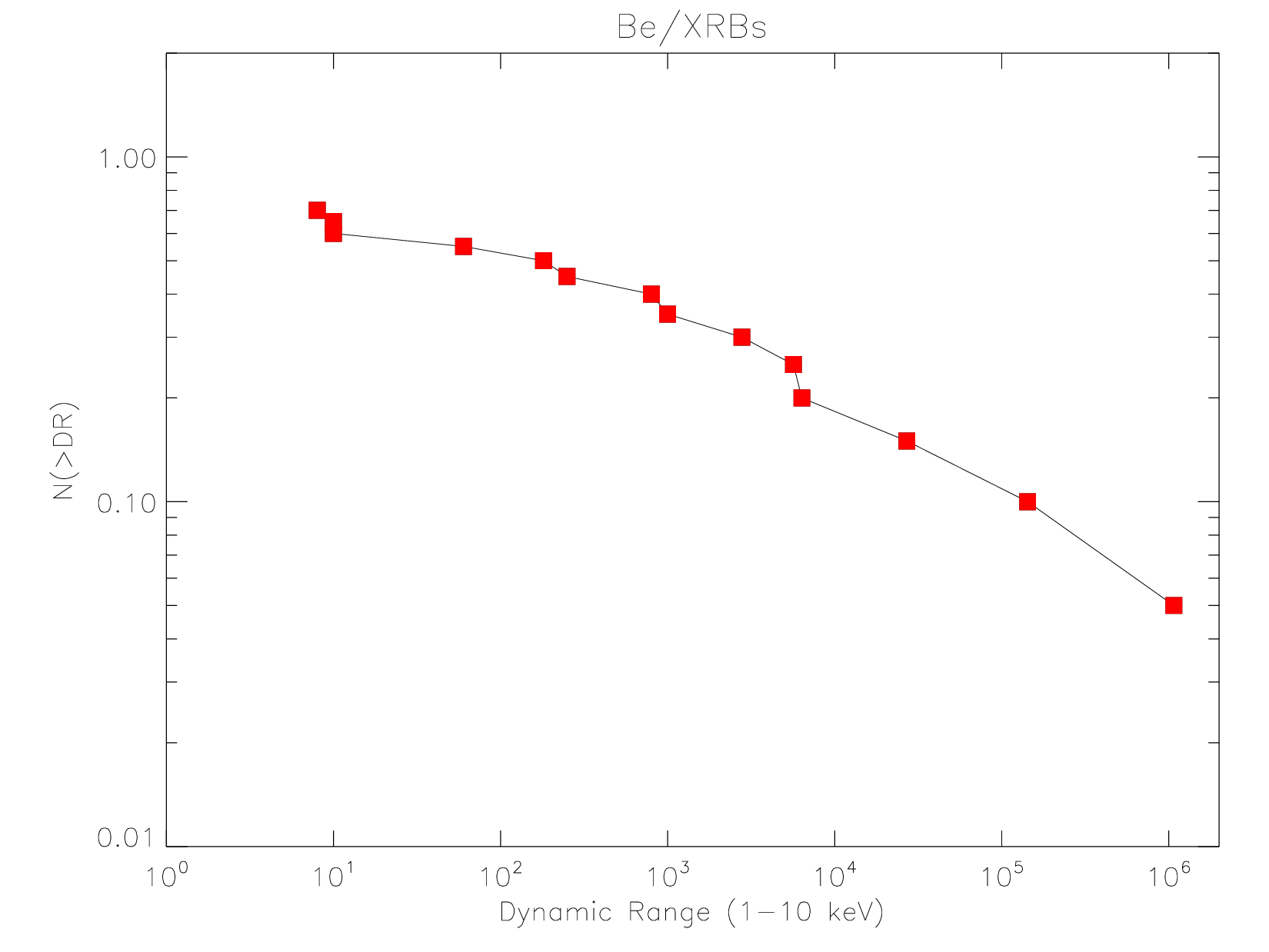} \\
\includegraphics[height=6.3cm, angle=0]{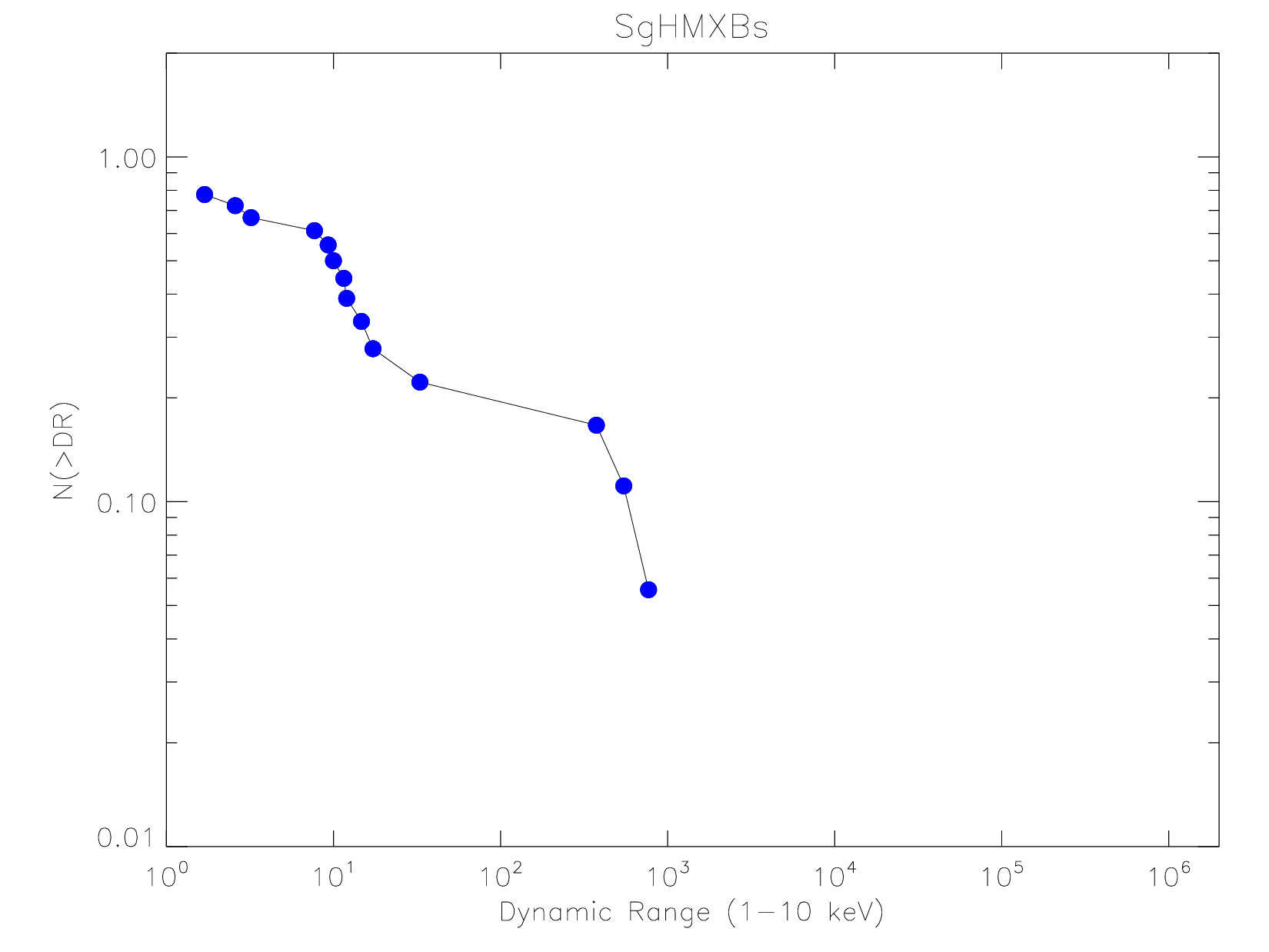} & 
\includegraphics[height=6.3cm, angle=0]{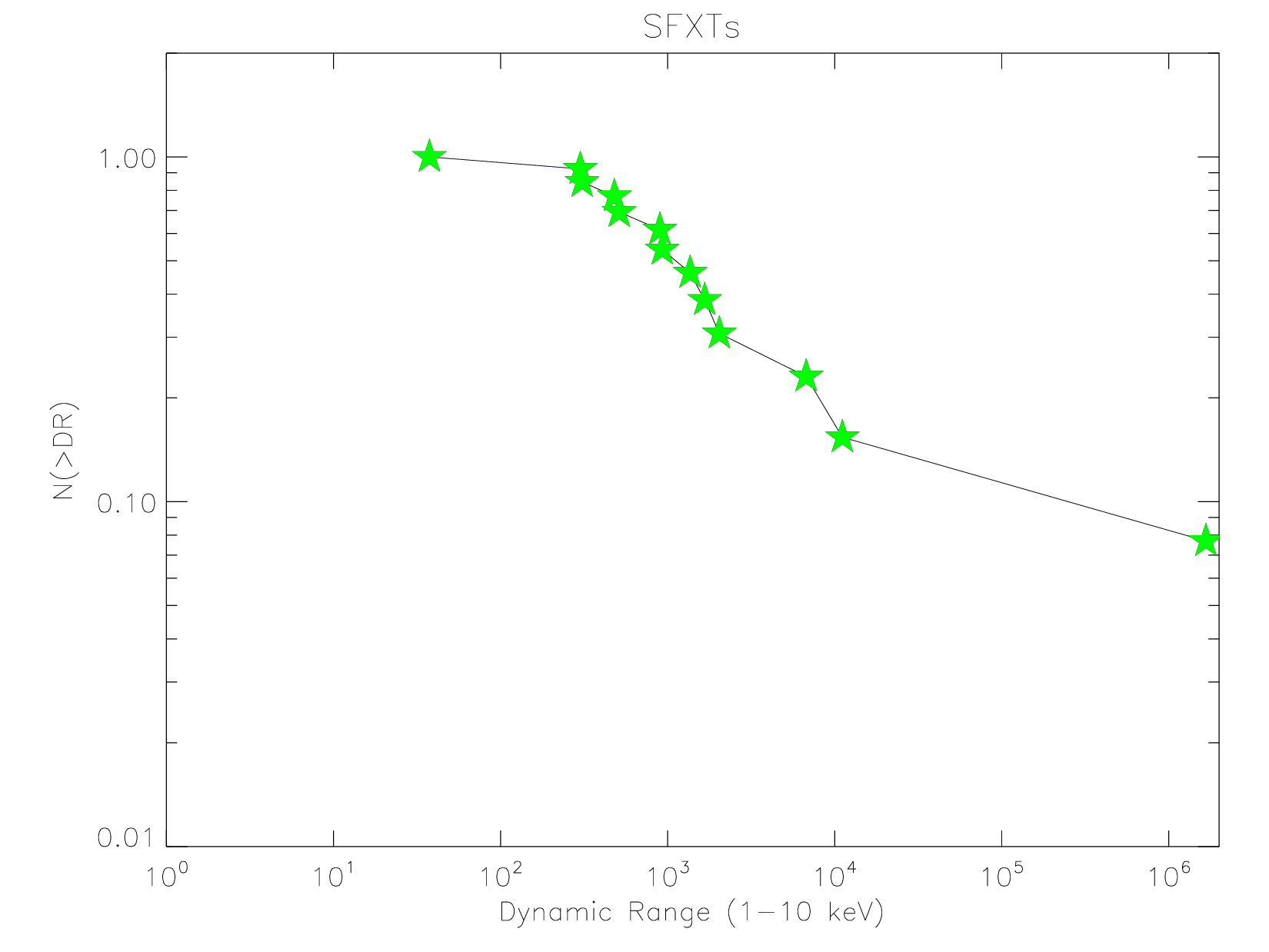}
\end{tabular}
\caption{Normalized cumulative distributions of the 1--10\,keV dynamic ranges, 
for all sources (black) and for the three sub-classes of Be/XRBs (red), SgHMXBs (blue) and SFXTs (green).}
\label{lsfig:cum_dyn_1_10}
\end{figure*}

We found an ample range of values also for the DR$_{1-10~keV}$ (Fig.~\ref{lsfig:cum_dyn_1_10}), covering
six orders of magnitude. The most extreme transients are among the SFXTs (IGR~J17544-2619) and the Be systems (EXO~0331+530),
while when looking at the global distribution, a gap is apparent around DR$_{1-10~keV}$=100,
a value that might be  used as a  threshold  distinguishing (almost) persistent from transient sources.
Again, we unveil an overlapping region of dynamic ranges where some members of the SgHMXB class show a larger variability than
the bulk of the class (DR$_{1-10~keV}$$>$100 in IGR~J18027-2016, H1907+097 and IGR~J19140+0951).
On the contrary, the SFXT IGR~J16465-4507 displays a low DR$_{1-10~keV}$ of $\sim$38. 
The  visual inspection of the cumulative distribution of DR$_{1-10~keV}$ in SFXTs  reveals a power-law-like  shape.
Therefore, it is simple to draw apart both the extreme values
of  IGR~J16465--4507 (overlapping with SgHMXB) and of IGR~J17544--2619. In this latter source, a different physical mechanism might have
produced a very high luminosity (i.e., the formation of a transient accretion disc; \citealt{Romano2015giant}).
In Be/XRBs a large range of DR$_{1-10~keV}$ is covered, demonstrating that the sub-class of Be sources includes many different
X--ray behaviours, with no preferred  variability amplitude.
%
%%%%%%%%%%%%%%%%%%%%%%%%%%%%%%%%%%%%%%%%%%%%%%%%%%%%%%%%
%%%%%%%%%%%%%%%%%%%%%%%%%%%%%%%%%%%%%%%%%%%%%%%%%%%%%%%%

\begin{figure*}
\centering
\begin{tabular}{cc}
\includegraphics[height=6.5cm, angle=0]{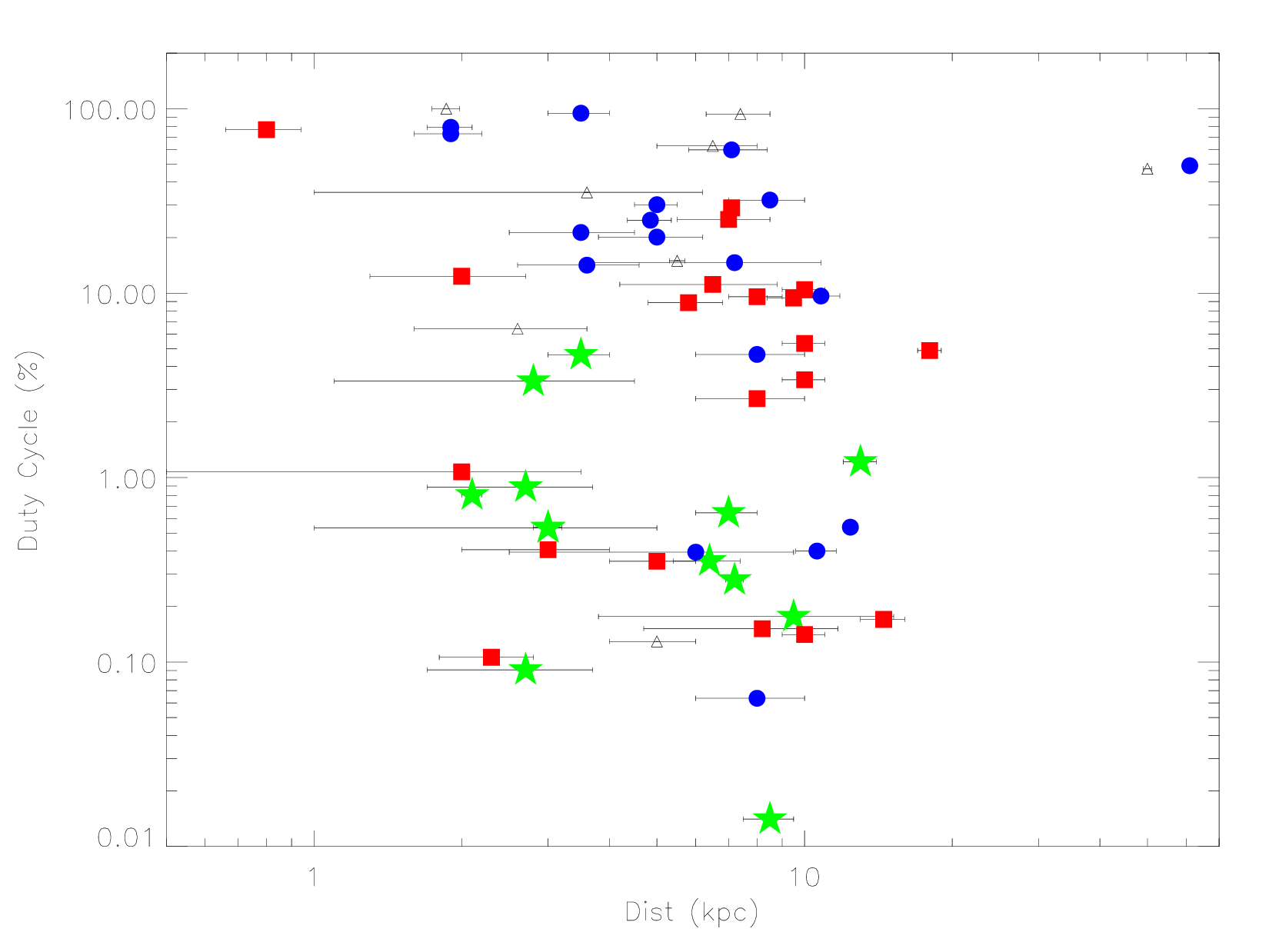} & 
\includegraphics[height=6.5cm, angle=0]{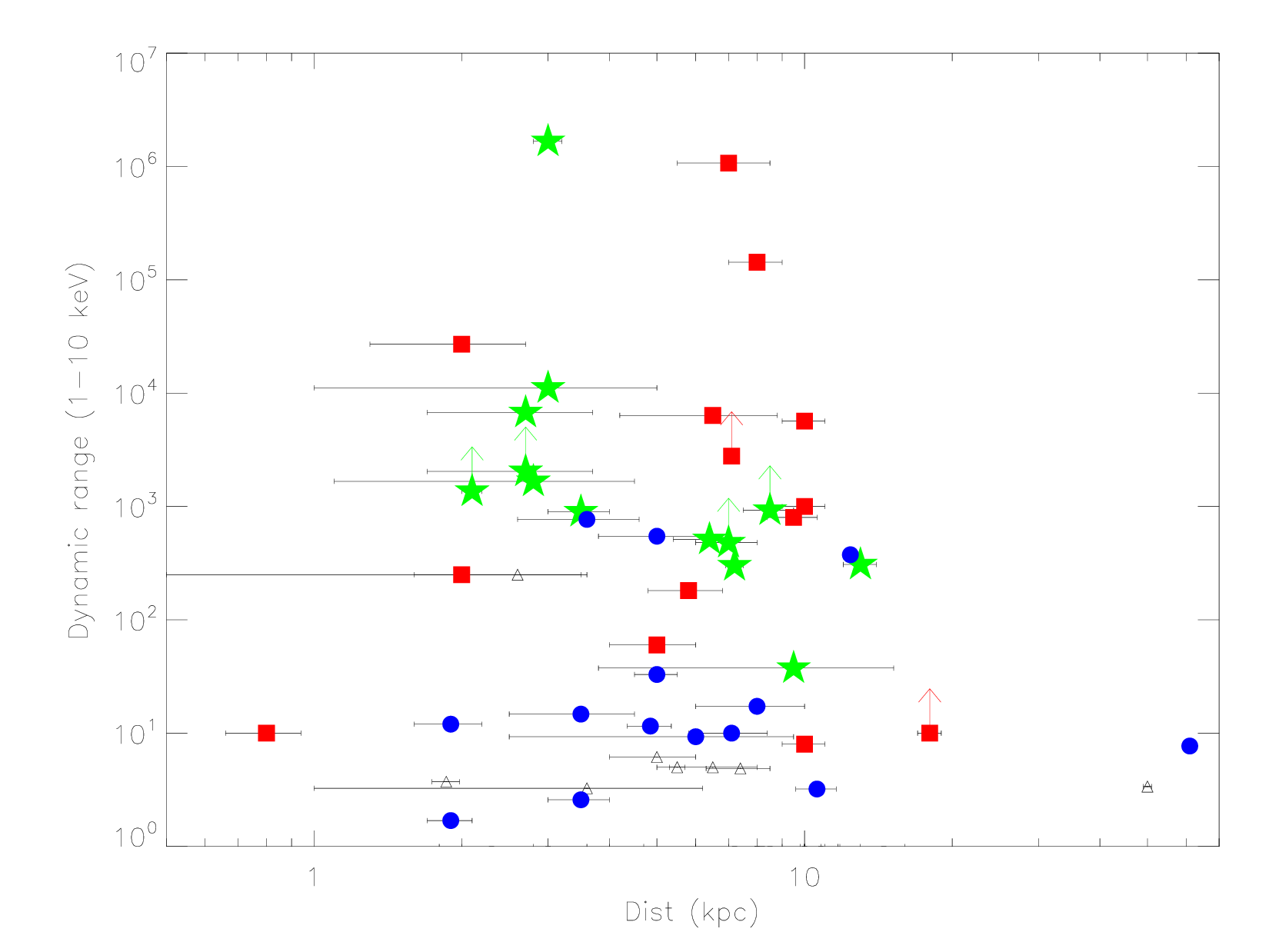}
\end{tabular}
\caption{Dependence of the hard X-ray duty cycles (left panel) and of the soft X--ray  dynamic ranges (right panel) with the source distance. 
Error bars indicate the uncertainty on the source distances, 
as reported in Table~\ref{tab:literature}. When Table~\ref{tab:literature} does not list
any uncertainty  (because it is not reported in the literature), we assumed $\pm{1}$~kpc. 
Green stars mark SFXTs, blue circles indicate SgHMXBs, red squares Be/XRBs;  empty triangles mark sources that do not belong to these three subclasses
and are listed in Table~\ref{tab:literature} as ``giant HMXBs'', ``other HMXBs'' and the ``symbiotic binary''. 
}
\label{lsfig:dyn_1_10_dist}
\end{figure*}

There is no correlation between the DC$_{18-50~keV}$ 
and the source distance (Fig.~\ref{lsfig:dyn_1_10_dist}, left panel), implying that  
extreme transient sources (like SFXTs) cannot  be explained as SgHMXBs located at larger distances. 
In the right panel of the same figure the dependence of the source dynamic range with the distance is displayed, 
showing no apparent trend.
%%%%%%%%%%%%%%%%%%%%%%%%%%%%%%%%%%%%%%%%%%%%%%%%%%%%%%%%
%%%%%%%%%%%%%%%%%%%%%%%%%%%%%%%%%%%%%%%%%%%%%%%%%%%%%%%%

\subsubsection{DC$_{18-50~keV}$  and other source properties}

\begin{figure*}
\centering
\begin{tabular}{cc}
\includegraphics[height=6.5cm, angle=0]{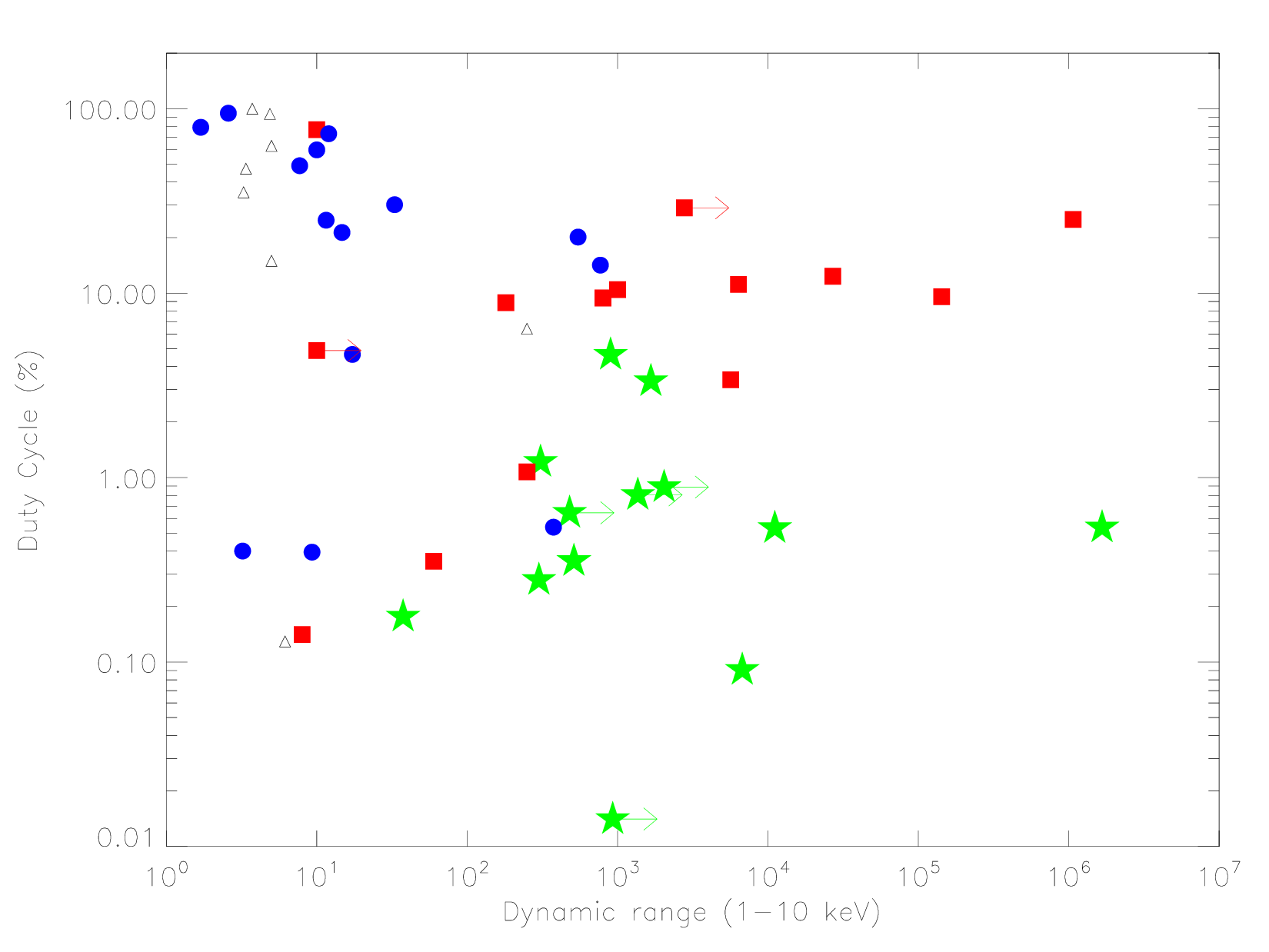} & 
\includegraphics[height=6.5cm, angle=0]{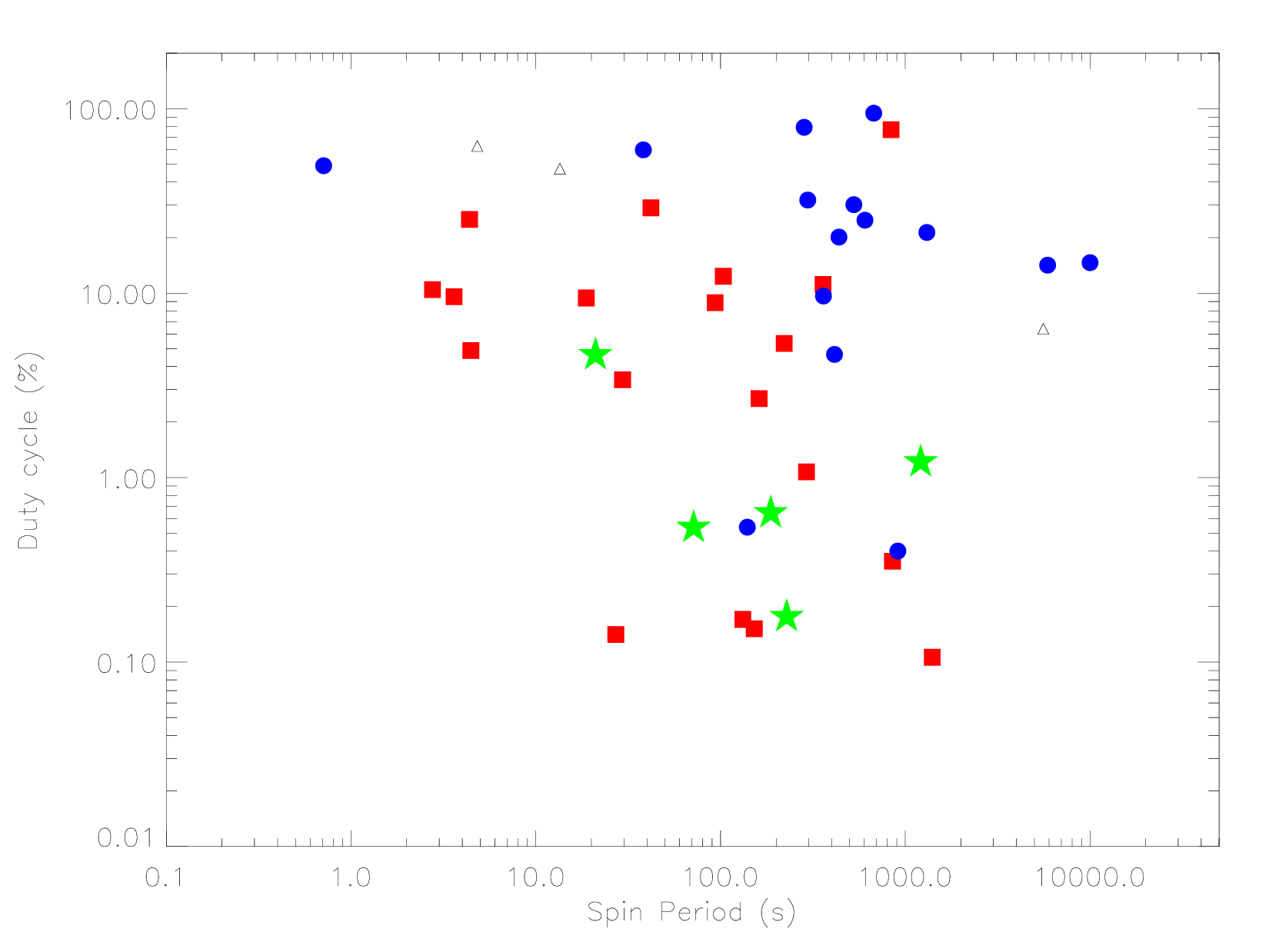} \\
\includegraphics[height=6.5cm, angle=0]{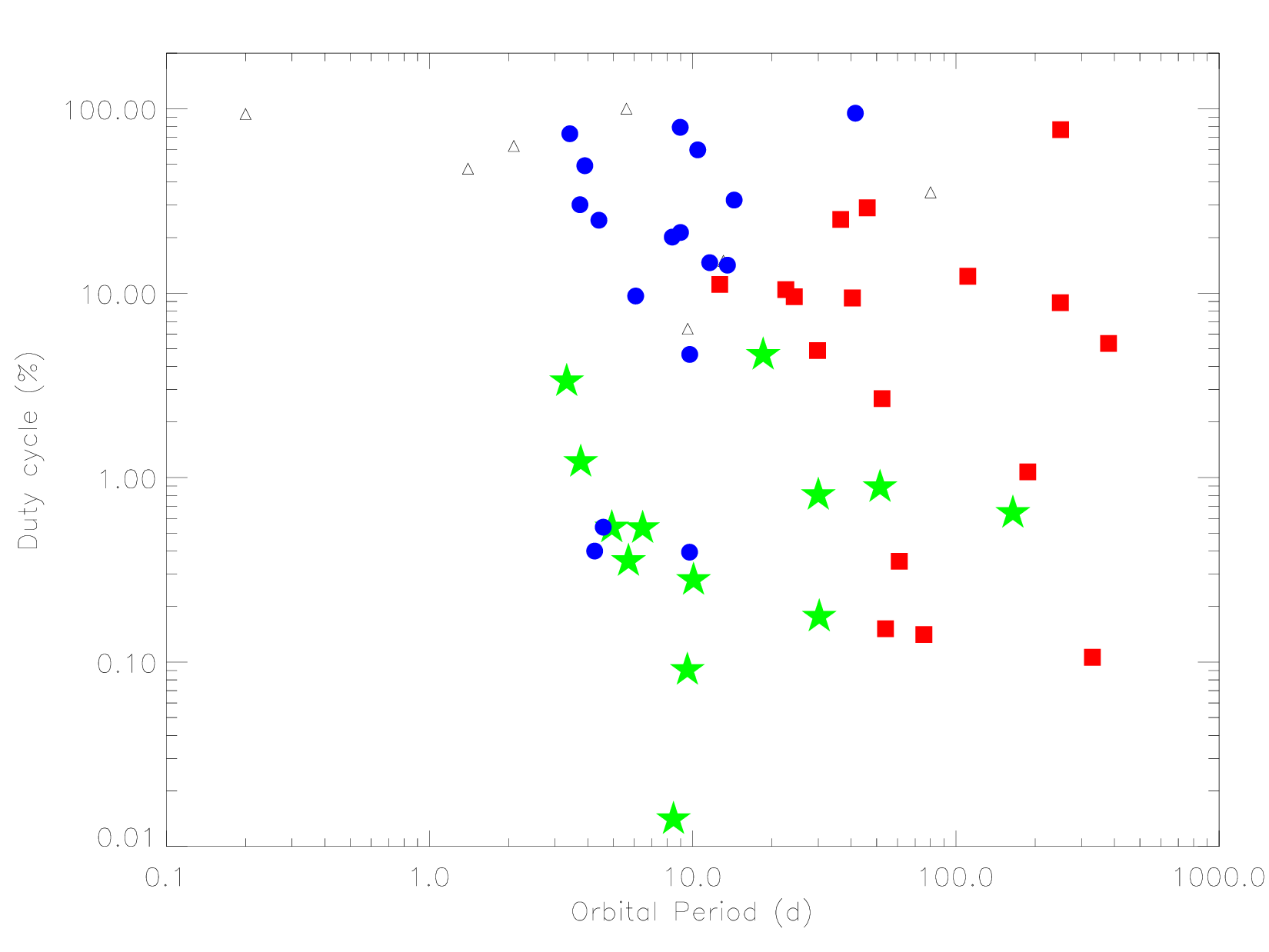} & 
\includegraphics[height=6.5cm, angle=0]{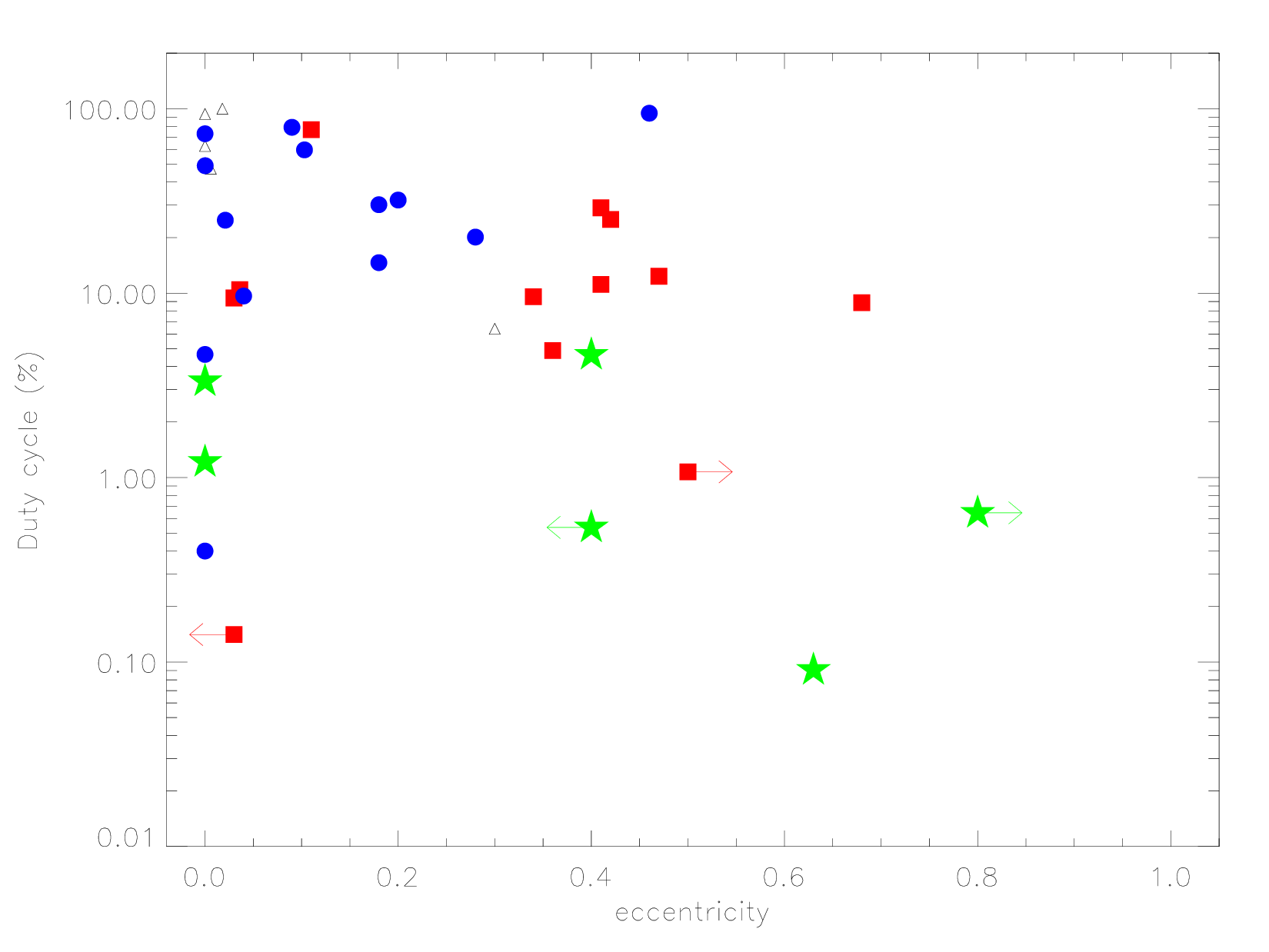} 
\end{tabular}
\caption{Properties of the sources in our sample, presented in Tables~\ref{tab:inte} and \ref{tab:literature}. 
The meaning of the symbols is the same as in Fig.~\ref{lsfig:dyn_1_10_dist}. Arrows mark lower or upper limits (according to Table~\ref{tab:literature}).
}
\label{lsfig:dc}
\end{figure*}

In Figure~\ref{lsfig:dc} we investigate if there is a dependence of the source DC$_{18-50~keV}$ with other properties.
In the first plot (DC$_{18-50~keV}$ versus DR$_{1-10~keV}$), 
persistent and less variable sources (like most of the SgHMXBs, marked by blue dots)
lie in the upper left part (high DC$_{18-50~keV}$, low DR$_{1-10~keV}$), while the most transient sources (SFXTs, marked by green stars), 
occupy the bottom, right part of the plot (low DC$_{18-50~keV}$, high DR$_{1-10~keV}$).
This 2D plot can be considered as a way to classify a source: 
sources with DR$_{1-10~keV}$$>$100 can be considered transients; among them, Be/XRTs cluster in the region with DC$_{18-50~keV}$$>$5 per cent, 
while the others with DC$_{18-50~keV}$$<$5 per cent are SFXTs. 
There are some rare exceptions (the same we mentioned before in each single plot).  
In particular, we suggest that IGR~J18027-2016, previously classified as a SgHMXB, is an SFXT, since it shows both 
a very low DC$_{18-50~keV}$ (0.54 per cent) and a high DR$_{1-10~keV}$ (375).
The region of the plane characterized by DR$_{1-10~keV}$$<$100 and DC$_{18-50~keV}$$<$1\%, 
is populated by SgHMXBs and Be/XRBs in equal number (plus the SFXT IGR~J16465--4507 and a symbiotic binary):
the two SgHMXBs  IGR~J16207--5129 and IGR~J16393-4643 and the Be sources 4U~1036-56 and XTE~J1543-568.

Interestingly, the peculiar source 3A~2206+543 (marked by the  empty triangle
at the location DC$_{18-50~keV}$=7\% and DR$_{1-10~keV}$$\sim$100) 
lies just in-between the three main locii occupied by the great majority of the members of the three sub-classes,
confirming its anomalous phenomenology, difficult to classify (not only in the optical, but also in X--rays).
%
%%%%%%%%%%%%%%%%%%%%%%%%%%%%%%%%%%%%%%%%%%%%%%%%%%%%%%%%
%%%%%%%%%%%%%%%%%%%%%%%%%%%%%%%%%%%%%%%%%%%%%%%%%%%%%%%%

In Figure~\ref{lsfig:dc}, the two plots of DC$_{18-50~keV}$ versus the pulsar spin and the orbital periodicities is meant to enlarge the
transient versus persistent phenomenology shown by the Corbet diagram  (see below) also to sources where one or the other are unknown.
The fourth plot with DC$_{18-50~keV}$ versus the orbital eccentricity reveals two branches: low-eccentricity sources  spanning a large range
of hard X--ray activity, and sources drawing an anticorrelation that starts from the top left part of the graph 
with persistent SgHMXBs plus the low-eccentricity, Be system X~Per, to the bottom right zone, occupied by 
two SFXTs with the most eccentric orbits (IGR~J08408--4503 and IGR~J11215-5952).
This suggests that the orbital eccentricity might play a role in producing transient X--ray emission in some HMXBs.

%%%%%%%%%%%%%%%%%%%%%%%%%%%%%%%%%%%%%%%%%%%%%%%%%%%%%%%%%%%%%%%%%%%%%%%%%%%%%%%%%%%%%%%%%%%%%%%
\subsubsection{Orbital geometry and pulsar spin period} 

\begin{figure}
\begin{center}
\centerline{\includegraphics[width=9.cm]{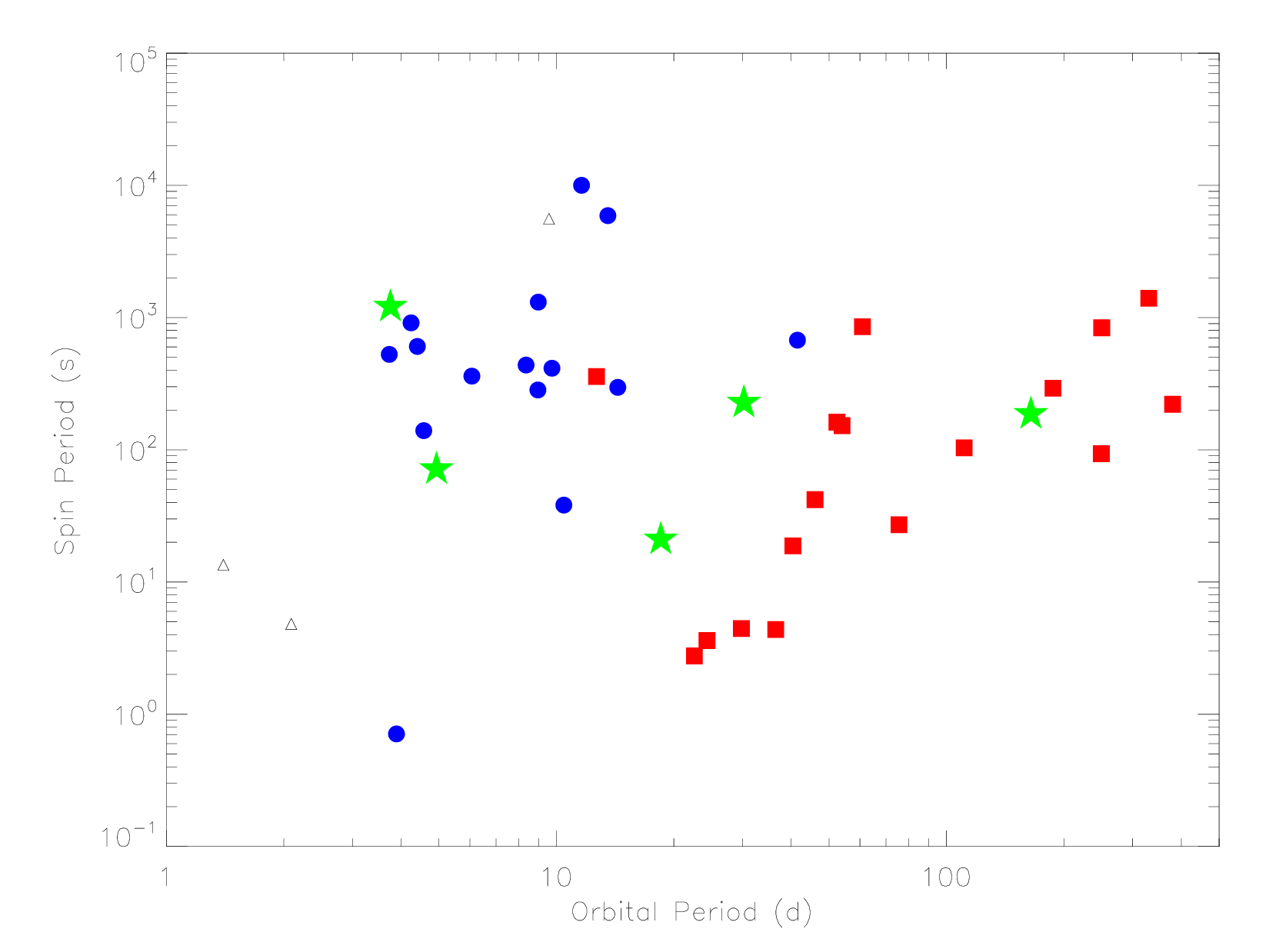}}
\caption{Pulsar rotational period versus orbital period (the so-called Corbet diagram) is shown for the HMXBs in our sample.
The meaning of the symbols is the same as in Fig.~\ref{lsfig:dyn_1_10_dist}.
}
\label{lsfig:corbet}
\end{center}
\end{figure}

In Fig.~\ref{lsfig:corbet} we display the so-called Corbet diagram \citep{Corbet1986} for our sample, 
where the pulsar spin period is plotted versus the orbital period of the system.  
The segregation of different source types  is well known \citep{Corbet1986}, 
together with the more recent findings that SFXTs appear
to bridge the two locii where SgHMXBs and Be/XRBs are clustered \citep{Sidoli2017review}. 
However note that, among SFXTs, the spin period of IGR~J17544--2619 (i.e., 71~s) needs a confirmation,
since it was derived with a collimator ($RXTE$/PCA), so it is possible that 
pulsed X--rays actually come from a different transient source 
within the field of view. 
The three RLO-systems are shown in the 
bottom left region of the plot (LMC~X--4, SMC~X--1 and Cen X--3). 
%
%%%%%%%%%%%%%%%%%%%%%%%%%%%%%%%%%%%%%%%%%%%%%%%%%%%%%%%%%%%%%%%%%%%%%%%% 
%%%%%%%%%%%%%%%%%%%%%%%%%%%%%%%%%%%%%%%%%%%%%%%%%%%%%%%%%%%%%%%%%%%%%%%%
%
%%%%%%%%%%%%%%%%%%%%%%%%%%%%%%%%%%%%%%%%%%%%%%%%%%%%%%%%
\begin{figure*}
\centering
\begin{tabular}{cc}
\includegraphics[height=6.5cm, angle=0]{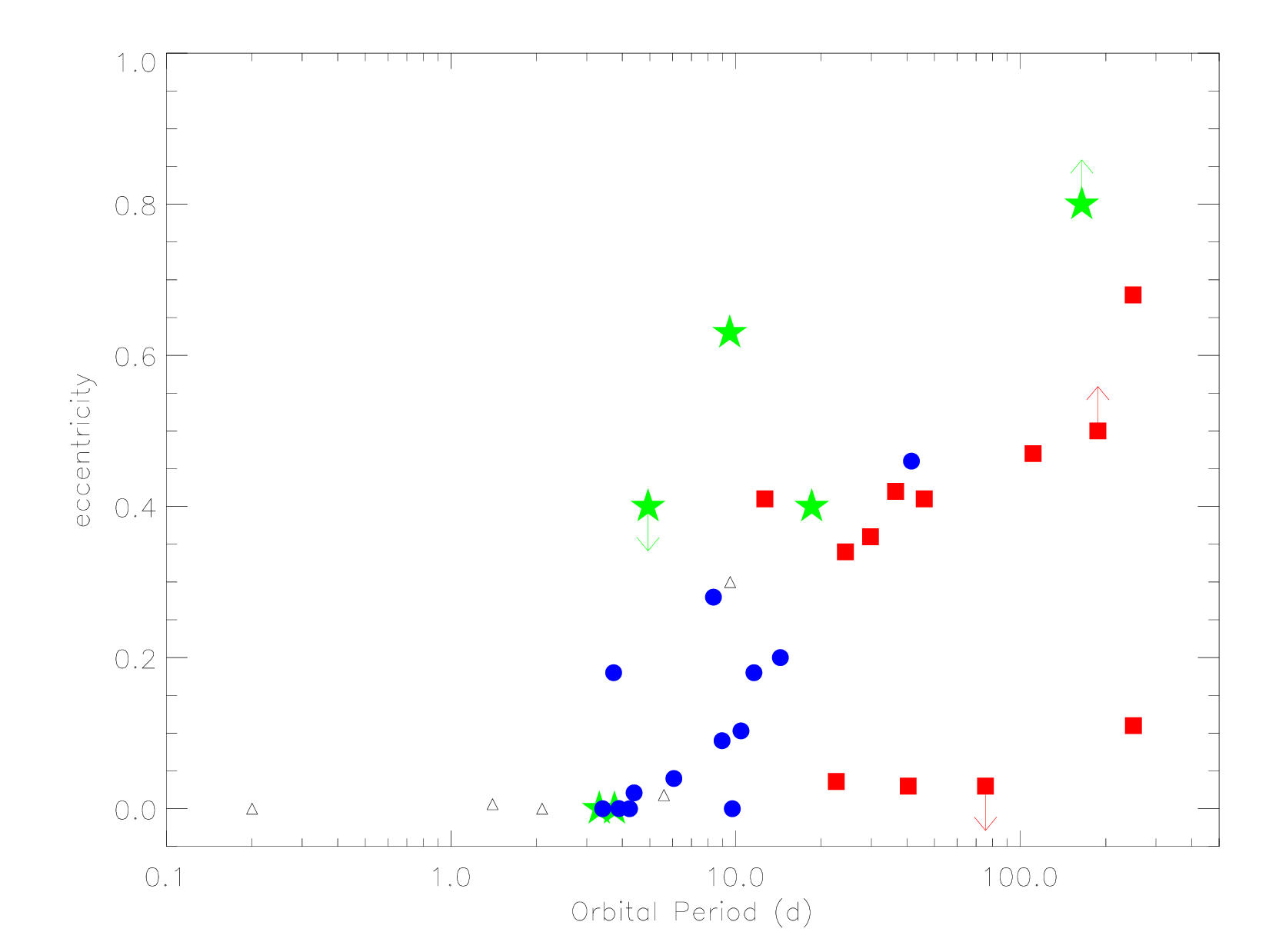} &
\includegraphics[height=6.5cm, angle=0]{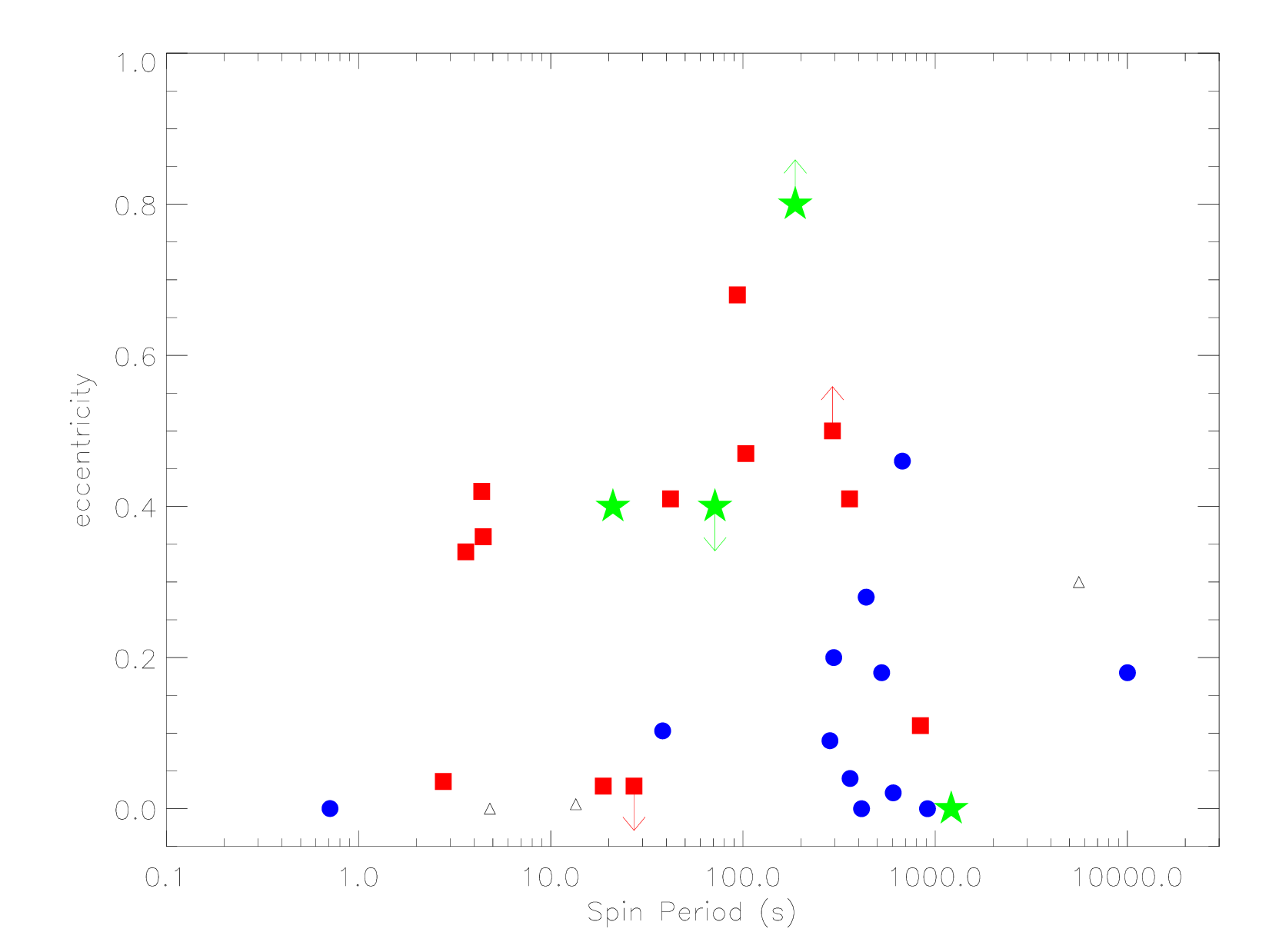} 
\end{tabular}
\caption{Properties of the sample of HMXBs, in terms of orbital eccentricity versus orbital period ({\em left panel}) and
of eccentricity versus pulsar rotational period ({\em right panel}). 
The meaning of the symbols is the same as in Fig.~\ref{lsfig:dyn_1_10_dist}  and \ref{lsfig:dc}. 
}
\label{lsfig:ecc}
\end{figure*}
%%%%%%%%%%%%%%%%%%%%%%%%%%%%%%%%%%%%%%%%%%%%%%%%%%%%%%%%

In the left panel of Fig.~\ref{lsfig:ecc}, we plot the source eccentricity against the orbital period, 
obtaining two trends:
for the large majority of HMXBs,  the longer the period, the  more eccentric the orbit; 
there are however four low-eccentric Be/XRBs with orbital periods longer than 20~days (starting 
from the shorter orbital period, they are 4U~1901+03, KS~1947+300, XTE~J1543--568 and X~Per). 
This plot was previously investigated in detail  by \citet{Townsend2011} on a larger sample of HMXBs (also including a few SMC pulsars), 
so we will not discuss this further, but we  remind here the arguments put forward by these authors to explain the two trends: 
first of all, the low-eccentricity, wide-orbit, Be systems 
represent a separate population of Be/XRBs where the NS experienced a smaller 
natal kick with the first supernova explosion than the other, more eccentric, Be/XRBs \citep{Pfahl2002}.
Second, the significant correlation of eccentricity with the period in all other HMXBs is actually made of two separate sub-classes,
the supergiant (at lower orbital periods) and the Be binaries (at higher orbital periods), as is also evident in our graph. 
\citet{Townsend2011} suggested that this latter correlation, though significant taken as a whole, 
might be simply explained by the superposition of the two sub-classes (SgHMXBs and Be/XRBs) lying in two 
different regions of this parameter space. 
However, it is important to note  a new remarkable finding in our version of this plot:
the presence of the SFXTs (not considered by  \citealt{Townsend2011}):
although the eccentricity of two SFXTs (at e$\sim$0.4) are quite uncertain (and only derived from the amplitude of their X-ray orbital modulation), 
the eccentricities (and orbital periods) of the most eccentric SFXTs IGR~J08408-4503 (e=0.63$\pm{0.03}$, P$_{orb}$=9.5436$\pm{0.0002}$, \citealt{Gamen2015}) 
and IGR~J11215-5952 (P$_{orb}$=164.6~days, \citealt{Sidoli2007, Romano2009}; e$>0.8$, \citealt{Lorenzo2014}) appear better determined.
This allows HMXBs with supergiant companions to overlap with Be systems at large eccentricities and orbital periods.

The plane of the orbital eccentricity versus the pulsar spin period is more complex (Fig.~\ref{lsfig:ecc}, right panel).
In order to discuss this plot, for the sake of clarity, we distinguish three different regions. 
The eccentricity e$\sim$0.3 seems to divide the plot into two main regions. 
The upper part is populated by the more transient systems (Be/XRTs and SFXTs) plus the SgHMXB GX301--2,
while the region with e$<$0.3 can be divided into two more parts, 
depending on the spin period (around P$_{spin}$=100~s): there are no pulsars (in our HMXB sample) with 
rotational periods in the range 40--200~s and low eccentricities (e$<$0.3). 
Considering only the Be/XRBs, this recalls the result obtained by \citet{Knigge2011}, 
on a sample of Be/XRB pulsars (including Magellanic Cloud's systems), 
who found that short (long) spin periods 
are preferentially located in low (high) eccentric binaries, and interpreted it as the results 
of two distinct types of NS forming supernovae.
An alternative explanation was proposed by \citet{Cheng2014}, who 
ascribed the bimodal spin distribution to different accretion modes in the Be/XRBs: shorter spin periods are 
present in Be/XRBs where the NS can efficiently accrete matter (producing a Type~II outburst)
from the warped, outer Be stellar disc, forming an accretion disc that spins up the pulsar.
In Be/XRBs that are either persistent or that experience mostly Type I outburst, the accretion is suggested to be quasi-spherical
and the transfer of angular momentum to the pulsar inefficient, resulting in longer spin period Be/XRB pulsars  \citep{Cheng2014}.
Three are the Be/XRBs in this region of the eccentricity-spin period plane: 4U~1901+03, KS~1947+300 and XTE~J1543-568. 
This warping mechanism might explain why these low-eccentric Be/XRBs do show outbursts, instead of being persistent (like X~Per): 
in the scenario proposed by \citet{Cheng2014} they might display the
right combination of values for their eccentricity and orbital periods (see Fig.~4 in \citealt{Cheng2014}).
The lower-right region of this plane (long spin periods and low-eccentric orbits) 
is occupied by wind-fed SgHMXBs (plus X~Per, one SFXT and the peculiar wind accretor 3A~2206+543), 
where the wind accretion is unable to efficiently spin up the pulsar.
Systems with supergiant and giant companions at low eccentricities and spin period shorter than 40~s, 
can be explained by disc accretion: these sources are the three RLO-systems together with OAO~1657--415.

\subsubsection{Minimum and maximum luminosity (1-10 keV)}

%%%%%%%%%%%%%%%%%%%%%%%%%%%%%%%%%%%%%%%%%%%%%%%%%%%%%%%%
\begin{figure*}
\centering
\begin{tabular}{cc}
\includegraphics[height=6.5cm, angle=0]{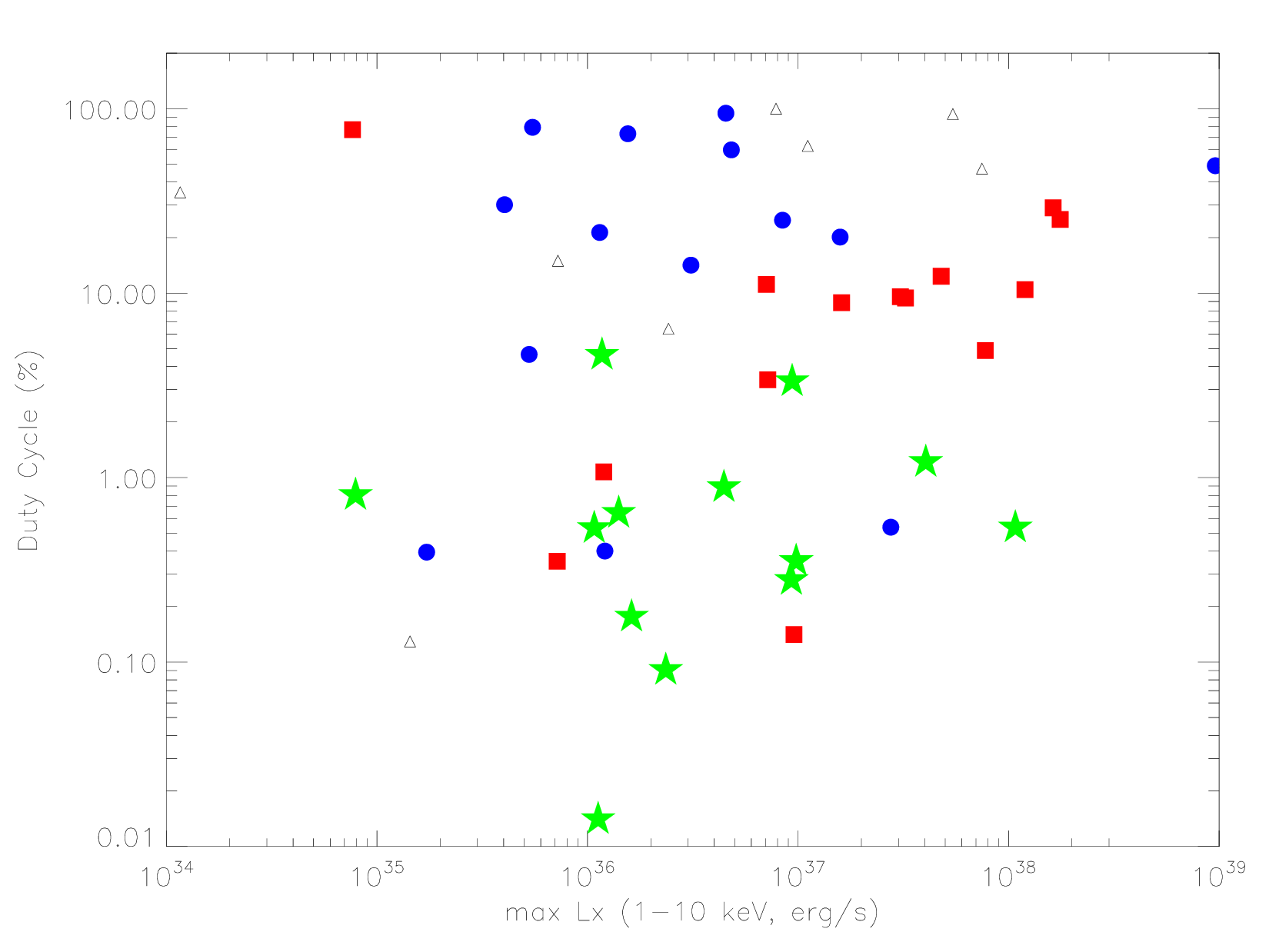} & 
\includegraphics[height=6.5cm, angle=0]{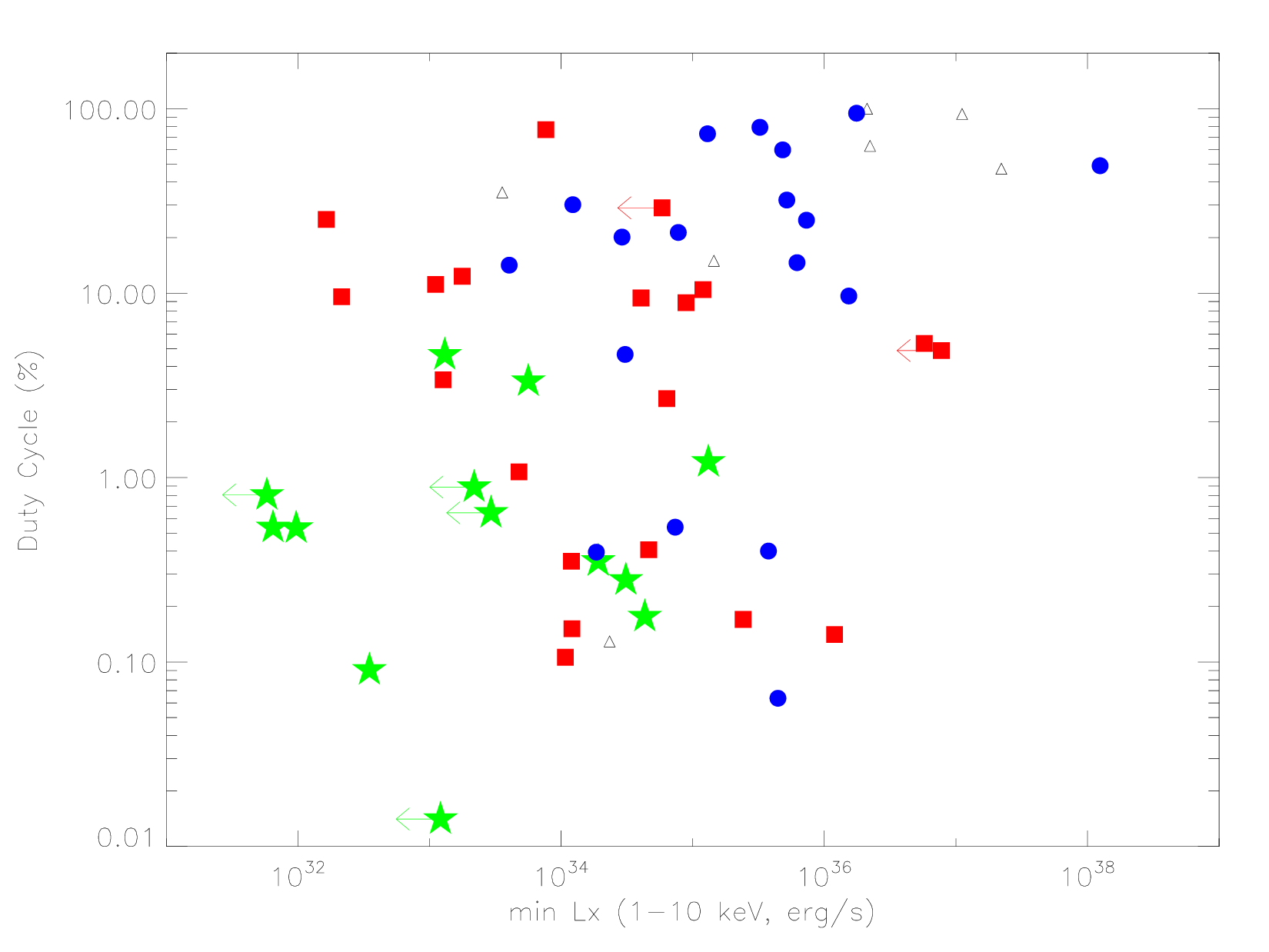} \\
\includegraphics[height=6.5cm, angle=0]{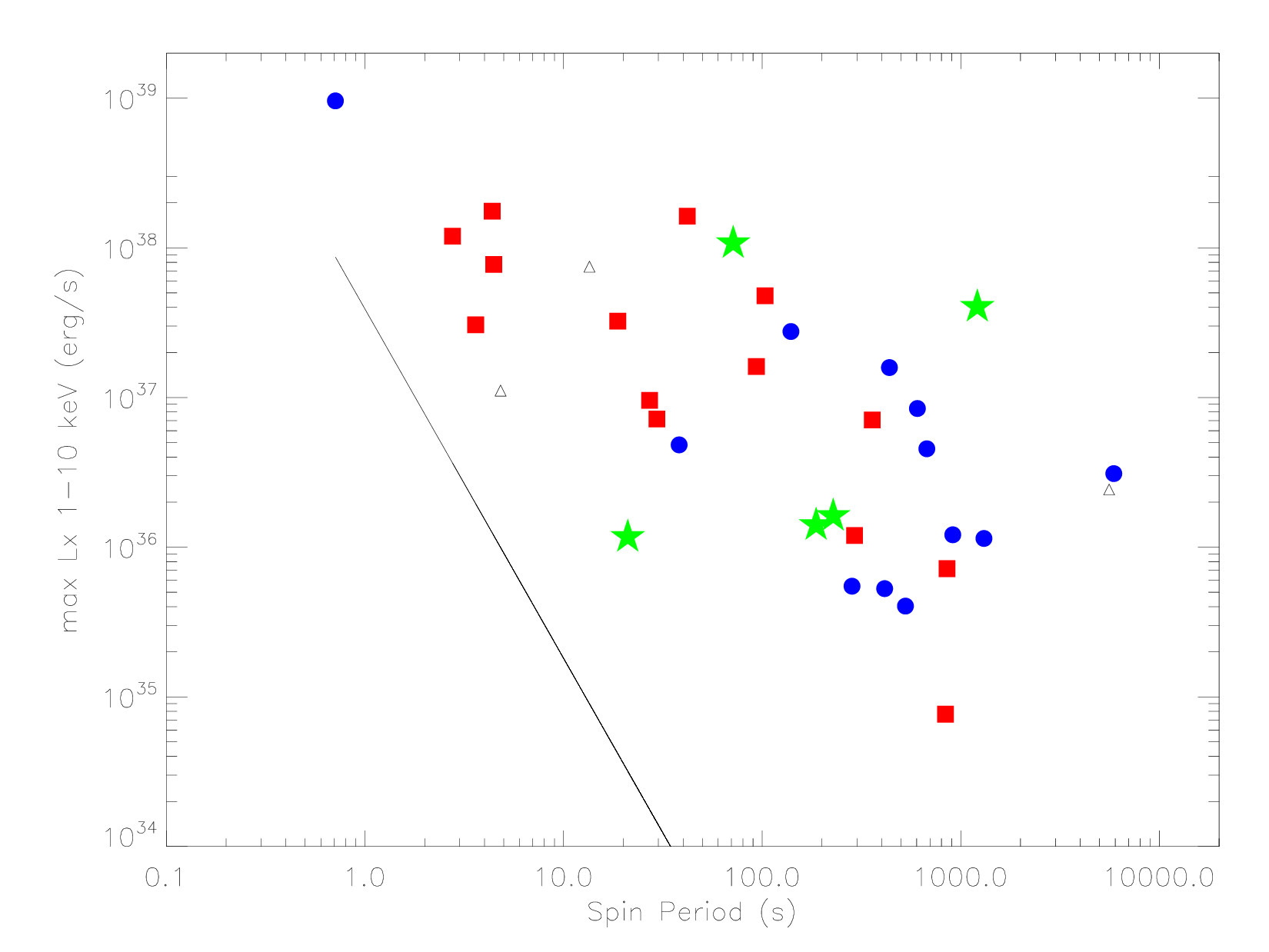} &
\includegraphics[height=6.5cm, angle=0]{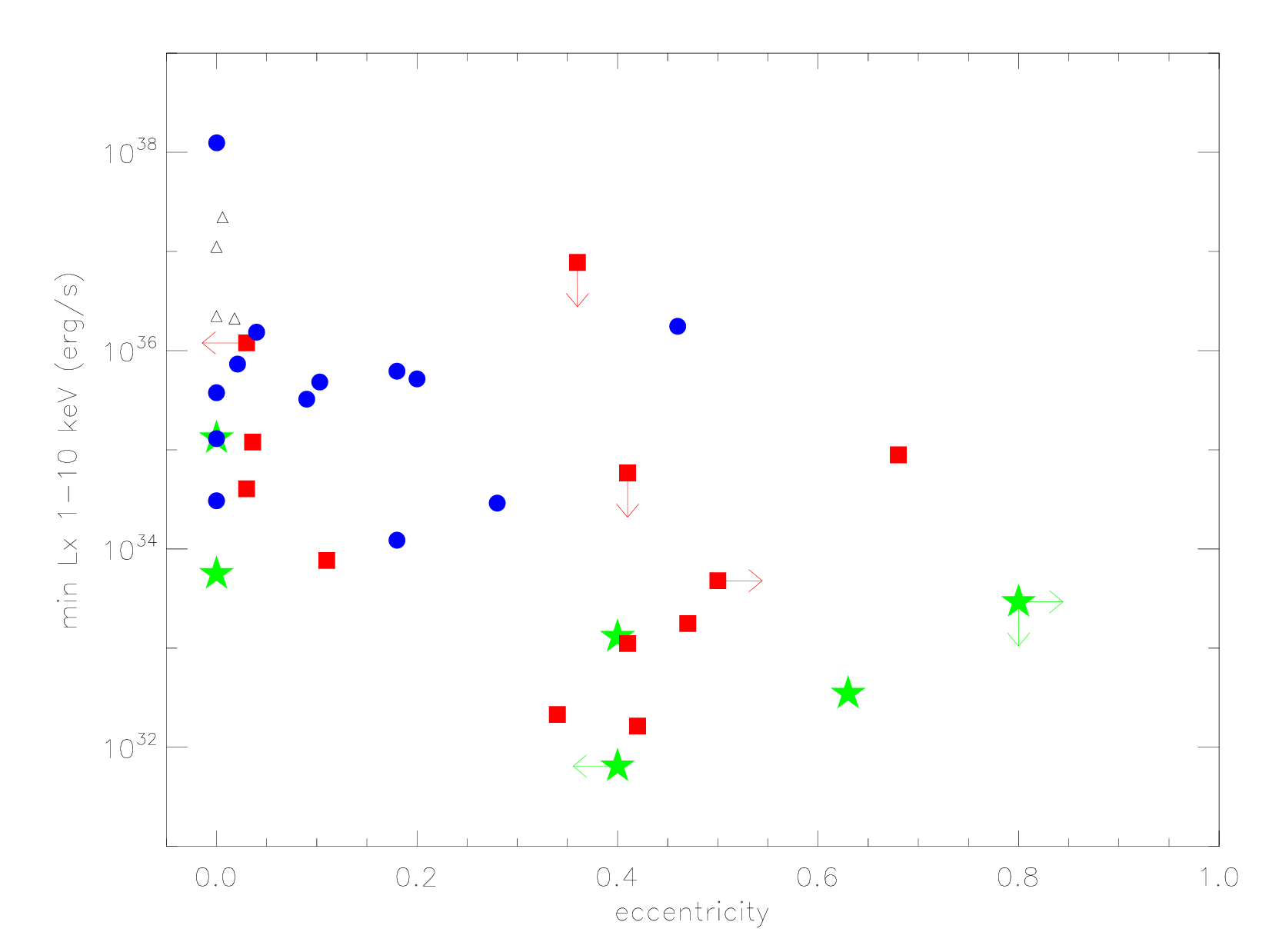} \\
\includegraphics[height=6.5cm, angle=0]{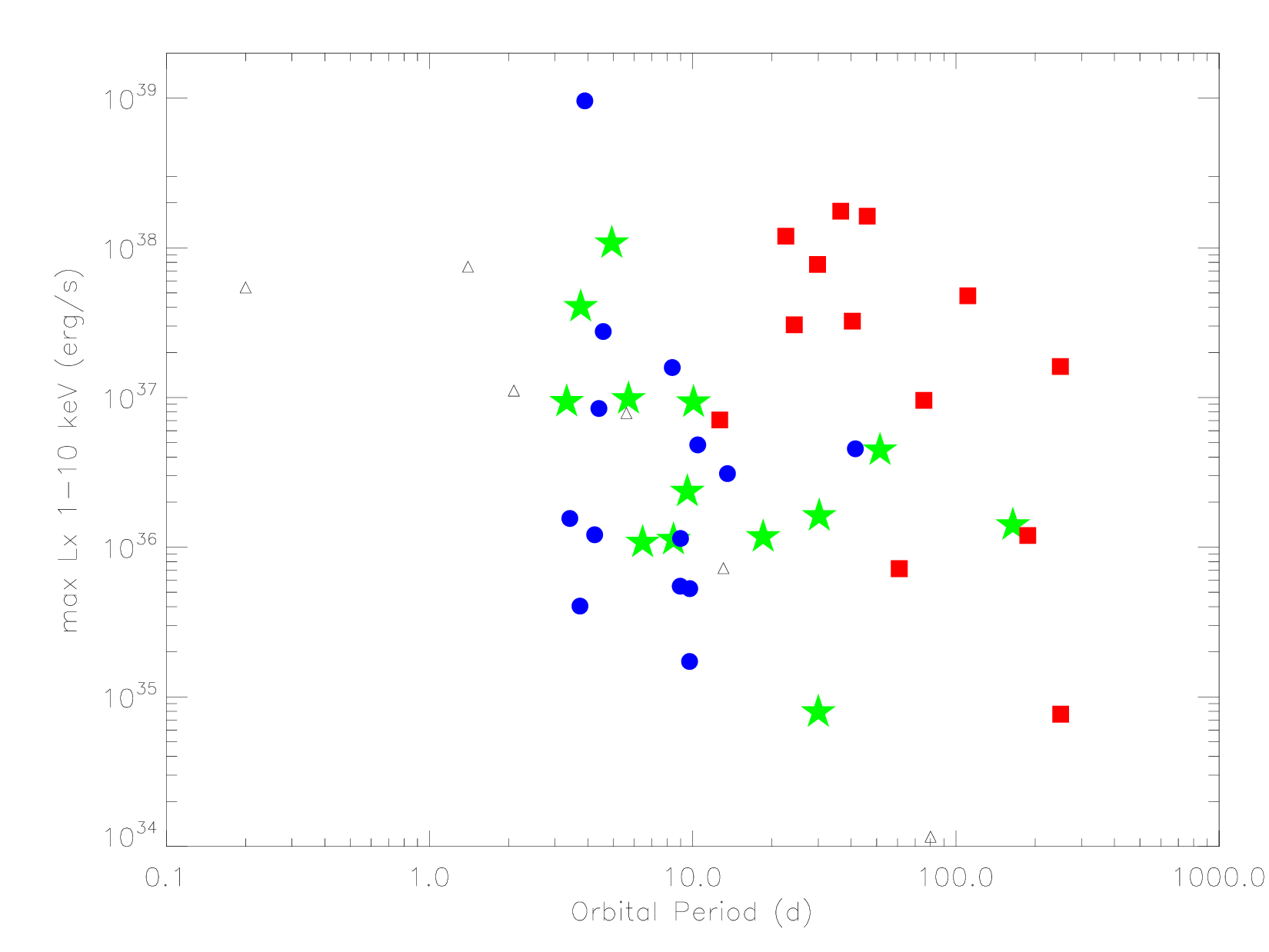} & 
\includegraphics[height=6.5cm, angle=0]{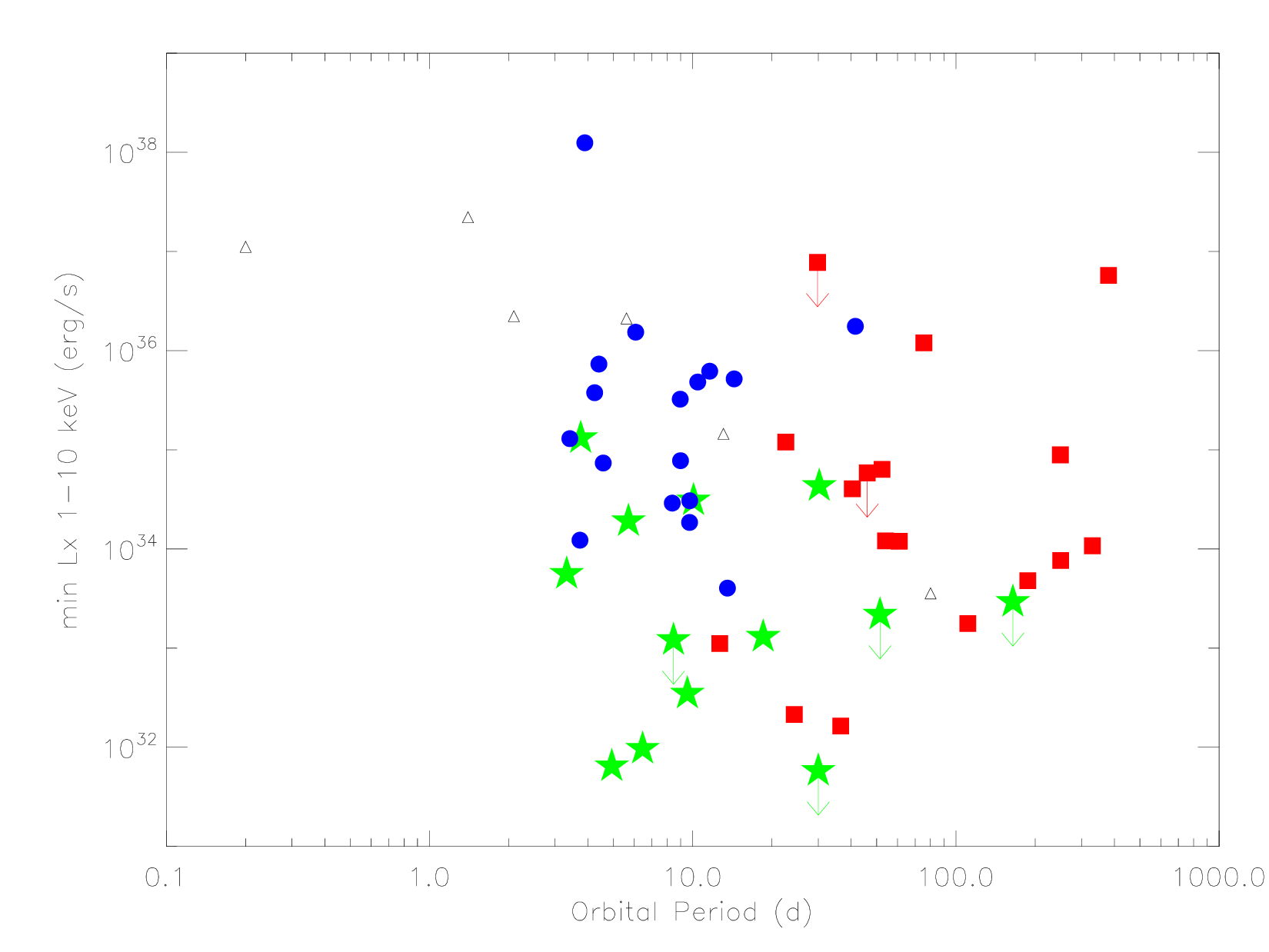} 
\end{tabular}
\caption{Properties of the sources in our sample, presented in Tables~\ref{tab:inte} and \ref{tab:literature}.
The meaning of the symbols is the same as in Figs.~\ref{lsfig:dyn_1_10_dist}  and \ref{lsfig:dc}. 
In the plot of the maximum luminosity (1--10\,keV) against the spin period, the minimum 
luminosity before the on-set of the propeller effect is drawn (solid line), 
following Eq.~\ref{eq:propeller} (assuming a NS magnetic field of 10$^{12}$~G and $\xi=1$).
}
\label{lsfig:softx}
\end{figure*}
%%%%%%%%%%%%%%%%%%%%%%%%%%%%%%%%%%%%%%%%%%%%%%%%%%%%%%%%
%
In Fig.~\ref{lsfig:softx} we show the dependence of the soft X--ray properties with the hard X--ray duty cycle and 
other quantities like the spin period and the orbital geometry.
There is no apparent correlation between the DC$_{18-50~keV}$ and the maximum soft X--ray luminosity (first upper panel), 
while it seems that (although with a large scatter) the lower the DC$_{18-50~keV}$, the lower the minimum soft X--ray luminosity (second upper panel).
The third panel indicates that accreting pulsars with shorter rotational periods avoid low values of their maximum luminosities (e.g., \citealt{Stella1986}). 
Although  we do not plot here necessarily the X--ray luminosity of the observations where  a spin period was measured, 
we can ascribe this trend to the on-set  of the propeller effect  (the solid line is drawn from Eq.~\ref{eq:propeller}).
The same trend is not present (and we do not show it here) when the minimum soft X--ray luminosity is plotted against the spin period, since for
X-ray transients it is taken from quiescence.
In the forth panel we show  the dependence of the minimum soft X--ray luminosity from the orbital eccentricity, where we found that 
in more eccentric systems a lower luminosity level can be reached, because the NS can orbit farther away from the donor.
In the bottom  panels of Fig.~\ref{lsfig:softx} we show, for completeness, maximum and minimum soft X--ray luminosities versus  the orbital period.
In these graphs, the sub-classes cluster in different regions, reflecting the trend of the Corbet diagram (that is, 
persistent SgHMXBs have shorter orbital period than Be/XRBs, while SFXTs bridge the two regions).

%%%%%%%%%%%%%%%%%%%%%%%%%%%%%%%%%%%%%%%%%%%%%%%%%%%%%%%%
\begin{figure*}
\centering
\begin{tabular}{cc}
\includegraphics[height=6.5cm, angle=0]{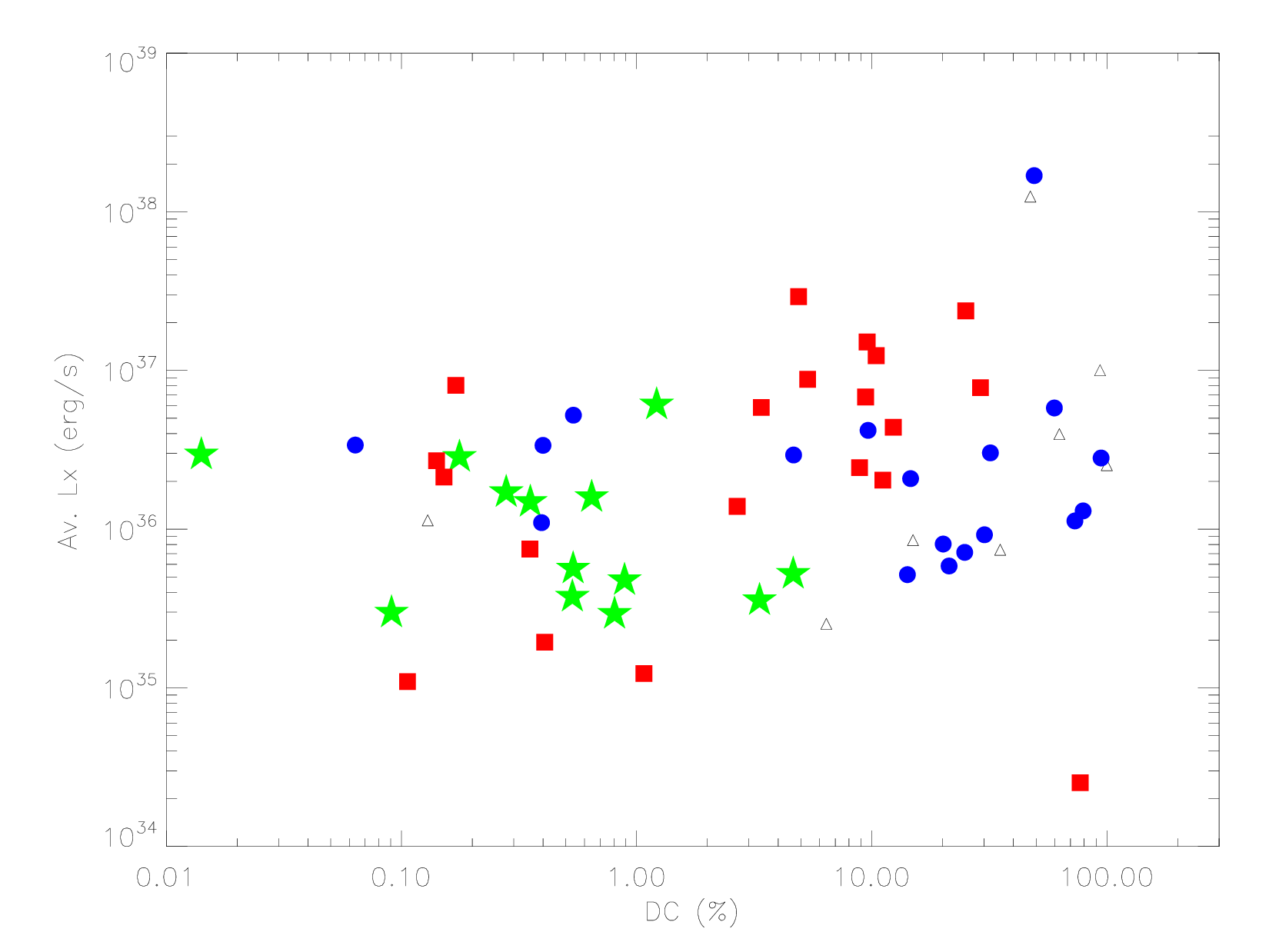} & 
\includegraphics[height=6.5cm, angle=0]{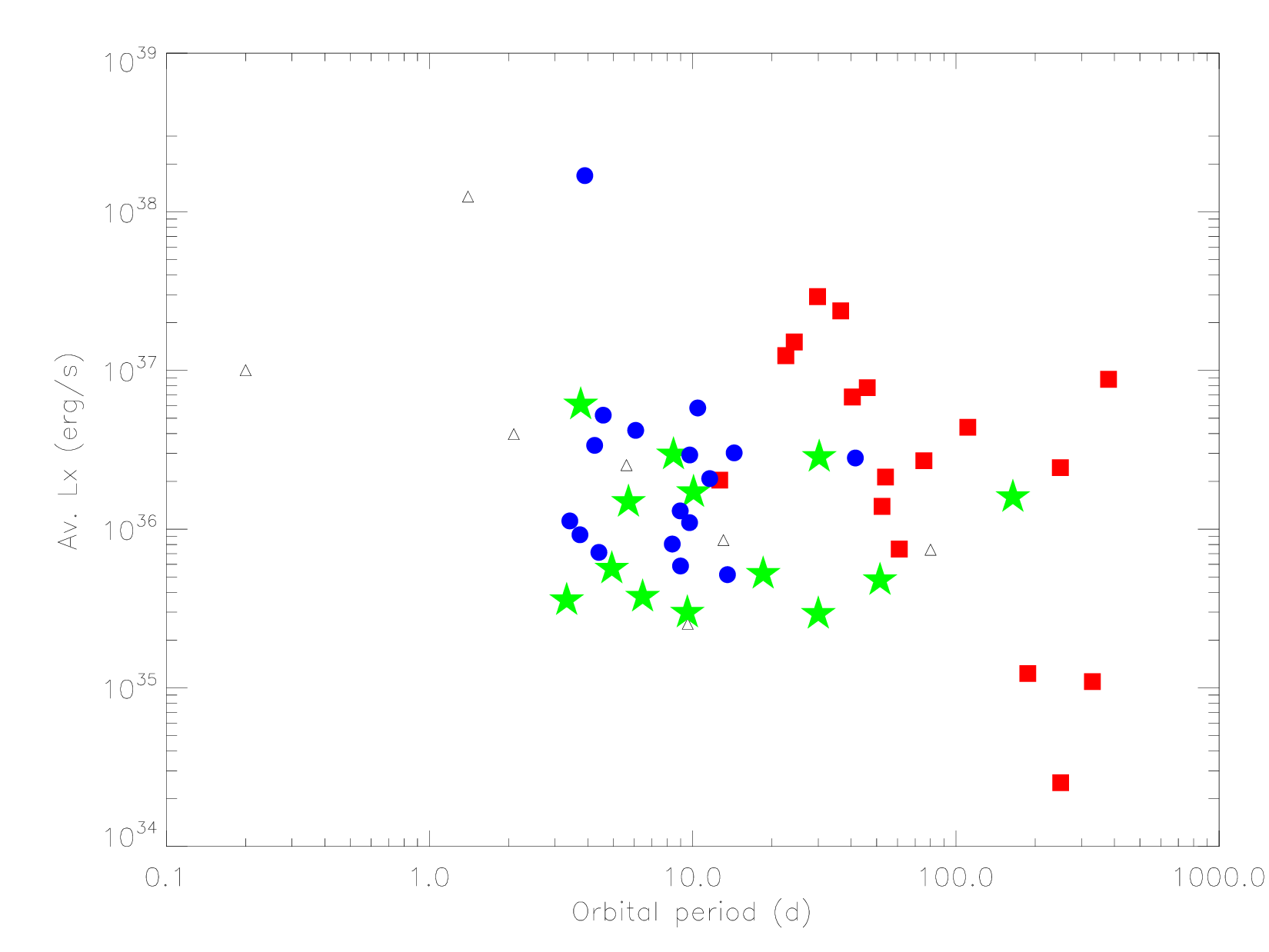}  \\
\includegraphics[height=6.5cm, angle=0]{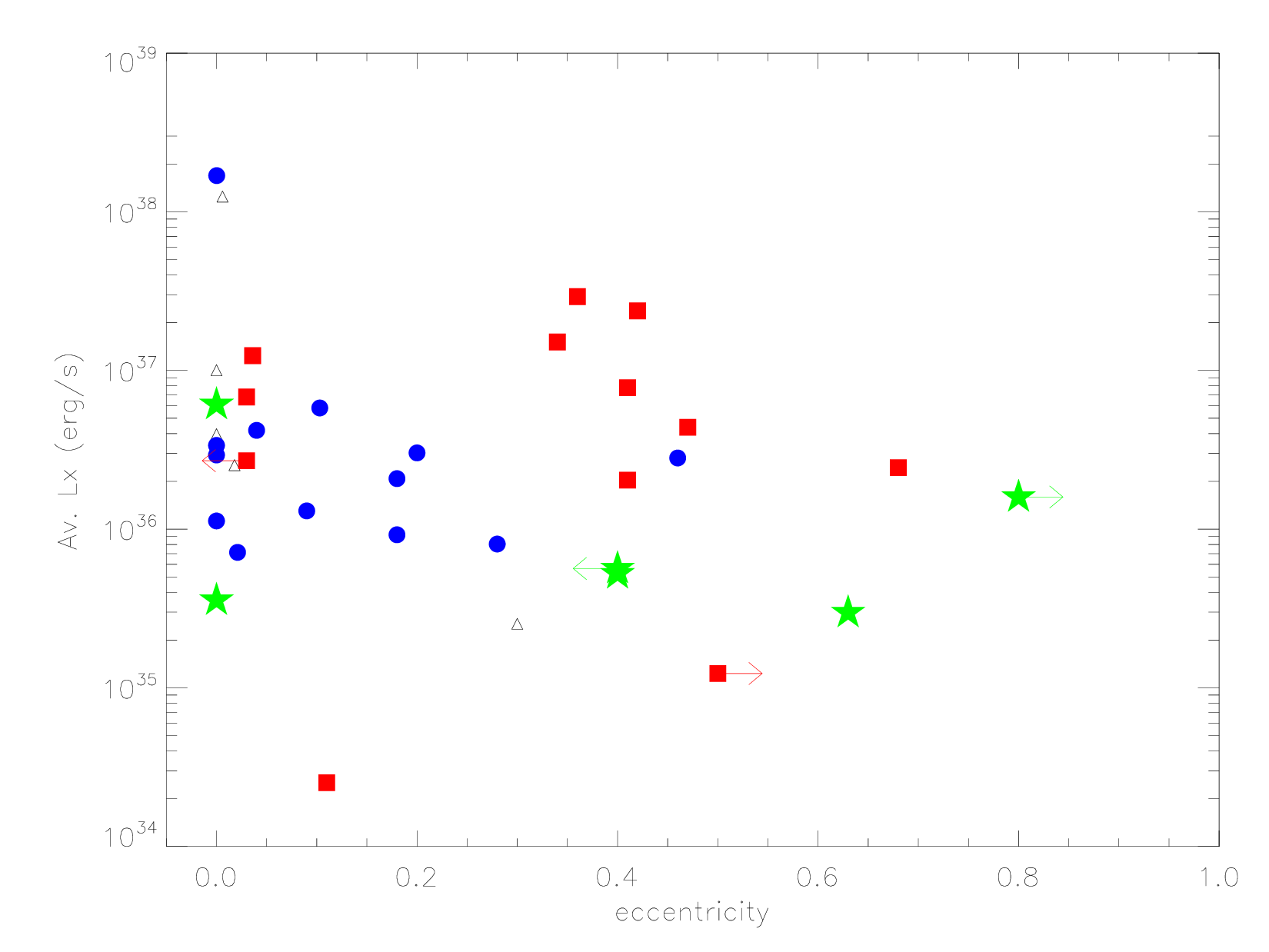} & 
\includegraphics[height=6.5cm, angle=0]{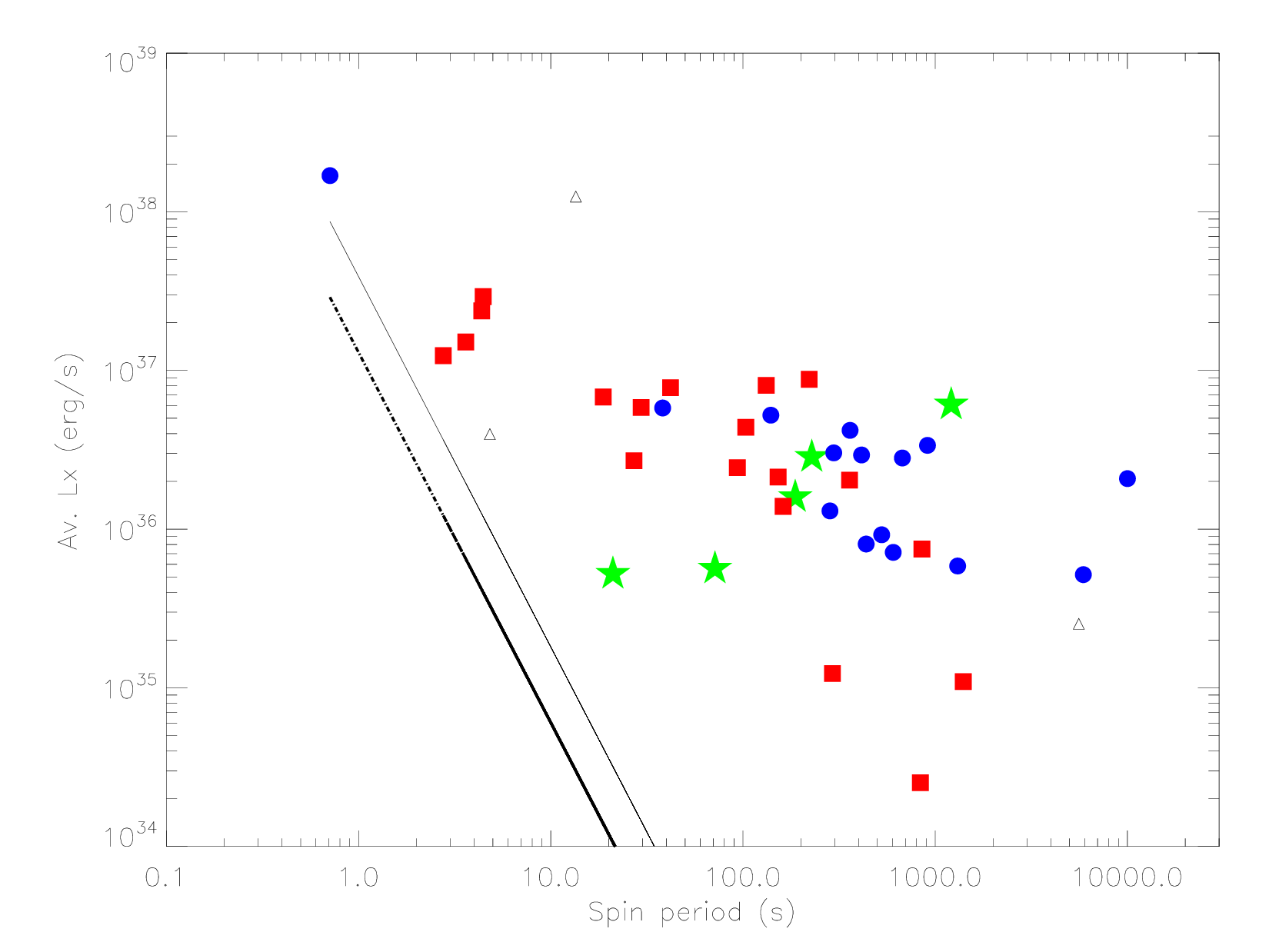}  
\end{tabular}
\caption{Properties of the sources in our sample, presented in Table~\ref{tab:inte} and Table~\ref{tab:literature}.
The meaning of the symbols is the same as in Figs.~\ref{lsfig:dyn_1_10_dist}   \ref{lsfig:dc}. 
The average luminosity  is at hard X--rays (18--50\,keV). 
In the last panel, where we have plotted the average luminosity versus  the spin period, the thin black line 
marks the minimum luminosity before the on-set of the propeller effect, following Eq.~\ref{eq:propeller}, and  
assuming a NS magnetic field of 10$^{12}$~G and $\xi=1$ (solid line). The thicker line is the same, but 
divided by a factor of three (assumed to be a realistic conversion factor between bolometric and 18--50\,keV luminosity, see the text for details). 
}
\label{lsfig:avlx}
\end{figure*}
%%%%%%%%%%%%%%%%%%%%%%%%%%%%%%%%%%%%%%%%%%%%%%%%%%%%%%%%

\subsubsection{Average hard X-ray luminosity}

In Fig.~\ref{lsfig:avlx} we investigate the behaviour of the average luminosity (18--50 keV)
against the source duty cycle  and other source properties. 
In the first panel, three types of HMXBs lie in three different regions of the plane DC$_{18-50~keV}$---L$_{X}$,
with the SgHMXBs on the right, the SFXTs on the left (low duty cycles but average luminosities in outburst  similar to  the classical SgHMXBs),
and most of the Be/XRTs showing higher average luminosities (in outburst) and intermediate duty cycles (around 10 per cent).
This suggests another  way to distinguish the three kinds of HMXBs from the hard X--rays properties (see below, when discussing about the source dynamic range).
In Fig.~\ref{lsfig:avlx}, the plot of  the average luminosity (18--50\,keV) against the orbital period shows an anticorrelation. 
However, it is important to remind that for transient sources (both Be and SFXTs), 
the averange luminosity plotted here is the average value during outbursts, as observed by \inte.
It is  known that persistent, wind-fed, SgHMXBs hosting NSs  show a trend of L$_{X}$ with the orbital period 
(\citealt{Stella1986, Bhadkamkar2012, Lutovinov2013} and references therein),
as L$_{X}\propto$P$_{orb}^{-4/3}$. In particular, Lutovinov et al. (2013) proposed a model that permitted them 
to obtain an allowed region in the plane P$_{orb}$--L$_{X}$ for 
a NS accreting from the wind of a supergiant. 
The most luminous, short period, SgHMXBs do not fit their model, being powered by RLO. 
We overplot here also the Be/XRBs: among them, the three
low luminosity, long-period, systems (X~Per, H~1145--619 and  RX~J0146.9+6121) seem to follow the same anticorrelation as supergiant systems. 
This can be explained by a similar accretion mechanism (from the polar wind of their Be companions,
instead from the Be decretion disc). The other Be systems (transient ones) do not follow the same trend as supergiant systems, 
since they accrete from completely different winds (the Be decretion disc),
forming, in most cases, an accretion disc around the NS. 
The last panels in Fig.~\ref{lsfig:avlx} reports on the behavior of the hard X--ray average luminosity 
with the orbital eccentricity (on the left) and with the pulsar spin period (on the right). 
There is no strong correlation between luminosity and orbital eccentricity (note that for transient sources 
the luminosity plotted here is in outburst).
The forth graph in Fig.~\ref{lsfig:avlx} indicates an anticorrelation, the hard X--ray analogue of the one already 
discussed about the soft X--ray, maximum, luminosity \citep{Stella1986}.
Moreover, it confirms what found by \citet{Cheng2014} in Be/XRBs, where systems with spin period shorter than 40~s 
show more luminous (Type II) outbursts than other Be/XRBs.
This appears to be a confirmation of the warping mechanism, suggested to produce the giant outbursts in Be sources.

We note here that \citet{Lutovinov2013} studied the properties of the population of persistent HMXBs in the Milky Way with \inte, investigating their luminosity function and the spatial density distribution over the Galaxy. The authors used the 9 year \inte\ Galactic plane survey by \citet{Krivonos2012}, focusing only on the persistent, wind-fed, SgHMXBs. 
On one side, this resulted in a sample that does not include all the transient HMXBs studied here, but on the other, given the different aims of the two works, their flux-limited sample had a sensitivity that is about one order of magnitude better than ours (F$>$10$^{-11}$~erg~cm$^{-2}$~s$^{-1}$ compared with a few 10$^{-10}$~erg~cm$^{-2}$~s$^{-1}$ for our \inte\ survey). In this respect, our sample of persistent wind-fed, accreting SgHMXBs with neutron stars represents a sub-sample of the HMXBs studied by Lutovinov et al. (2013).

\subsubsection{ DR$_{1-10\,keV}$   and other source properties}

%%%%%%%%%%%%%%%%%%%%%%%%%%%%%%%%%%%%%%%%%%%%%%%%%%%%%%%%
\begin{figure*}
\centering
\begin{tabular}{cc}
\includegraphics[height=6.5cm, angle=0]{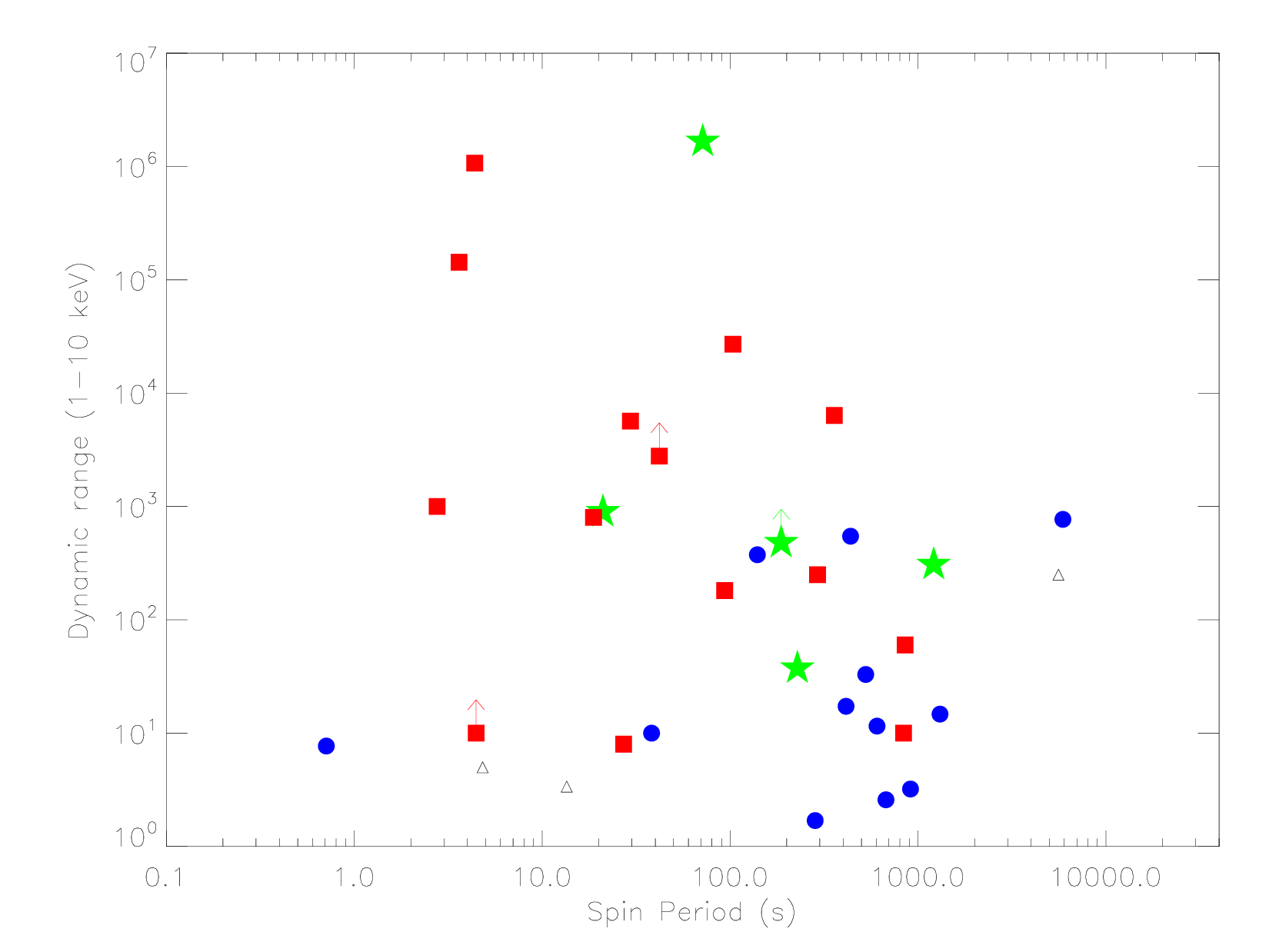} & 
\includegraphics[height=6.5cm, angle=0]{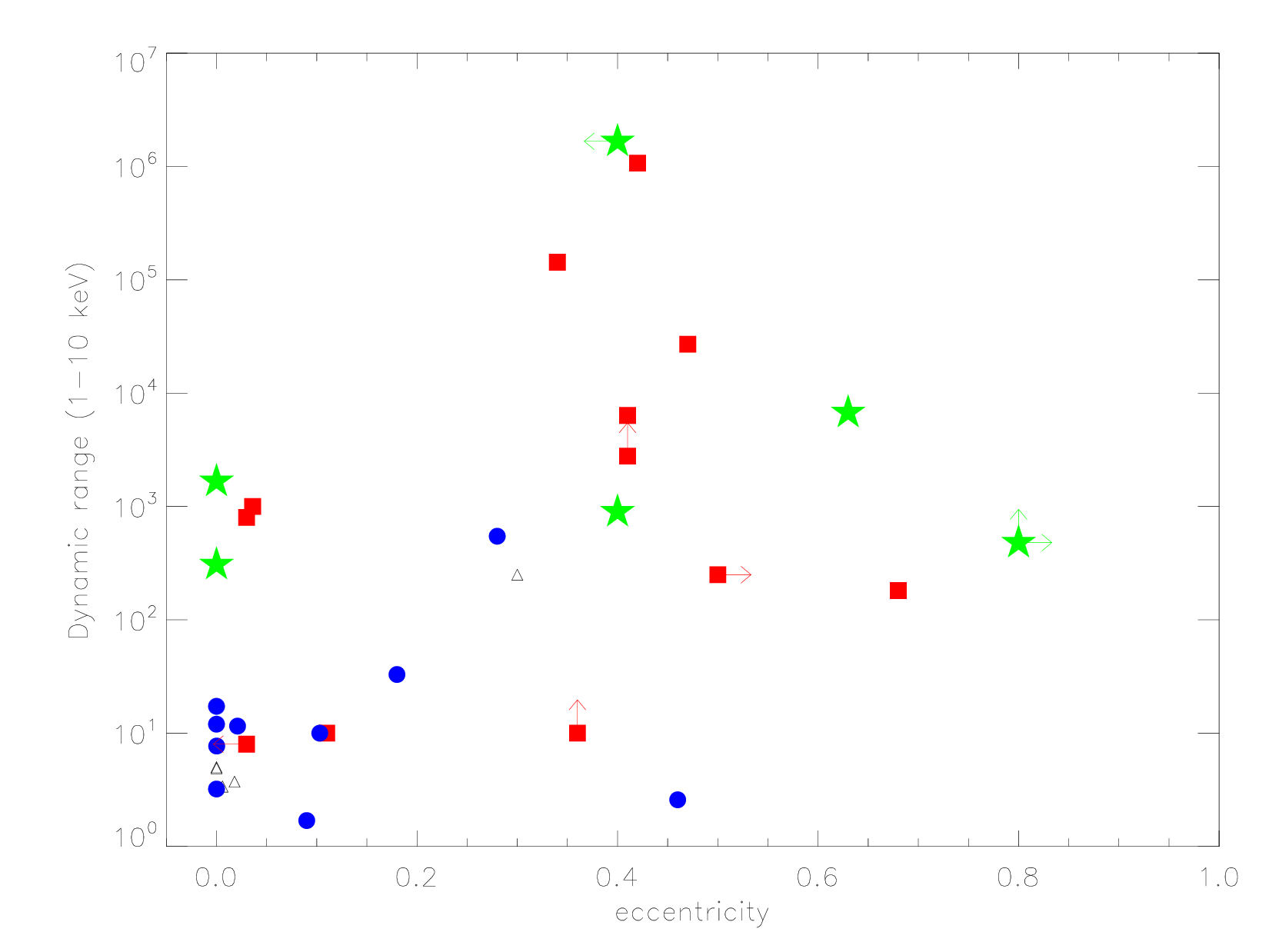} \\
\includegraphics[height=6.5cm, angle=0]{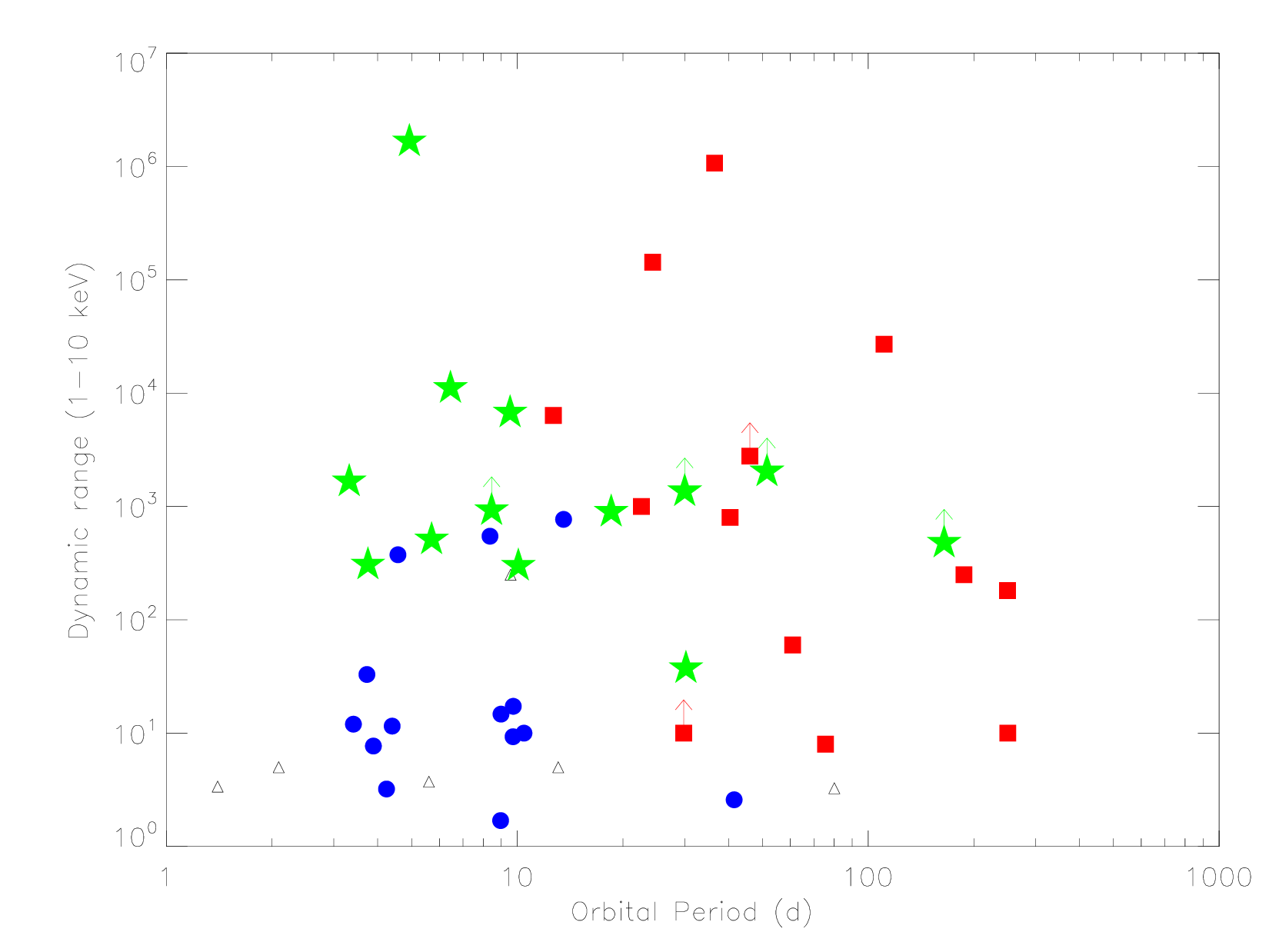} &  
\includegraphics[height=6.5cm, angle=0]{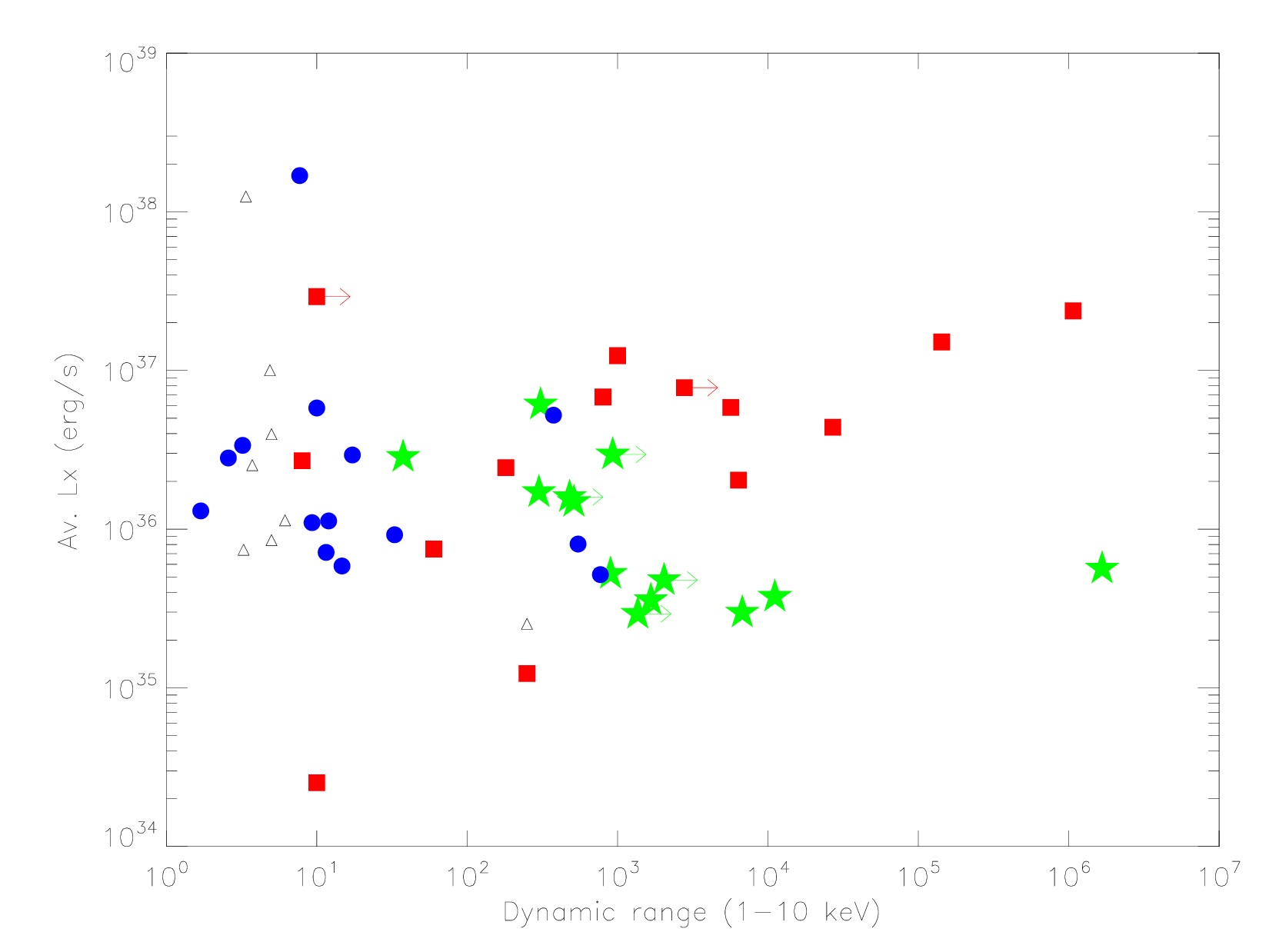}  
\end{tabular}
\caption{Properties of the sources in our sample, presented in Table~\ref{tab:inte} and Table~\ref{tab:literature}.
The meaning of the symbols is the same as in Figs.~\ref{lsfig:dyn_1_10_dist}  and \ref{lsfig:dc}. 
}
\label{lsfig:dyn_1_10}
\end{figure*}
%%%%%%%%%%%%%%%%%%%%%%%%%%%%%%%%%%%%%%%%%%%%%%%%%%%%%%%%
%
In Fig.~\ref{lsfig:dyn_1_10} we show the behaviour of the variability amplitude in soft X--rays (DR$_{1-10\,keV}$) depending on other 
source properties. 
The dependence of  DR$_{1-10\,keV}$ from the spin period appears twofold: sources with a very 
low variabiliy are the RLO systems (SMC~X--1, in blue; Cen X--3 and LMC~X--4, marked by empty triangles),
together with the low-eccentric Be/XRB XTE~J1543--568 and the SgHMXB OAO1657--415 (that alternates disc to wind accretion). All other systems seem to show an
anticorrelation with the pulsar rotational period (although note that the spin period of the SFXT IGR~J17544-2619, $\sim$71~s, is uncertain).
The largest range of variability is shown by sources with a significant orbital eccentricity, 
while for what concerns the orbital period there is no evidence for a trend, as a whole. However, different sub-classes
occupy different regions of the DR$_{1-10\,keV}$--P$_{orb}$ plane: this simply reflects the 
fact that, already mentioned before, most SgHMXBs have shorter orbital periods than Be/XRBs, while $\sim$40 per cent 
of SFXTs have orbital periods larger than 10~days, and overlap with both regions.
When the average hard X--ray luminosity is  plotted against the DR$_{1-10\,keV}$, the SgHMXBs are the less variable sources, 
with an average luminosity $\sim$10$^{36}$~erg~s$^{-1}$,
while the transient ones, with a dynamic range larger than 10$^3$, divide into two completely different regions: 
the less luminous SFXTs (with an average hard X-ray luminosity in outburst
around $\sim$5$\times$10$^{35}$~erg~s$^{-1}$), whereas the Be/XRBs display  luminosities 
higher than 10$^{36}$~erg~s$^{-1}$ (note that these luminosities are average values, while 
during Type II outbursts they reach much larger values). Once again, the average luminosity 
appears a good indicator of the sub-class of HMXBs, for a range of high DR$_{1-10\,keV}$.
Other less variable (low-eccentricity, like X~Per) Be systems overlap with SgHMXBs.

\subsubsection{A global view of HMXBs}

In summary, we are now able to obtain an overview
of the HMXB sample reported in this paper, from the inspection of both  Fig.~\ref{lsfig:dc} and Fig.~\ref{lsfig:avlx} (first panels), where the three 
characterizing quantities of the source duty cycle DC$_{18-50~keV}$, 
the average hard X-ray luminosity (in outburst, for transient sources) and the soft X--ray dynamic range, DR$_{1-10~keV}$ are reported.

We can  characterize the global behaviour of the different sub-classes, as follows: 
\begin{itemize}

\item supergiant (not RLO) HMXBs show low DR$_{1-10~keV}$ ($<$ 40), low average X--ray luminosity ($\sim$10$^{36}$~erg~s$^{-1}$), high source duty cycles (DC$_{18-50~keV}$$>$10 per cent);

\item SFXTs show high DR$_{1-10~keV}$ ($>$100),  low average  X--ray luminosity in outburst ($\sim$10$^{36}$~erg~s$^{-1}$), low source duty cycles (DC$_{18-50~keV}$$<$5 per cent);

\item Be/XRTs show high DR$_{1-10~keV}$ ($>$100), high average  X--ray luminosity in outburs ($\sim$10$^{37}$~erg~s$^{-1}$), intermediate source duty cycles (DC$_{18-50~keV}$$\sim$10 per cent);

\end{itemize}

A few sources do not fit in this picture: the RLO HMXBs (with DR$_{1-10~keV}$$<$10, 
high average X--ray luminosity - reaching the Eddington luminosity,  high DC$_{18-50~keV}$),
the persistent Be system X~Per and the three low-eccentric Be/XRBs; the reasons for their difference are clear and have been already discussed.
However, there are also other systems that do not fit into this scheme, like the very variable SgHMXBs IGR~J18214--1318, IGR~J18027--2016,
 H~1907+097 and IGR~J19140+0951.
Since IGR~J18027--2016 displays both a high DR$_{1-10~keV}$ and a low DC$_{18-50~keV}$, 
overlapping with the range of parameters shown by SFXTs, we propose to re-classify it
as an SFXT. The source IGR~J18214--1318 is detected only a few times by \inte\ and deserves further investigation, 
while the highly variable source  IGR~J19140+0951 is much more clearly characterized.
It can be considered as an intermediate system between persistent SgHMXBs and SFXTs \citep{Sidoli2016}, 
similar to other SgHMXBs that have been already suggested 
in the literature as intermediate systems, as well \citep{Doroshenko2012}.

As discussed by \citet{Negueruela2001c} many years ago, there is a number of massive X--ray binaries  
that does not fit into the traditional division of HMXBs sub-classes. At that time, these authors 
referred to SgHMXBs and Be/XRBs, but we can extend this argument including SFXTs, and the intermediate systems we found 
during our HMXB survey with \inte.
This suggests that, although members of different sub-classes appear to cluster in
different regions of the parameter space, 
a number of massive X--ray binaries display intermediate properties. 
This can also be due  to the properties of the optical counterpart (as in the ``peculiar'' wind-fed system 3A~2206+543).

%%%%%%%%%%%%%%%%%%%%%%%%%%%%%%%%%%%%%%%%%%%%%%%%%%%%%%%%%
\section{Conclusions} \label{sec:concl}
%%%%%%%%%%%%%%%%%%%%%%%%%%%%%%%%%%%%%%%%%%%%%%%%%%%%%%%%%

We performed the analysis of 14 years of  \inte\  observations (18--50 keV; bin time of 2~ks) of a sample of 58 HMXBs 
belonging to different sub-classes (SgHMXBs, SFXTs, Be/XRBs, plus a few systems which do not fit into these three types).
This \inte-driven sample represents about a half of the total number of HMXBs in our Galaxy \citep{Liu2006}.
We extracted integrated quantities (where the temporal information is lost) from this long-term dataset, to obtain an overall, representative and quantitative 
view of their phenomenology at hard X--rays.

Then, we collected from the literature the published values for the source properties (distance, orbital period and eccentricity, pulsar spin period) together with
their minimum and maximum flux (1--10\,keV), from which we have calculated their ratio. 
This led to the compilation of an updated catalogue of the dynamic ranges at soft X--rays for the HMXBs in our sample.
We used all this information to characterize the sources and to disentangle between the properties of each sub-class.

The main results of our investigation can be summarized as follows:

\begin{itemize}

\item We built the long-term hard X-ray CLDs of HMXBs, from which it is possible to quantify, for each source, 
the percentage of time spent in different luminosity states,
the limiting luminosity (18--50\,keV), the duty cycle of their activity DC$_{18-50~keV}$ (transient versus  persistent behaviour), 
the variability amplitude when observed by \inte.
Moreover:

\begin{enumerate}

\item the CLDs show  different shapes in the three types of massive X--ray binaries:
power-law-like in SFXT flares, unimodal (and differently skewed) in supergiant and giant HMXBs hosting NSs (included RLO systems), 
multi-modal in BeXRTs (where the existence of different kind of outbursts, Type I vs Type II, can be immediately recognized 
from the step-like shape of their CLD); \\

\item in SgHMXBs (plus giant systems) the shape of the CLDs appears  steeper for sources
 with high DC$_{18-50~keV}$ and high median luminosity, and 
flatter at lower DC$_{18-50~keV}$ and lower luminosity. In fact, the skewness of the 
luminosity distributions shows an anti-correlation with the median luminosity 
and a correlation with the spin period, with the fastest SgHMXB pulsars residing in the most luminous sources, 
where the luminosity distribution is more symmetric and peaked. 
This is likely due to  disc-fed versus wind-fed  accretion. A by-product of this investigation is that the SgB[e] IGR~J16318-4545 
shows a CLD similar to  wind-fed SgHMXBs.

\end{enumerate}

%---------------------
\item We found that  the members of the three sub-classes  (SFXTs, SgHMXBs and Be/XRTs) tend to cluster around different regions defined by the following parameters: 
DR$_{1-10~keV}$, DC$_{18-50~keV}$, average 18--50\,keV luminosity (in outburst for transient sources).
In particular:

\begin{enumerate}

\item supergiant (not RLO) HMXBs show: DR$_{1-10~keV}$$<$40, DC$_{18-50~keV}$$>$10 per cent and L$_{18-50~keV}$$\sim$10$^{36}$~erg~s$^{-1}$ \\

\item SFXTs show:  DR$_{1-10~keV}$$>$100, DC$_{18-50~keV}$$<$5 per cent and low average X--ray luminosity during flares around L$_{18-50~keV}$$\sim$10$^{36}$~erg~s$^{-1}$  \\

\item Be/XRTs show:  DR$_{1-10~keV}$$>$100, DC$_{18-50~keV}$$\sim$10 per cent and a high average X--ray luminosity in outburst around L$_{18-50~keV}$$\sim$10$^{37}$~erg~s$^{-1}$  \\
 
\end{enumerate}

We note however that one of the results of the present study is that a number of sources shows intermediate properties, 
suggesting  smoother edges (gradual transitions?) between the three currently known HMXBs sub-classes.
With the term ``transition'' we do not allude to any  evolutionary meaning, here.
For instance, some sources classified in the literature as SgHMXBs appear to overlap with SFXTs, showing  
low DC$_{18-50~keV}$ and high DR$_{1-10~keV}$ (IGR~J18027-2016 is a striking example), 
suggesting a more appropriate re-classification.
There are also some classical systems, like the SgHMXBs GX~301-2 or H~1907+097, that show peculiar properties with respect to SgHMXBs
(as already reported in the literature). More examples are discussed in Sect.~\ref{sec:disc}. \\

%---------------------
\item Albeit the \inte-driven source sample, we would like to note here that in the plot of the eccentricity versus the orbital period 
(already discussed in previous literature), 
we were able to add the values of some SFXTs,  allowing supergiant systems to extend at larger eccentricities and orbital periods. \\

%---------------------
\item We performed a thorough research in the literature and compiled the most up to date catalogue of 
published properties of the  HMXBs in our sample, including: distance, orbital period, eccentricity, spin period, 
self-consistently extrapolated  minimum and maximum 1--10\,keV flux and dynamic range, 
as well as the complete list of relevant references.

\end{itemize}
%---------------------

In conclusion, this study about HMXBs puts together the long-term \inte\ public archive spanning 14 years, 
the soft X--ray fluxes collected from the literature, and other important source properties (spin periods, orbital geometries), 
offering an interwoven overview of these sources from an observational - bird's eye -  point of view.

%%%%%%%%%%%%%%%%%%%%%%%%%%%%%%%%%%%%%%%%%%%%%%%%%%%%%%%%%
\section*{Acknowledgments}
%%%%%%%%%%%%%%%%%%%%%%%%%%%%%%%%%%%%%%%%%%%%%%%%%%%%%%%%%

%----------
Based on observations with \textit{INTEGRAL}, an ESA project
with instruments and science data centre funded by ESA member states
(especially the PI countries: Denmark, France, Germany, Italy,
Spain, and Switzerland), Czech Republic and Poland, and with the
participation of Russia and the USA. 
We acknowledge financial contribution from ASI/INAF n.2013-025.R1 contract and from PRIN-INAF 2014 grant 
``Towards a unified picture of accretion in High Mass X-Ray Binaries'' (PI Sidoli).
This research has made use of the NASA's Astrophysics Data System to 
access the scientific literature, of the \textsc{simbad} database, operated at CDS, Strasbourg, France,
and of the software  \textsc{WebPIMMS} provided by the High Energy Astrophysics Science Archive Research Center (HEASARC), 
which is a service of the Astrophysics Science Division at NASA/GSFC and the 
High Energy Astrophysics Division of the Smithsonian Astrophysical Observatory.
We are grateful to A.J. Bird for the useful discussions.
We thank the anonymous referee for the careful reading and constructive report that greatly helped improve the manuscript.

\bibliographystyle{mn2e} 
\bibliographystyle{mnras}

\bsp

\label{lastpage}

\end{document}